\documentclass{aa}
\usepackage{graphicx}
\usepackage{lscape}
\usepackage[varg]{txfonts}
\usepackage[colorlinks=True,linkcolor=blue,citecolor=blue, urlcolor=blue]{hyperref}
\usepackage{natbib,twoopt}
\usepackage{comment}
\usepackage{caption}

\bibpunct{(}{)}{;}{a}{}{,}             

\newcommand{\hii}{\ion{H}{ii}\xspace}
\newcommand{\htco}{H$\rm _2$CO\xspace}
\newcommand{\phtco}{\textit{p}$-$H$\rm _2$CO\xspace}%

\newcommand{\tkin}{$T_{\rm kin}$\xspace}

\newcommand{\nh}{$n\rm(H_2)$\xspace}
\newcommand{\cd}{$N$(\phtco)\xspace}
\newcommand{\fad}{$X$(\phtco)\xspace}
\newcommand{\hcd}{$N\rm(H_2)$\xspace}

\newcommand{\kms}{$\rm km\,s^{-1}$\xspace}

\newcommand{\lal}{($\rm3_{0,3}-2_{0,2}$)\xspace}

\begin{document} 
	
	\title{Exploring the interplay between molecular and ionized gas in \hii regions} 
	
	\author{S.~Khan	\inst{1,\thanks{Member of the International Max Planck Research School (IMPRS) for Astronomy and Astrophysics at the Universities of Bonn and Cologne.}},
        A.~M.~Jacob \inst{2,1},
        M.~R.~Rugel \inst{3,4},
        J.~S.~Urquhart \inst{5},
        S.~Neupane \inst{1},
        F.~Wyrowski \inst{1},
        A.~Brunthaler \inst{1},
        J.~D.~Pandian\inst{6},
        Y.~Gong \inst{7},
        I.~Barlach Christensen \inst{1} \and K. M. Menten\inst{1}\fnmsep\thanks{Deceased}
        }
	
	\institute{Max-Planck-Institut f\"{u}r Radioastronomie, Auf dem H\"{u}gel 69, 53121 Bonn, Germany \\
		\email{skhan@mpifr-bonn.mpg.de}
        \and
	    I. Physikalisches Institut, Universit\"{a}t zu K\"{o}ln, Z\"{u}lpicher Str. 77, 50937 K\"{o}ln, Germany
        \and
        Deutsches Zentrum f\"{u}r Astrophysik, Postplatz 1, 02826 G\"{o}rlitz, Germany
        \and
        National Radio Astronomy Observatory, P.O. Box O, 1003 Lopezville Road, Socorro, NM 87801, USA
		\and 
        Centre for Astrophysics and Planetary Science, University of Kent, Canterbury, CT2 7NH, UK
		\and
        Department of Earth $\&$ Space Sciences, Indian Institute of Space Science and Technology, Trivandrum 695547, India
		\and
        Purple Mountain Observatory, and Key Laboratory of Radio Astronomy, Chinese Academy of Sciences, 10 Yuanhua Road, Nanjing 210023, PR China
    }

	\date{Received ; accepted }
	
	
	\abstract
	{Massive stars strongly impact their natal environments and influence subsequent star formation through feedback mechanisms including shocks, outflows and radiation. \hii regions are key laboratories for studying this impact. To understand such feedback, it is crucial to characterize the physical conditions of the dense molecular gas in which these regions are embedded.}
	{We aim to constrain the kinetic temperature and $\rm H_2$ volume density of massive star-forming clumps associated with \hii regions using multiple \phtco transitions. In addition, we investigate the interplay between ionized gas, molecular gas, and dust to probe how massive stars influence their parental clumps. } 
    {We observed the $J_{K_{\rm a}K_{\rm c}}$ transitions of \phtco (within its $J$ = 3$-$2 and 4$-$3 states) with the Atacama Pathfinder EXperiment (APEX) 12\,m submillimeter telescope using the nFLASH230 and SEPIA345 receivers towards a sample of 61 \hii regions. Spectral line parameters are derived via multi-component Gaussian fitting, which was then used to constrain the physical conditions determined using PyRADEX, a non–local thermodynamic equilibrium (LTE) radiative transfer code in combination with Markov Chain Monte Carlo (MCMC) analysis.}
	{The non-LTE analysis yielded kinetic temperatures (\tkin) ranging from 33.7\,K to 265\,K and $\rm H_2$ densities (\nh) between $\rm 0.8 \times 10^{4}$ to $\rm 1.05 \times 10^{7}\,cm^{-3}$, providing a detailed characterization of the dense molecular gas contained in these clumps. In addition to the \phtco emission arising from the targeted clump a large fraction (57\%) of the sources exhibit multiple \phtco components, with the secondary components being characterized by higher \tkin and broader linewidths. Investigating the nature of the secondary component revealed its association with supersonic non-thermal motions and turbulent gas. When comparing the physical properties of the molecular gas and dust components with those of the ionizing gas, we find that parameters directly linked to the central high-mass star such as bolometric luminosity ($L\rm _{bol}$) and Lyman continuum photon rate ($N\rm _{Lyc}$), show stronger and more systematic correlations. Emphasizing the role of the central star in governing the interplay between the molecular and ionized gas. In our sample of \hii regions, the pressure of the neutral gas systematically exceeds that of the ionized gas. This suggests that the surrounding neutral molecular medium can hinder or slow down the expansion of \hii regions due to its higher pressure. However, given the limited spatial resolution concluding the role of molecular gas in confining \hii regions is still pending high resolution observations. }
	{}
	
	\authorrunning{S. Khan et al.}
	\titlerunning{Ionized and molecular gas associated with \hii regions}
	\maketitle
    \nolinenumbers
        \section{Introduction} \label{sec:intro}
        High-mass stars ($>8~M_\odot$) are fundamental drivers of galactic evolution and dynamics through their strong interaction with the surrounding interstellar medium (ISM). These stars play a crucial role in regulating the galaxy's energy balance through feedback mechanisms, including strong stellar winds, and intense ionizing ultraviolet (UV) radiation throughout their lifetimes, as well as powerful supernova explosions at the end of their life cycles. The UV photons emitted by high-mass stars have enough energy to ionize the surrounding neutral hydrogen, resulting in the formation of \hii regions. Because these stars have short lifetimes, of only a few million years \citep[see][and reference therein]{2018ARA&A..56...41M}, and are deeply embedded in their natal molecular clouds, \hii regions observed via their radio free-free emission or radio recombination lines (RRLs) serve as strong indicators of recent high-mass star formation (HMSF) activity in the galaxy.

        As \hii regions evolve, they expand under the combined influence of internal pressure, stellar radiation, and winds from central OB stars. This expansion can compress the surrounding molecular gas, potentially triggering secondary star formation along the ionization fronts \citep{2010A&A...523A...6D, 2011EAS....51...45E, 2012MNRAS.421..408T}. Young, embedded \hii regions profoundly affect their environments through feedback mechanisms such as shocks, outflows, and radiation, thereby shaping the conditions for subsequent star formation \citep{1977ApJ...214..725E, 2003A&A...399.1135D, 2011EAS....51...45E}. However, the extent to which these regions modify the initial conditions or trigger new episodes of HMSF within their parental molecular clouds remains uncertain. A comprehensive understanding of these processes requires accurate measurements of the molecular gas temperature and density near expanding \hii regions, along with the physical characteristics of the associated ionized gas and the relationships between them.

        In the past, extensive radio continuum and recombination-line surveys have unveiled a large number of Galactic \hii regions
	    \citep[e.g.,][]{1970A&A.....4..357R,1979A&AS...35...23A,1989ApJS...71..469L,1997ApJ...488..224K,2007A&A...461...11U,2009A&A...501..539U,2013MNRAS.435..400U,2011ApJS..194...32A,2014ApJS..212....1A,2018A&A...615A.103K,2019A&A...623A.105G}. Recently, the GLObal view on STAR Formation in the Milky Way survey \citep[GLOSTAR\footnote{\url{https://glostar.mpifr-bonn.mpg.de/glostar/}};][]{2019A&A...627A.175M,2021A&A...651A..85B} cataloged a large population of new and known \hii regions using highly sensitive Karl G. Jansky Very Large Array (VLA) continuum data in D-- \citep{2019A&A...627A.175M, 2024A&A...689A.196M} and B--configuration \citep{2023A&A...670A...9D, 2023A&A...680A..92Y} at spatial resolutions of 18\arcsec~and 1\arcsec, respectively. \citet[][hereafter Paper I]{2024A&A...689A..81K}, compiled a catalog of Galactic \hii regions detected with GLOSTAR RRLs data observed with the VLA in D-configuration and analyzed their ionized gas properties. This study revealed that approximately 50\% of these \hii regions are associated with sources identified in the Atacama Pathfinder EXperiment (APEX) Telescope Large Area Survey of the GALaxy \citep[ATLASGAL; ][]{2009A&A...504..415S}, which provided a comprehensive view of HMSF at 870\,$\mu$m-- a wavelength particularly sensitive to cold dust emission tracing dense star-forming clumps. While the ATLASGAL survey determined properties of the dust, the gas density and temperature of the clump play a crucial role in regulating their chemistry and star formation activity, as well as influencing the stellar initial mass function. A precise measurement of the dense gas density and temperature in the vicinity of \hii region are essential to deepen our understanding of how \hii regions impact their parental molecular clouds and surrounding  environments.
        
        \htco, which has been ubiquitously detected in the ISM, is primarily formed via the hydrogenation of CO on dust grain surfaces \citep{2020A&A...634A..52S} and is subsequently released into the gas phase through processes such as UV irradiation or shocks. As a result, it is commonly associated with star-forming and \hii regions \citep{1980A&AS...40..379D, 1983A&A...127..388H, 2011A&A...532A.127D, 2011ApJ...736..149G, 2023A&A...678A.130G}, where its abundance remains relatively stable throughout different stages of the star formation cycle \citep{1990ApJ...348..542M, 2018A&A...611A...6T}. Due to its large dipole moment \citep[2.33\,Debye;][]{1977JChPh..67.1576F}, \htco exhibits rotational transitions across the centimeter and sub-millimeter wavelength range \citep{mangum}. Spectroscopically the rotational states of this asymmetric and prolate molecule are described by the total angular momentum and its projection quantum numbers, $J_{K_{\rm a}K_{\rm c}}$. Furthermore, due to the nuclear spin of the hydrogen nuclei, \htco exists in two spin isomers: ortho ($K_{\rm a} =$ odd) and para ($K_{\rm a} =$ even). However, one of the most valuable properties of \htco for astrophysical studies is its characteristic `$K$-doublet' splitting, which arises from the coupling between vibrational and rotational angular momentum states. In its rotational spectrum, this results in closely spaced pairs of spectral lines (or `K-doublets') with opposite parity, rather than a single transition.
       
        With advancements in receiver technology and increased access to sub-mm windows, astronomers have expanded the utility of \htco as a diagnostic tool for characterizing the physical conditions in molecular gas, even using its higher lying sub-mm transitions \citep[][ and references therein]{2018A&A...609A..16T,2018A&A...611A...6T,2021A&A...655A..12T}. These properties make \htco one of the very few molecular species that act as both a thermometer and densitometer for dense gas, particularly in environments where other tracers like NH$\rm _3$ or CH$\rm _3$CN may be less accessible. For these reasons observations of the `K-doublet' transitions of \htco serve as a great tool to study the physical properties of the molecular medium and allow observations with a variety of radio telescopes. 

        To summarize, detailed study of a large sample of embedded \hii regions and surrounding molecular clouds are required to probe the underlying physical processes, physical properties of the ambient dense molecular cloud into which they are expanding, and the effect of HMSF on the parent molecular clouds. 

        To measure the kinetic temperature and gas density of massive star-forming clumps identified in the GLOSTAR \hii region and ATLASGAL surveys we here analyze the rotational transitions of \phtco ($J$ = 3--2 and 4--3). Using these results, we aim to: (a) explore possible correlations between the characteristics of molecular gas, ionized gas, and dust in these regions; (b) examine the temperature and density structure of molecular gas surrounding \hii regions; and (c) investigate whether the pressure of the surrounding medium is sufficient to confine \hii regions. This paper is structured as follows: Sect.~\ref{sec:observation} describes the source sample studied and the selection criteria used, the \htco observations, and data reduction process. In Sect.~\ref{sec:method}, we outline the methods used to analyze and model the observed spectral line profiles of the \phtco transitions studied. Section~\ref{sec:result} presents the results, followed by a discussion in Sect.~\ref{sec:discussion}. Finally the main conclusions are summarized in Sect.~\ref{sec:conclusion}.

        \section{Sample, observations and data reduction} \label{sec:observation}
        \subsection{Source selection}\label{sec:source_selection}
        We identified 244 Galactic \hii regions in Paper~I, using detected stacked (H98$\alpha$--H114$\alpha$) RRL within the GLOSTAR survey coverage ($-$2$\degr \leq$ $\ell$ $\leq$ 60$\degr$ \& |\textit{b}| $\leq$ 1$\degr$ and 76$\degr \leq$ $\ell$ $\leq$ 83$\degr$ \& $-$1$\degr \leq$ \textit{b} $\leq$ 2$\degr$). These authors derived physical properties of identified \hii regions such as the electron temperature, electron density, emission measure, to characterize the associated ionized gas. By comparing the \hii region catalog with dust continuum data, it was found that nearly half of the identified \hii regions are spatially associated with dust clumps from the ATLASGAL survey. \citet{2018MNRAS.473.1059U, 2022MNRAS.510.3389U} previously classified these clumps as \hii regions and provided key dust properties such as the bolometric luminosities ($L_\mathrm{bol}$), the clump masses ($M_\mathrm{clump}$), and the dust temperatures ($T_\mathrm{dust}$). Such a combined dataset offers insights into both the ionized gas and dust components associated with \hii regions. This study extends the analysis in Paper~I by investigating in addition to the ionized gas and dust, the properties of the molecular gas surrounding \hii regions and exploring the their interplay. 
        
        The targets for this study are selected based on the following criteria: (1) \hii regions with the ionized gas properties already derived (as cataloged in Paper~I) (2) association with 870\,$\mu$m continuum emission taken from ATLASGAL catalog, with dust properties derived in \citet{2018MNRAS.473.1059U, 2022MNRAS.510.3389U} (3) deconvolved full width half maximum (FWHM) of the \hii region (from Paper I) is comparable to or smaller than that of the corresponding ATLASGAL clump (4) has well characterized kinematics. We excluded \hii regions located toward the Galactic center ($-$2$\degr \leq$ $\ell$ $\leq$ 2$\degr$) and the Cygnus~X complex due to significant uncertainties in their kinematic distances. 
        
        These criteria ensure that the targeted \hii regions are likely to remain embedded within their natal molecular clouds, enabling an investigation of the physical link between the ionized and molecular components during massive star formation. Based on these criteria, we selected 61 \hii regions from Paper I for the \phtco spectral line observations. Figure~\ref{fig:rgb_image} presents three--color mid--infrared composite images (8, 24, and 70\,$\mu$m) with contours marking the GLOSTAR 5\,GHz radio continuum emission, highlighting the selected \hii regions in this study.  
        
        Among them, only G30.720$-$0.082 is unresolved in the GLOSTAR RRL data at 25\arcsec~resolution (see Fig.~\ref{fig:atlas_hii_size}). Six sources (G7.177+0.088, G10.440+0.011, G24.470+0.490, G25.159+0.060, G43.169+0.008, and G43.149+0.013) exhibit extended radio emission surrounding compact bright cores, resulting in sizes larger than their associated ATLASGAL clumps (see Fig.~\ref{fig:atlas_hii_size}). We retained these sources in the sample to preserve diversity in the population. The selected regions span Galactic longitudes from 3$\degr$ $\leq$ $\ell$ $\leq$ 53$\degr$, encompassing several prominent high-mass star-forming complexes, including W31, W33, W43, and W51.

\begin{figure}
            \includegraphics[width=0.9\linewidth]{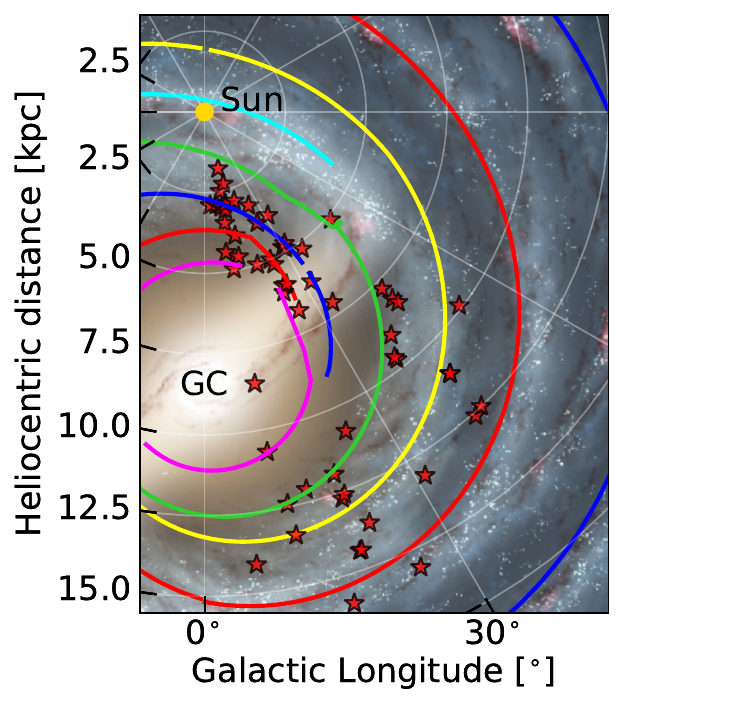}
            \caption{Galactic distribution of the source sample, where the locations of the targets are denoted by red stars on an artistic rendering of the Milky Way. The distances to the observed \hii regions and their associated clumps are taken from Paper I (and references therein). The spiral arms are marked following the description presented in \citet{2019ApJ...885..131R} where the individual spiral arms are colored as follows: the 3 kpc in magenta, the Norma-Outer arm in red, the Scutum-Centaurus arm in blue, the Sagittarius-Carina arm in green, the local arm in cyan, and the Perseus arm is in yellow.}
                \label{fig:dist}
        \end{figure}

        Figure~\ref{fig:dist} presents the Galactic distribution of the sources in this study, overlaid on an artist's impression of the Milky Way\footnote{\url{https://science.nasa.gov/photojournal/tracing-the-arms-of-our-milky-way-galaxy/}}. The sources span distances from 1.8 to 20.4\,kpc, with values for each \hii region and its associated clump taken from Paper I (and references therein). The distribution reveals two distinct populations: one clustered near 4\,kpc and another around 12\,kpc. Figure~\ref{fig:size_component} shows the distributions of the angular and physical sizes of the selected \hii regions. The angular sizes are concentrated, with a median of 15.6\arcsec, reflecting our selection criterion based on source angular size. The physical size distribution, with a median of 0.52\,pc, is more scattered, due to scatter in distance. This scatter in distance introduces no significant bias in the \hii region properties but rather ensures a diverse and representative sample in terms of size and spectral type of central OB stars of \hii region as discussed in Appendix~\ref{sec:dist_eff}. 

        \subsection{APEX 12\,m observations of \htco}
        The single pointing observations were carried out from March to April 2023 using the Atacama Pathfinder EXperiment \citep[APEX;][]{2006A&A...454L..13G}\footnote{This publication is based on data acquired with the APEX telescope (\url{https://www.apex-telescope.org/ns/}). APEX telescope is operated and maintained by the Max-Planck-Institut f\"{u}r Radioastronomie, Bonn, Germany.} 12\,m sub-mm telescope located in the Chajnantor plateau, Chile. Table~\ref{tab:transitions} summarizes the key spectroscopic parameters of the observed \phtco transitions alongside the receiver setups used and the detection rates within the sample. All three transitions of \phtco within the $J$=3--2 states near 218\,GHz were observed using the dual sideband nFLASH230 receiver. At this frequency, the observations were carried out with a half-power beam width of $\sim$27$\arcsec$. In addition to these \phtco transitions, the receiver was tuned such that it also covered the H30$\alpha$ line at 231~GHz in its upper sideband. Total integration times between 7\,mins and 16\,mins were spent in this setup. The nFLASH230 receiver was connected to a MPIfR-built high-capacity fast Fourier transform spectrometer \citep[FFTS;][]{2012A&A...542L...3K} backend with two sidebands. Each sideband has two spectral windows with 4\,GHz bandwidth, providing both orthogonal polarizations and resulting in a total bandwidth of 8\,GHz. This provided a spectral resolution of 61\,kHz, resulting in a velocity resolution of 0.08\,\kms at 218\,GHz. 
        
        The three transitions of \phtco within the $J$=4$-$3 levels were observed with the SEPIA345 receiver \citep{2022A&A...668A...2M} with a beam size of $\sim$20$\arcsec$ and integration times between 8\,mins to 20\,mins. Like the nFLASH230 receiver, the SEPIA345 receiver is a dual polarization, dual sideband receiver which covers a total frequency range from 272\,GHz to 376\,GHz. Each sideband (and polarization) of the SEPIA345 receiver output is recorded using two FFTS spectrometer units, each of them sampling 4\,GHz, with a spectral resolution of up to 64\,k channels per sideband, this results in velocity resolutions of 0.04\,\kms. Both sets of observations were performed in position-switching mode with off positions offset from the on position of the sources by ($\pm$600\arcsec, $\pm$600\arcsec) in ($\ell,b$).
        
        The data reduction was performed using the CLASS software from the GILDAS package\footnote{\url{https://www.iram.fr/IRAMFR/GILDAS}}. The spectra were converted from antenna temperature ($T_{\rm A}$) scales to main-beam brightness temperature ($T_{\rm MB}$) scales using a factor of ${\rm \eta_{F}}/\eta_{MB}$, where ${\rm \eta_{F}}$ is the forward efficiency and $\rm \eta_{MB}$\footnote{\url{https://www.apex-telescope.org/telescope/efficiency/?yearBy=2022}} is the telescope main beam efficiency. In our observations, ${\rm \eta_{F}}$ was set to 0.95 and $\rm \eta_{MB} =$ 0.79 $\pm$ 0.07 for the data collected using both nFLASH230 and SEPIA354 (based on observations of Mars). To enhance the signal-to-noise ratios (S/N) in individual channels for both sets of observations, the data were smoothed to a uniform channel spacing of 0.6\,\kms. The typical line width of \htco transitions toward \hii regions tends to be of the order of a few \kms, so the channel smoothing has no adverse impact on our results.

        \begin{table*}[ht!]
	    \caption{Summary of the \phtco transitions observed in this study. }
	    \label{tab:transitions}
	    \centering
	    \begin{tabular}{cccccc}
		    \hline \hline
		    Transition & Frequency & $E_{\rm u}$ & Receiver & HPBW & Detection rate \\
		      \phtco  & [GHz]     & [K]     &          &   [\arcsec] &  [\%]\\
		    \hline \hline

            $\rm3_{0,3}-2_{0,2}$ & 218.222&20.96&nFLASH230&27.63&100 (61/61)\\
            $\rm3_{2,2}-2_{2,1}$ & 218.476&68.09&nFLASH230&27.60&85 (51/61)\\
            $\rm3_{2,1}-2_{2,0}$ & 218.760&68.11&nFLASH230&27.56&86 (52/61)\\
            $\rm4_{0,4}-3_{0,3}$ & 290.623&34.90&SEPIA345&20.75&98 (60/61)\\
            $\rm4_{2,3}-3_{2,2}$ & 291.237&82.07&SEPIA345&20.70&83 (51/61)\\
            $\rm4_{2,2}-3_{2,1}$ & 291.948&82.12&SEPIA345&20.65&85 (52/61)\\
   		\hline
	    \end{tabular}
        \tablefoot{The columns are from left to right: The spectral line transitions studied, their rest frequencies, upper level energies, $E_{\rm u}$, the receiver used, its Half Power Beam-width (HPBW) and the detection rate (i.e., above 3$\sigma$ of rms noise of 15 $\sim$ 20\,mK). The frequencies and spectral line parameters are taken from the Cologne Database for Molecular Spectroscopy \citep[CDMS;][]{2001A&A...370L..49M, 2005JMoSt.742..215M, 2016JMoSp.327...95E}.}
    \end{table*}

        \section{Analysis}\label{sec:method}
        In this section, we describe the multi--component Gaussian fitting of the \phtco spectral lines observed in this study. We then use the resulting line-fitting parameters to constrain the non--local thermodynamic equilibrium (non--LTE) analysis and derive the kinetic temperature (\tkin), gas density \nh, and the \phtco column density \cd. 

        \subsection{Spectral line fitting}\label{sec:line_properties}
        
           We modeled the observed spectral line features using Gaussian profiles that were fitted using the MODEL function within the LMFIT\footnote{\url{https://lmfit.github.io/lmfit-py/}} package in Python, that provided as output: the velocity-integrated line intensity ($I = \int{T_{\rm mb}dv}$), the local standard of rest (LSR) velocity ($\upsilon_{\rm LSR}$), the FWHM of the Gaussian line, and the peak brightness temperature on main-beam temperature scales ($T_{\rm peak}$).

           We began our analysis by fitting single-component Gaussian to the \phtco spectra. However, this approach yields well constrained physical parameters for only 25 out of 61 targets (40\%). To improve the modeling, we performed multi-component Gaussian fits to better model the observed \phtco line profiles. As the brightest transition, \phtco ($\rm3_{0,3}-2_{0,2}$) was detected toward all sources and served as the reference for identifying the number of velocity components and their centroid velocities. These centroid velocities were then fixed for all other \phtco transitions to ensure consistency across lines. Each \phtco ($\rm3_{0,3}-2_{0,2}$) spectrum was initially fitted with one Gaussian component, if the residuals are larger than 3$\sigma$, a second or even a third component was added, until the standard deviation in the residual is minimized. We verified all fits interactively using the \texttt{LINE/MINIMIZE} task in \texttt{CLASS/GILDAS}. To verify that the \htco lines exhibit multiple components and are not affected by optical depth effects like self-absorption, we compared their spectra with those of the optically thin C$\rm^{18}$O line. We did not find any evidence of self-absorption in our sample. We found no preferential occurrence of multiple components in sources located at either near or far distances.
           
\begin{figure*}
                \centering
                \includegraphics[width=1\linewidth]{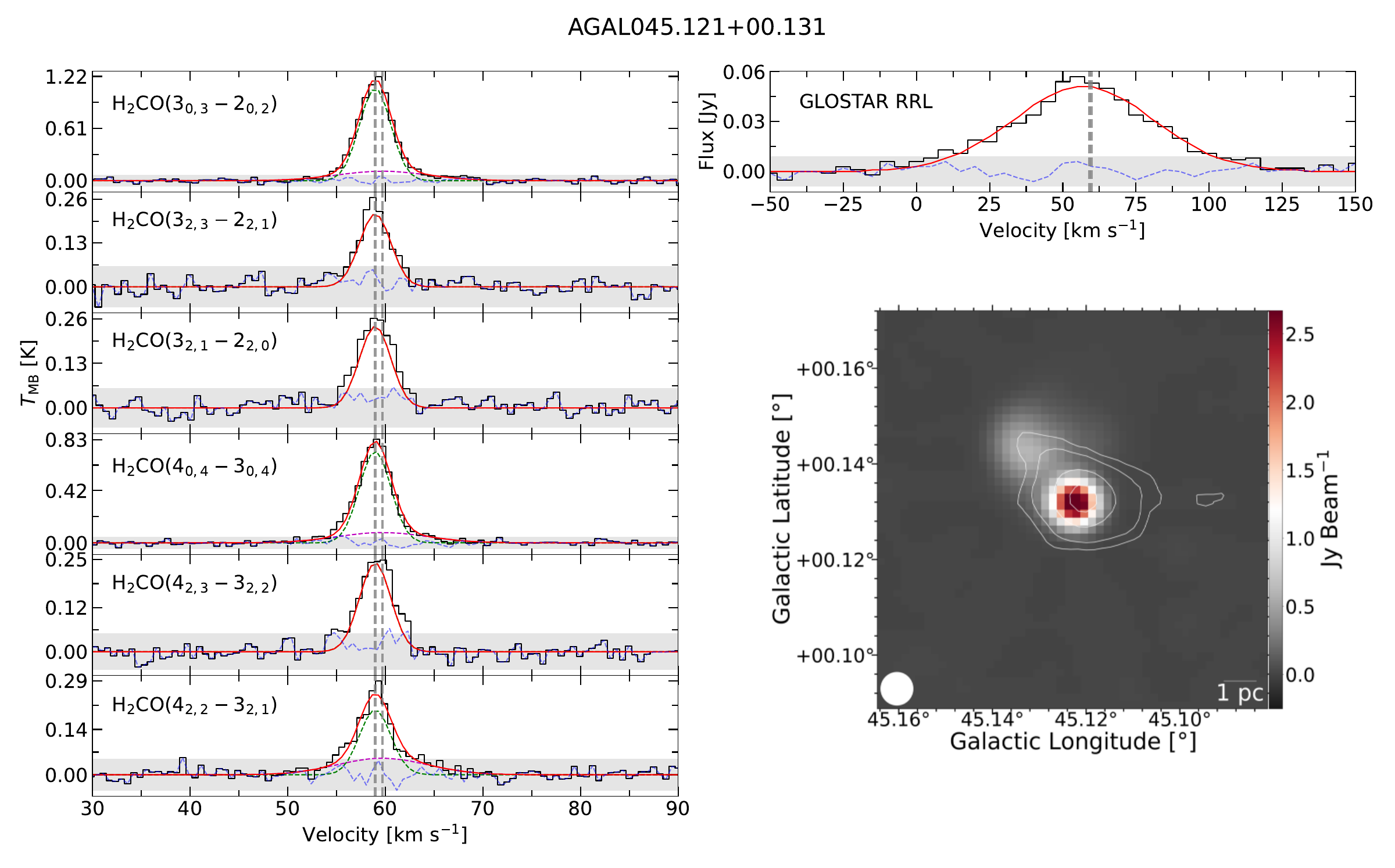}
                \caption{Spectra of the different \phtco and RRL transitions studied in this work toward AGAL045.121+00.131. \textit{Left panels:} Spectra of all six of the observed \phtco transitions (black) along with the multi-component Gaussian fitting (green and magenta) and residual of the fitting (dashed blue). The total fit is displayed in red and the gray shaded region demarcates the 3$\sigma$ level of the rms noise. The vertical dashed gray lines mark the centroid velocities of each fitted Gaussian component. \textit{Right panel:} Spectra of the stacked GLOSTAR $\langle$Hn$\alpha \rangle$ RRL in black alongside the Gaussian fit in red. Also displayed is the 5.3~GHz radio continuum emission map obtained with the GLOSTAR VLA D-configuration toward this source and overlaid with white contours is the ATLASGAL 870\,$\mu$m. The white filled circle shows the 25$\arcsec$ GLOSTAR beam.}
                \label{fig:plot_eg}
            \end{figure*}

           Among the 61 observed sources, the \phtco spectra toward 34 of them displayed multiple Gaussian components, resulting in a total of 105 identified components in the entire sample. Since all six transitions are expected to trace gas with similar kinetic temperature and dynamic motion, we expect them to have similar line widths. However, this need not always be true especially when fitting additional components arising from surrounding envelopes or outflows. Therefore, the line widths were allowed to vary within 1\,$\sigma$ of the value constrained by the linewidth of the $\rm3_{0,3}-2_{0,2}$ line of \phtco, whereas the velocity centroids of the components were fixed. Any component with peak brightness temperatures below 3$\sigma$ level of the rms noise at 0.6\,\kms or line width less than 3$\times\Delta v$ (1.8\,\kms), is flagged before further analysis. An example of the \phtco multi-component Gaussian fitting is shown in Fig.~\ref{fig:plot_eg} (left panel) toward one of the targets in our study, AGAL045.121+00.131 (also listed as G045.122+0.131 in the GLOSTAR \hii region catalog).

           If a source exhibits multiple velocity components, then the component associated with the source at a \phtco velocity close to the systemic velocity of the corresponding ATLASGAL clump -- determined from multiple molecular line surveys and reported by \citet[][see their Table 1]{2018MNRAS.473.1059U} -- is considered the main component and will hereafter be referred to as `\htco Main'. All secondary components are then considered as sub-components and will hereafter be referred to as `\htco Sub'. These secondary components may either be associated with the source itself via outflows/inflows, envelops, or may arise from distinct molecular clumps covered in our beam. However, owing to the limited spatial resolution, the exact origin of the `\htco\ Sub' components remains uncertain. 
           
            The derived line parameters for \phtco are then used as input to constrain the column densities and the gas densities. However, as we observed two different sets of transitions ($J = 3$--2 and $J=4$--3) at different frequencies (218\,GHz and 291\,GHz) using two different receivers (nFLASH230 and SEPIA345), with varying HPBWs 27.6\arcsec~and  20.7\arcsec, the observations  resulted in slightly different spatial coverages. Therefore, to ensure uniformity, we further correct for beam filling effects.  Previous studies of \htco \citep{2014A&A...572A..63I, 2018A&A...609A..16T, 2018A&A...611A...6T} have demonstrated a tight correlation between the integrated intensities of the \phtco transitions and flux density of the dust emission at 870\,$\mu$m. This suggests that the dense gas traced by \htco is associated with dust emission at 870\,$\mu$m. Based on these findings, we assume that the size of the \htco emitting area toward the targets of this study are equivalent to the FWHM source sizes of 870\,$\mu$m dust emission, as derived by \citet{2014A&A...565A..75C}. Following \citet{2018A&A...611A...6T}, we corrected for beam dilution as follows,  $T_{\rm MB}/\eta_{\rm bf}$ where $\eta_{\rm bf}$ is the beam filling factor, given by, $\eta_{\rm bf} = \rm \theta^2_S/(\theta^2_S +\theta^2_{beam})$. Here, $\rm \theta_{S}$ and $\rm \theta_{beam}$ denote the source and beam size, respectively. The corrected \phtco line intensities with the respective beam filling factors used and other line fitting parameters are summarized in Tables~\ref{tab:nflash_line} and \ref{tab:sepia_line}.

            \subsection{Non-LTE modeling with Pyradex}\label{sec:radex}
\begin{figure}
                \centering
                \includegraphics[width=1\linewidth]{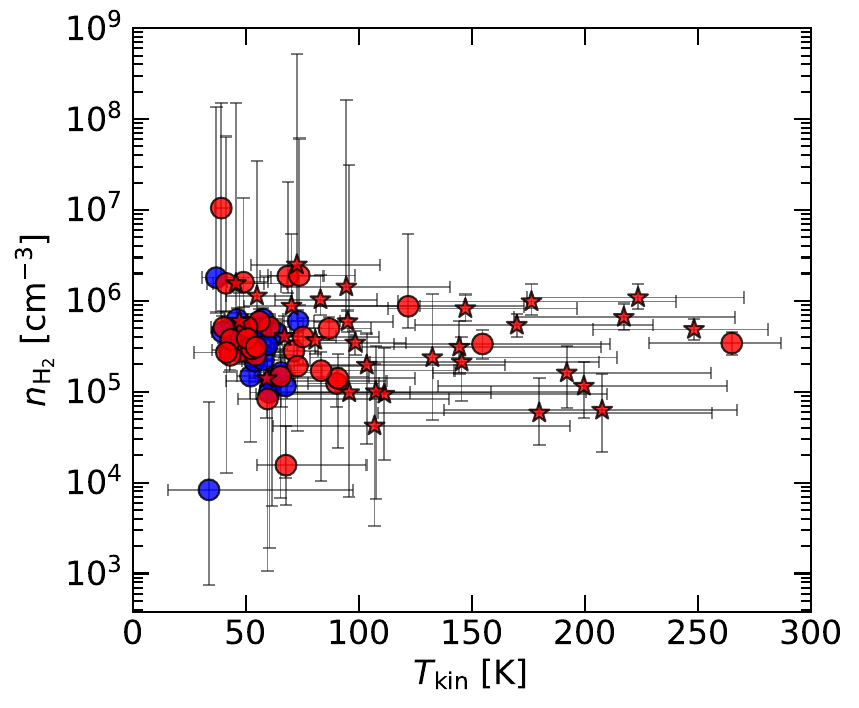}
                \caption{Distribution of the sources in the \nh–\tkin plane. Blue and red color represent sources with single and multiple component Gaussian profiles, respectively. Filled circles denote components classified as `\htco Main', while stars indicate those identified as `\htco Sub'. }
                \label{fig:nh2_tkin_space}
            \end{figure}

            We carried out non-LTE radiative transfer calculations to derive the physical properties of the clumps, including the gas kinetic temperature, \tkin, spatial density, \nh, and column density of \phtco, \cd. We employed the Python wrapper, PyRADEX, to interface with the non-LTE radiative transfer code RADEX \citep{2007A&A...468..627V}, in combination with the emcee\footnote{\url{https://emcee.readthedocs.io/en/stable/}} \citep{2019JOSS....4.1864F} package in Python, which implements a Markov Chain Monte Carlo (MCMC) algorithm as shown in \citet{2017A&A...608A.144Y} and Christensen et al. (in prep.). To reproduce the observed integrated line intensities of the varied \phtco transitions (as determined in the previous Sect.~\ref{sec:line_properties}), the PyRADEX models were run with constraints from the observed line widths using collisional rate coefficients computed for collisions between \phtco and \textit{ortho-}, and \textit{para}-H$_2$ \citep{2013MNRAS.432.2573W} and assuming the ortho-to-para ratio (OPR) of 3. For each velocity component toward a given source, the line width was fixed to the average value across the different \phtco transitions probed. This approach minimizes variations and allows us to simultaneously reproduce the physical conditions traced by all six transitions-- a valid constraint, since all transitions are expected to originate from the same gas. We further assumed a background temperature, $T_{\rm bkg}$, of  2.73\,K. We initialized the MCMC process using the \texttt{curve\_fit} function from the SciPy\footnote{\url{https://scipy.org/}} package in Python, then explored the parameter space within  \nh = 10 -- 10$^{\rm 10}$\,cm$^{\rm-3}$, \tkin = 10 -- 300\,K and \cd = 10$^{\rm 10}$ -- 10$^{\rm 17}$\,cm$^{\rm-2}$. For each component, we used 400 walkers, discarding the first 100 steps as burn-in and allowing the final 1000 steps to converge on the best-fit integrated line intensity (see also Christensen et al., in prep.). Using this method, we were able to successfully constrain the physical parameters toward 83 of the 105 targets  studied in this work. Figure~\ref{fig:corner_plot_eg} illustrates the results of the PyRADEX and MCMC computation, displaying the results in the form of a corner plot with one-dimensional (1D) histograms of the posterior distributions across the explored parameter space. Table~\ref{tab:phy_prop} lists the derived gas kinetic temperatures, volume densities, and column densities of \phtco. Figure~\ref{fig:nh2_tkin_space} displays the distribution of the sources in the \nh–\tkin parameter space. The `\htco Main' components of sources with single and multiple Gaussian profiles show a similar distribution, while the extreme values are associated with the `\htco Sub' components. The only exceptions are G31.412+0.308, which exhibits the highest \tkin, and G10.462+0.033, which shows the highest \nh. 

	\section{Results} \label{sec:result}
        In the following section, we present the results obtained from the joint PyRADEX and MCMC analyses described in Sect.~\ref{sec:radex}. These results include the derived kinetic temperature, gas density, \phtco column density, and the fractional abundance of \phtco, \fad, which we compare with values from previous studies. We also examine the optical depth of \phtco and assess its impact on our analysis (Appendix~\ref{sec:optical_depth}). 

            \begin{figure*}
                \centering
                
                \includegraphics[width=0.45\linewidth]{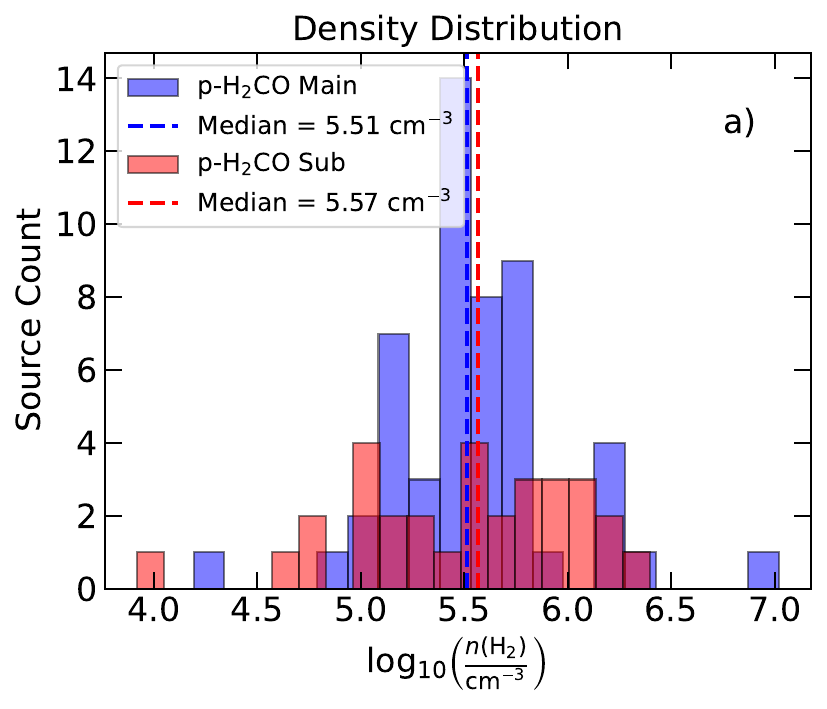}
                \includegraphics[width=0.45\linewidth]{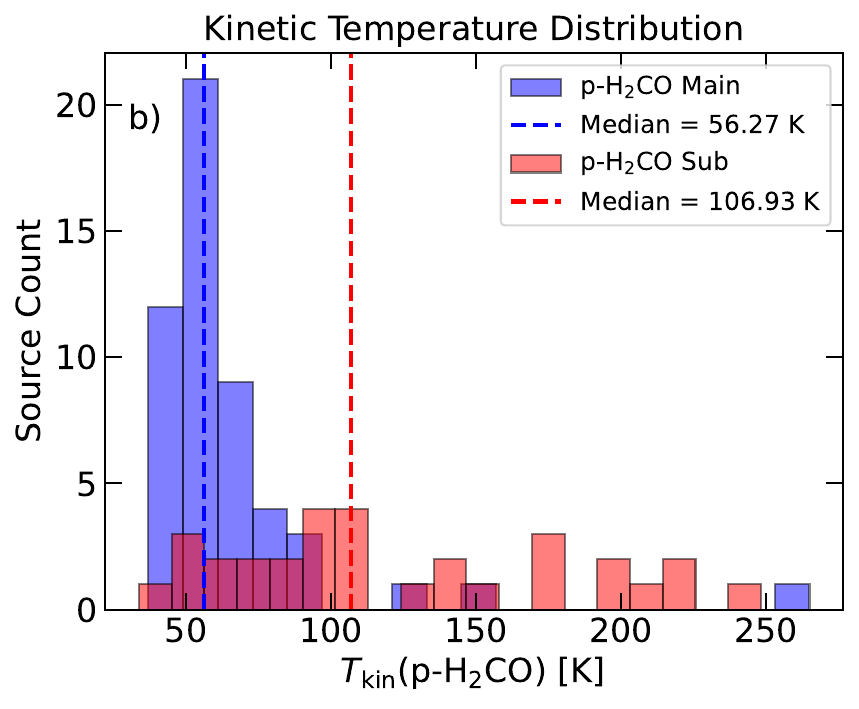}
                \includegraphics[width=0.45\linewidth]{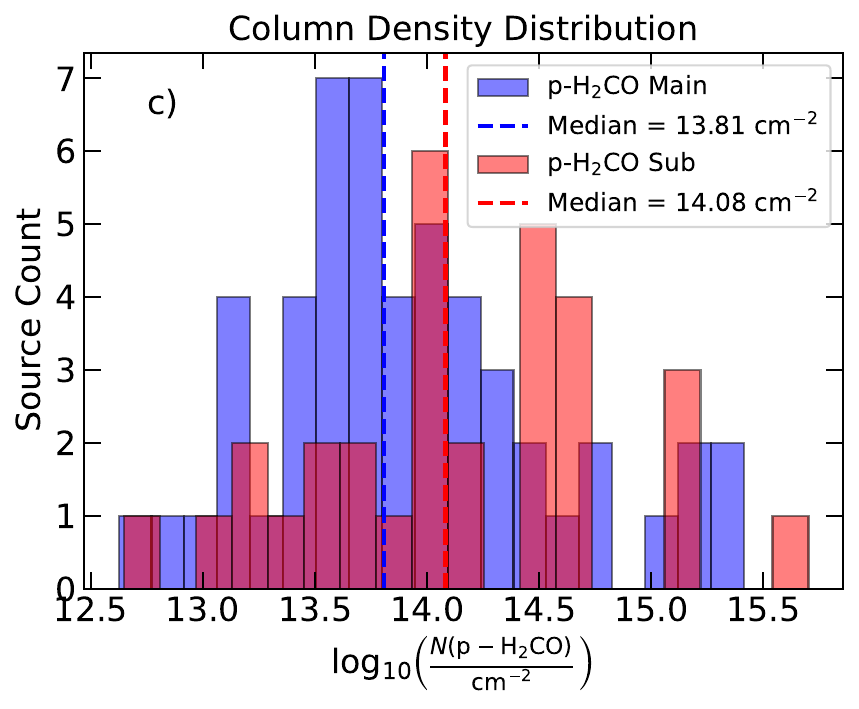}
                \includegraphics[width=0.45\linewidth]{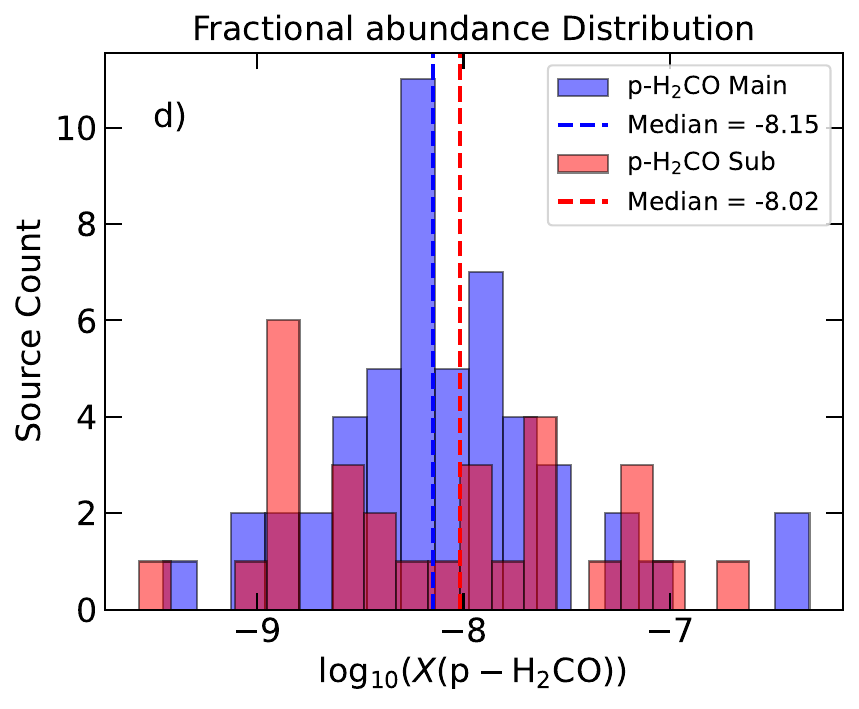}
                \caption{Distributions of the pyRADEX+MCMC derived (a) \nh, (b) $T_{\rm kin}$, (c) \cd, and (d) \phtco fractional abundance, \fad toward the different velocity components identified in our sample, as traced by the \phtco. The blue and red histograms represent the distributions for the `\htco Main' and `\htco Sub' components, respectively, while the corresponding dashed lines indicate the median values of each distribution.}
                \label{fig:hist}
            \end{figure*}
        Figure~\ref{fig:hist} shows the distributions of gas kinetic temperatures, gas densities, and \phtco column densities, jointly constrained from the PyRADEX and MCMC analyses discussed above while the results toward each source are listed in Table~\ref{tab:phy_prop}. In the following sections, we discuss these results in detail. 

        \subsection{Gas densities} \label{sec:dens}
\begin{figure}
                \centering
                \includegraphics[width=1\linewidth]{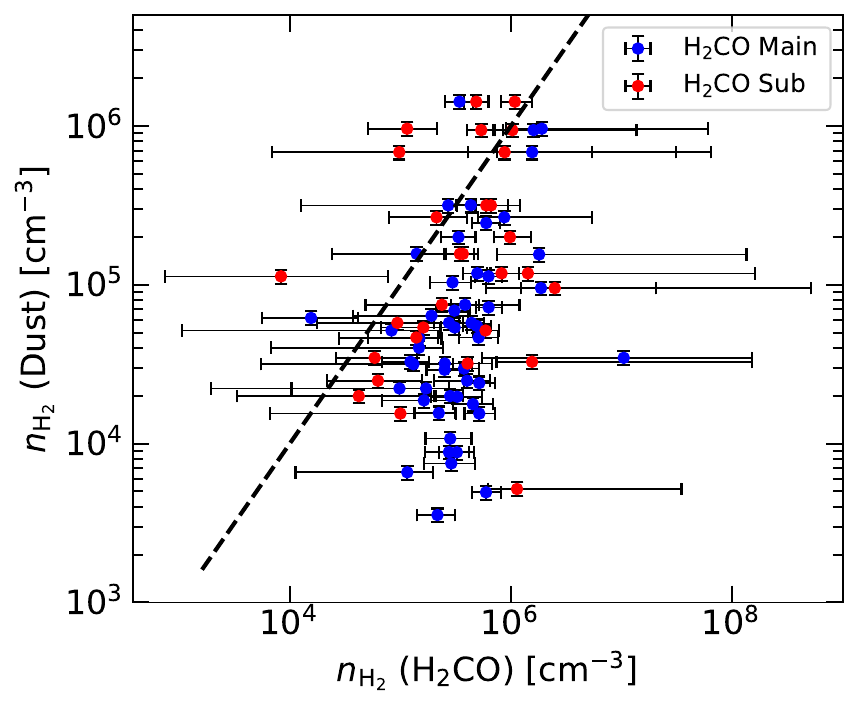}
                \caption{Comparison between \nh probed by 870\,$\mu$m emission from dust and \phtco as derived from our pyRADEX+MCMC models. The dashed black line displays a slope of unit value.}
                \label{fig:dust_vs_h2co_nH2}
            \end{figure}

        As shown by the distributions in Fig.~\ref{fig:hist} for our sample, the derived \nh values range between $\rm 8.32\,\times\,10^{3}$~cm$^{-3}$ and $\rm 1.05\,\times\,10^{7}\,cm^{-3}$ with a median\footnote{We report median values toward our sample throughout the paper, as the mean would be misleading given results at extremities that skew the distributions. This is most clearly seen in the distributions of the RADEX derived \tkin values as displayed in Fig.~\ref{fig:hist} (b).} value of $\rm (3.7\,\pm\,0.35)\,\times\,10^5$\,cm$^{-3}$. Our results are consistent with previous studies of star-forming regions \citep{mangum,2011ApJ...742...58M,2013ApJ...766..108M}, including the brightest ATLASGAL clumps associated with \hii regions, where densities of $6.4\,\times\,10^5$ -- $\rm 8.1\,\times\,10^6 \,cm^{-3}$ were reported, with a median value of $\rm 1.25\,\times\,10^5\,cm^{-3}$ \citep{2018A&A...611A...6T}. Furthermore, a comparison of our results with \nh values determined from 870\,$\mu$m dust emission \citep{2022MNRAS.510.3389U} reveals systematically higher densities within our sample (Fig.~\ref{fig:dust_vs_h2co_nH2}), indicating that these \phtco transitions trace denser gas than the dust continuum. This holds true for both the systemic velocity components and additional components arising from the envelopes or other expanding shells of these \hii regions. Figure~\ref{fig:hist}~(a) shows that the median gas density of the `\htco Sub’ component similar to that of the `\htco Main’.

        \subsection{Kinetic temperature} \label{sec:tkin}
           
        As with \nh and \cd, the PyRADEX+MCMC models yielded a range of gas temperatures across the sources in our sample, from 33.7\,K to >265\,K\footnote{For the components of AGAL031.412+00.307, the PyRADEX+MCMC did not converge to gas temperatures within the explored temperature range, resulting in lower limits set by the maximum gas temperature at which the collisional rate coefficients are computed.}, with a median value of $\rm 66\,\pm\,3$\,K. In contrast, the study by \citet{2018A&A...611A...6T}, which examined ATLASGAL dust clumps at various evolutionary stages, reports a median kinetic temperature of 67\,K across their entire sample, increasing to 104\,K for sources associated with only \hii\ regions. To assess whether the difference in kinetic temperature derived for the \hii region sample by \cite{2018A&A...611A...6T} and this work are statistically significant, we applied the Anderson-Darling (AD) test to the two distributions. The resulting $p$-value of approximately 0.001 indicates a statistical difference between the kinetic temperature distributions, providing evidence that \tkin do not arise from the same parent population. This discrepancy in kinetic temperatures may be due to the fact that we decompose the observed spectral features into multiple components likely originating from different environments, while \citet{2018A&A...611A...6T} carry out their analysis by considering only the integrated line intensities across the entire observed features. The latter may lead to higher line intensity ratios in the analysis conducted by \citet{2018A&A...611A...6T}, resulting in higher overall gas temperatures. Furthermore, we find the values of \tkin derived for the `\htco Sub' components (98.4\,K) to be higher than that for the `\htco Main' components (57.8\,K). This suggests that the `\htco Sub' components may originate from inflowing/outflowing gas surrounding the star, or from shock-heated molecular gas. Moreover, the lack of correlation between the kinetic temperature of the `\htco Sub' component and its velocity difference with both the `\htco Main' component and the RRL emission further supports a shock-heated origin for the `\htco Sub' components. 
        
        The \tkin\ values derived from the PyRADEX+MCMC analysis can be verified using the integrated intensity ratios between specific \phtco transitions, namely, $R_1 = I( 3_{21}-2_{20})/I(3_{03}-2_{02})$, and $R_2 = I( 4_{22}-3_{21})/I(4_{04}-3_{03})$ -- which provide an estimate of the \tkin assuming conditions of LTE. This method is applicable if the observed lines are optically thin (discussed further in Appendix~\ref{sec:optical_depth}) and originate from high-density regions. Following the approach outlined in Appendix A of \citet{mangum}, we computed the \tkin using these line ratios:
            \begin{equation}\label{eq:lte_temp1}
                T_{\rm LTE}(R_1) = \frac{47.1}{\ln{\left(0.556\frac{I(3_{03}-2_{02})}{I(3_{21}-2_{20})}\right)} } \; \text{K} \, ,
            \end{equation}
            and,
            \begin{equation}\label{eq:lte_temp2}
                 T_{\rm LTE}(R_2) = \frac{47.2}{\ln{\left(0.750\frac{I(4_{04}-3_{03})}{I(4_{22}-3_{21})}\right)} } \; \text{K} \, .
            \end{equation}            
            
\begin{figure}
                \centering
                \includegraphics[width=0.95\linewidth]{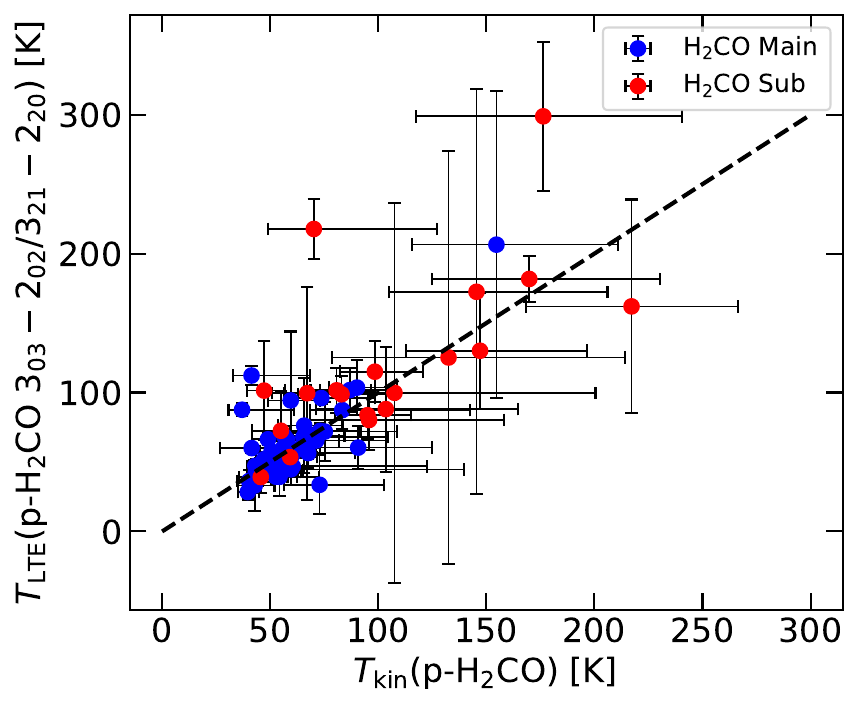}
                \includegraphics[width=0.95\linewidth]{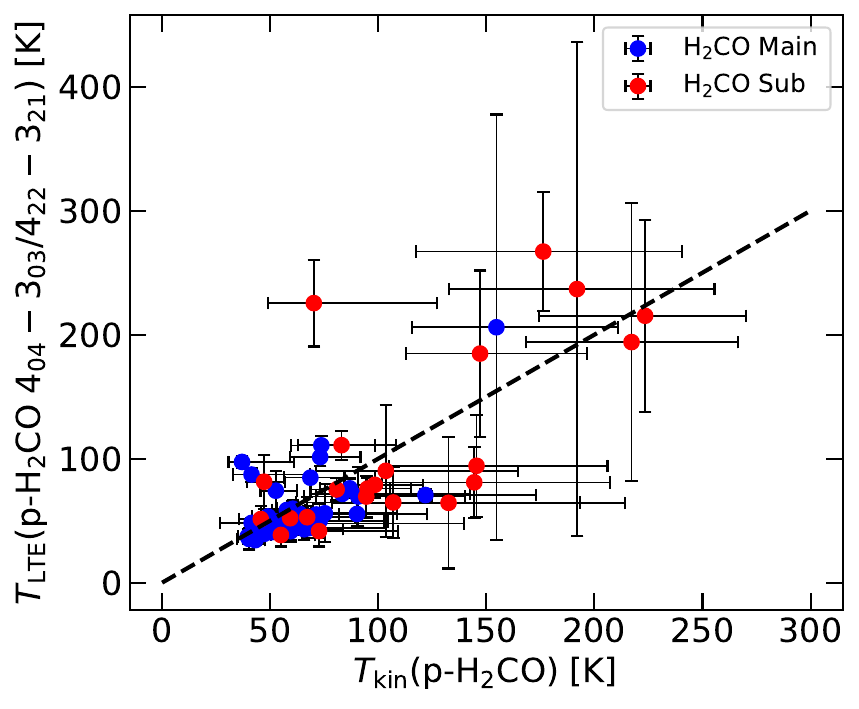}
                \includegraphics[width=0.95\linewidth]{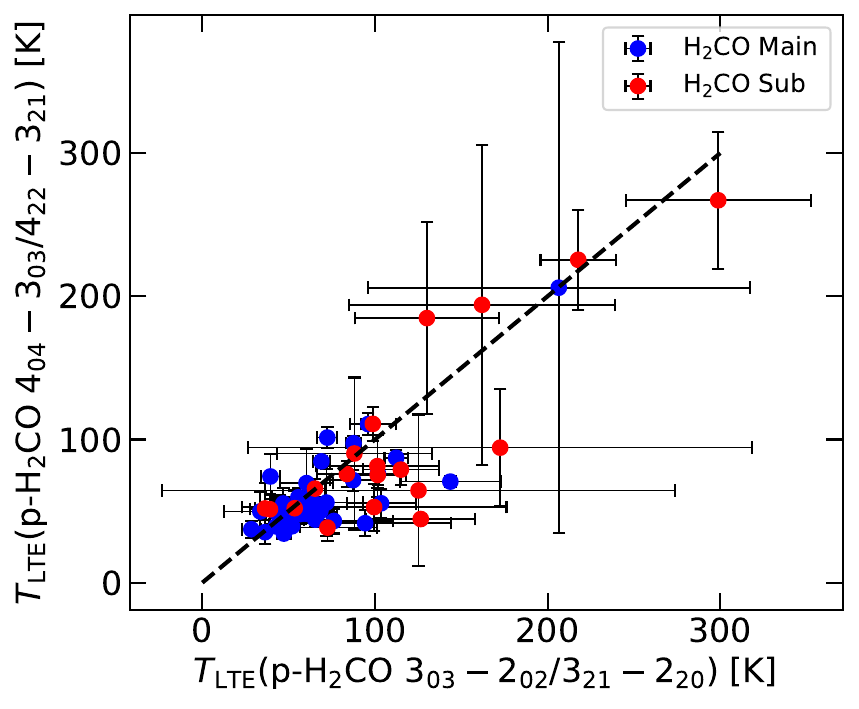}
                \caption{Comparison between the kinetic temperatures derived from pyRADEX+MCMC modeling and using the \phtco integrated line intensities ratios, $R_1$ (top panel) and $R_2$ (middle panel). The bottom panel compares the kinetic temperatures derived from $R_1$ and $R_2$. The LTE temperature uncertainties are obtained from uncertainties in the integrated intensities of the observed \phtco lines. The dashed black lines indicate a slope of unity.}
                \label{fig:lte_vs_nonlte}
            \end{figure}

        In Fig.~\ref{fig:lte_vs_nonlte}, we refer the temperatures derived using Eq.~\ref{eq:lte_temp1} and Eq.~\ref{eq:lte_temp2} as LTE temperatures, and find the results from both line ratios to be consistent with one another. If our assumption that the \phtco emission is optically thin is valid, then the kinetic temperatures obtained by this method have an uncertainty of $\lesssim$30\% \citep{mangum}. Comparisons between \tkin values derived from LTE and PyRADEX+MCMC non-LTE calculations show overall consistency, although some deviations appear at higher temperatures (Fig.~\ref{fig:lte_vs_nonlte}). More interestingly, the derived values of the kinematic temperature are significantly less scattered for the `\htco Main' velocity components than the `\htco Sub' component. Overall, these findings reaffirm the diagnostic power of \htco an excellent molecular thermometer.

        \subsection{Fractional abundance of \phtco}
        In this section, we compute the fractional abundance of \phtco with respect to H$_2$ (\fad = \cd/\hcd), where \hcd\ refers to the H$_2$ column density derived from the 870\,$\mu$m continuum emission by \citet{2018MNRAS.473.1059U}. The PyRADEX derived \cd values lie between $\rm 4\,\times\,10^{12}\,$cm$^{-2}$ and $\rm 5\,\times\,10^{15}\,cm^{-2}$ with a median value of $\rm 1.0\,\pm\,0.1\times\,10^{14}~$cm$^{-2}$ while \hcd is estimated from the peak flux densities of the clumps using:           
            \begin{equation}
               N({\rm H_2})  = \frac{{\rm D^2}\,F_{\rm 870}\,R}{B_{\rm 870}(T_{\rm d})\,\theta_{\rm beam}\,\kappa_{\rm 870}\,\mu_{\rm H_2}\,m_{\rm H}},
            \end{equation}
        where $D$ is the distance to the source; $F_{\rm 870}$ is the peak flux density taken from \citet{2014A&A...565A..75C}, $\theta_{\rm beam}$; the beam size of ATLASGAL survey \citep[= 19\farcs2;][]{2014A&A...565A..75C}, $\mu_{\rm H_2}$ is the mean molecular weight per hydrogen molecule assumed as 2.8, and $m_{\rm H}$ the mass of the hydrogen atom. $R$ is the gas-to-dust ratio assumed to be 100; $B_{\rm 870}(T_{\rm d})$ is the intensity of the blackbody at 870~$\mu$m at the dust temperature, $T_{\rm d}$, values for which are taken from \citet{2018MNRAS.473.1059U}, and $\kappa_{\rm 870}$ is the dust opacity at 870 $\mu$m taken to be 1.85~cm$^{2}$~g$^{-1}$, an average of all dust models presented in  \citet{1994A&A...291..943O}. This yields \hcd values ranging from $\rm 6\,\times\,10^{20}\,$cm$^{-2}$ to $\rm 8\,\times\,10^{22}\,cm^{-2}$ and \fad\ values from $\rm 2.7\,\times\,10^{-10}$ to $\rm 4.7\,\times\,10^{-7}$ (see Fig.~\ref{fig:hist} (d)), with a median value of $\rm 7.2\,\times\,10^{-9}$. These results are consistent with values reported for other star-forming regions \citep{2013A&A...550A.135A,2014A&A...563A..97G,2018A&A...609A..16T,2018A&A...611A...6T}. Figure~\ref{fig:hist}(c) shows that the median \cd of the `\htco Sub’ component is higher than that of the `\htco Main’, whereas the median \fad is similar for both components (see Fig.~\ref{fig:hist}(d)). This suggests that \htco can be at least as abundant --if not more so-- in the envelope/outflow components, underscoring its presence across diverse environments. However, the `\htco Sub' components may actually represent unresolved condensates, since their densities match those of the `\htco Main' components, suggesting that the secondary features likely comprise of multiple dense clouds at closely spaced velocities unresolved in our single dish observations.

   \section{Discussion}\label{sec:discussion}

    \subsection{Correlation between physical properties of ionized gas, molecular gas and dust}
          
\begin{figure}
                \centering
                \includegraphics[width=1.\linewidth]{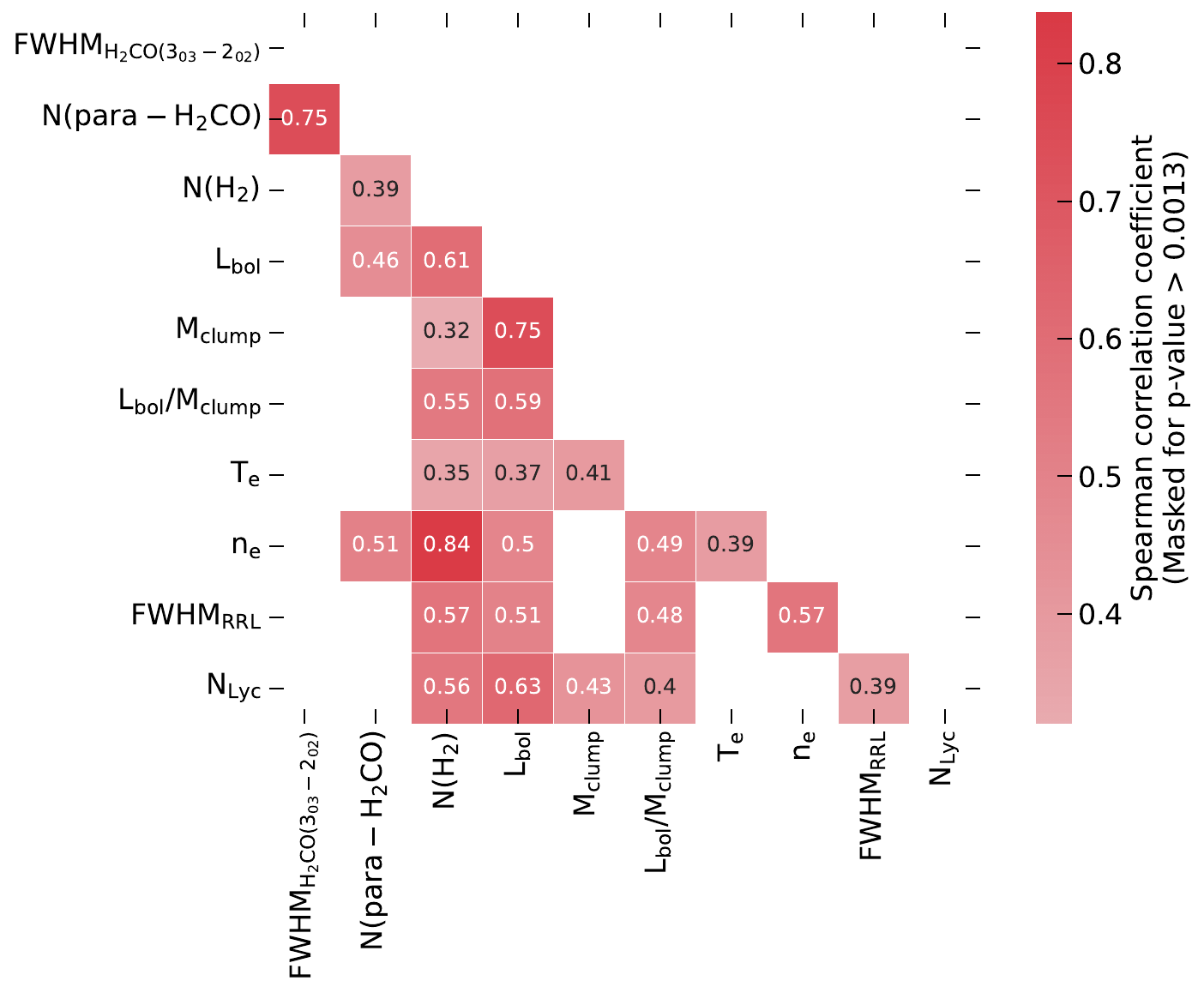}
                \caption{Correlation matrix showing the relationships between various physical properties of molecular gas, ionized gas, and dust clumps. The color bar and annotations indicate the Spearman’s correlation coefficients. We flagged and excluded correlations with $p-$values $>$ 0.0013 (3$\sigma$), thereby removing statistically insignificant relationships.}
                \label{fig:corr_mat}
            \end{figure}

            Variations in the physical properties of clumps and \hii regions associated with star formation offer valuable insights into the evolutionary stages and conditions of HMSF and its impact on the surrounding medium. For instance, over the years, the ratio of bolometric luminosity to clump mass has proven to be a reliable tracer of HMSF stages and has been widely used in numerous studies \citep{2008A&A...481..345M,2016ApJ...826L...8M,2017A&A...603A..33G,2022MNRAS.510.3389U}. Similarly, $n_{\rm e}$ and RRL linewidths have been extensively employed to characterize the evolutionary stages of \hii regions \citep[e.g., ][]{2004ApJ...605..285S, 2005IAUS..227..111K}. In order to investigate the relationship between the ionized, molecular gas, and dust in \hii regions, we compared the Spearman's correlation between the different physical properties of molecular gas traced by \phtco derived in the previous sections with those of dust, and ionized gas from Paper~I. The resulting correlations are presented in Fig.~\ref{fig:corr_mat}. Note that we have flagged the cross correlation for which the $p-$value is $>$ 0.0013.

            No strong correlation was found between \nh and the other ionized, molecular gas and dust properties (masked in Fig.~\ref{fig:corr_mat}). This suggests that, regardless of the specific characteristics of the HMSF clump, their molecular hydrogen densities typically remain around $\rm 10^{5-6}\,cm^{-3}$, in alignment with the findings of \cite{2018A&A...611A...6T}. 
            
            Observations of massive clumps using other molecular line tracers such as CO, NH$_3$, CH$_3$CN, CH$_3$CCH, and CH$_3$OH, \htco \citep{2012A&A...544A.146W, 2017A&A...603A..33G, 2018A&A...611A...6T} suggest that these clumps are heated by radiation from embedded massive stars. This implies that the \tkin values derived from these molecules should correlate with the $L_{\rm bol}$ of the clumps. \cite{2017A&A...603A..33G} reported such correlations using kinetic temperatures derived from CH$_3$CCH, and CH$_3$OH, with correlation coefficients of 0.68 and 0.54, respectively. Similarly, \cite{2018A&A...611A...6T} investigated the same relationship using \htco and found a somewhat weaker correlation, with a coefficient of $\sim$0.5. However, in this study, we find an even weaker correlation between \tkin and $L_{\rm bol}$, with a correlation coefficient of 0.32 (masked in Fig.~\ref{fig:corr_mat} due to $p-$value > 0.0013).
            
            Although the velocities of RRLs and \phtco lines are generally consistent, there is still a velocity offset between the two, with a standard deviation in the velocity offset of $\sim$4\,\kms and mean of --0.5\,\kms. For an \hii region of roughly 0.52\,pc in size, the inferred dynamical age is $t_{\rm dy}$ $\sim$ $\rm 1.3 \times 10^5$\,yr, which is comparable to the typical expansion timescales of \hii regions. Hence, a clump with a velocity difference of 4\,\kms is consistent with a molecular layer being located near the ionization front of the expanding \hii region. Assuming that this velocity difference serves as a proxy for the physical separation between the \hii region and the molecular clump, the correlation between the physical parameters is recomputed for sources where this velocity difference is less than 4\,\kms. For these closely associated sources, we find a stronger correlation between \tkin and $L_{\rm bol}$, with a coefficient of 0.49 ($p-$value < 0.0013). This result suggests that \tkin is more strongly influenced by bolometric luminosity when the molecular clump lies closer to the \hii region.

            The \phtco column density correlates strongly with the \phtco FWHM and more weakly with the clump bolometric luminosity. Similar but stronger trends were reported in massive star-forming regions by \citet{2014A&A...572A..63I} and \citet{2018A&A...611A...6T}. These correlations suggest that the \phtco abundance increases with the radiation field. This would indicate that motions like turbulence and shocks can contribute to the increase of \htco in the gas phase. A natural guess is that such motions help the desorption of \htco from dust grains.

            We also find that \cd correlates with the electron density of \hii regions. The electron density depends on the ionizing photon flux from the central massive star(s), as higher UV photon output generates more free electrons. In addition, the electron density shows a strong correlation with the clump hydrogen column density, \hcd. This indicates that young \hii regions with higher electron densities tend to be formed in clumps with higher $\rm{H_2}$ column densities. The electron temperature, $T_{\rm e}$, shows only a weak correlation with $n_{\rm e}$, with a Spearman’s correlation coefficient of 0.39.

            Bolometric luminosity, $L_{\rm bol}$, exhibits relatively strong correlations with dust, molecular, and ionized gas properties (Fig.~\ref{fig:corr_mat}). These results indicate that the central high-mass star significantly shapes the dynamics, chemistry, and physical conditions of its surrounding molecular and dusty environments. Furthermore, $N_{\rm Lyc}$, which directly traces the ionizing capability of the central star, also correlates with $L_{\rm bol}$.

   \subsection{Benchmarking the \htco thermometer in \hii regions}\label{sec:pdr_code}
            \subsubsection{Comparing gas kinetic temperature and linewidth}
            The values of \tkin, derived using the non-LTE radiative transfer analysis presented in Sect.~\ref{sec:tkin} are found to be consistently warmer than those derived from the dust emission. Additionally, 17 sources display blue- or red-skewed \phtco line profiles, indicating the presence of outflow or infall motions and associated shocked gas. The observed line profiles and derived gas temperatures of \htco suggest that its formation is likely driven by shock heating or outflows and is located close to the \hii regions within the parent molecular cloud \citep{2011ApJ...736..149G, 2017A&A...598A..30T, 2018A&A...611A...6T}.

\begin{figure}
                \centering
                \includegraphics[width=1\linewidth]{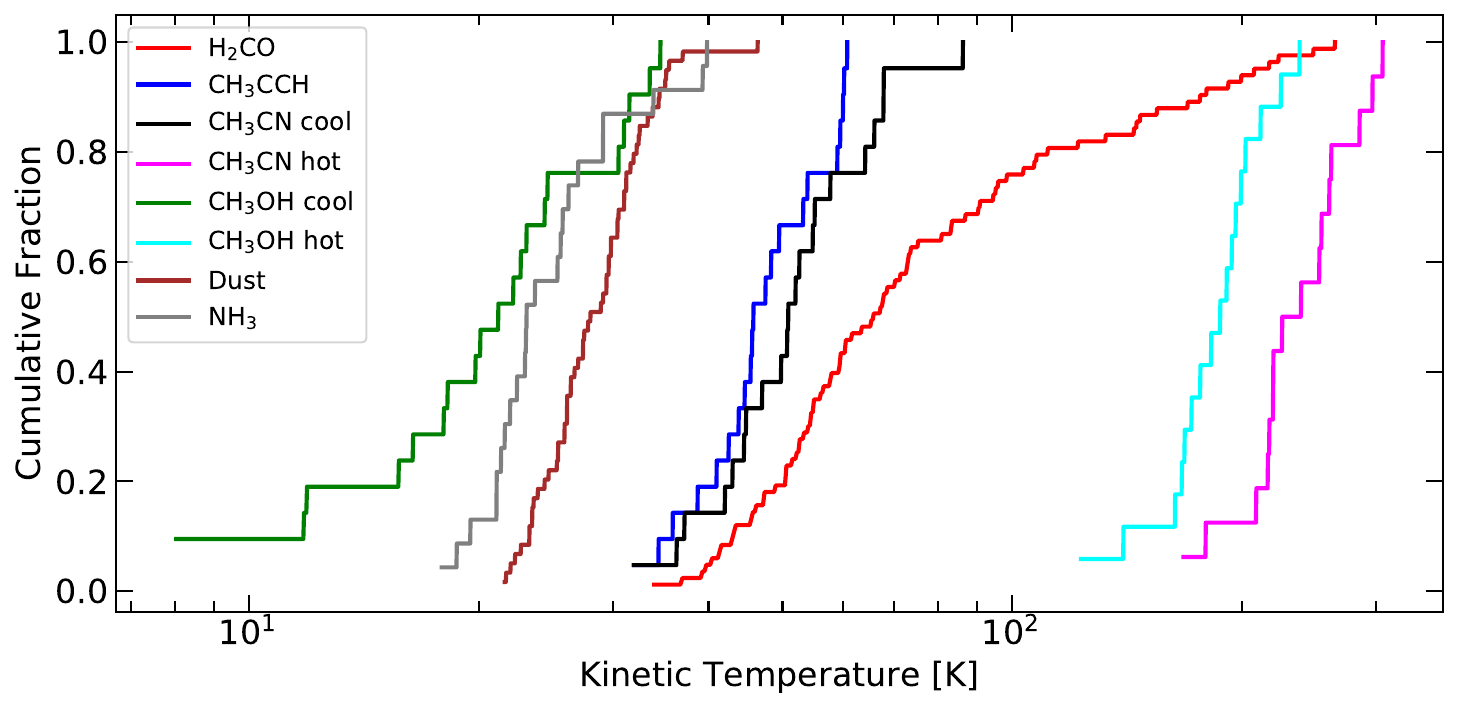}
                \includegraphics[width=1\linewidth]{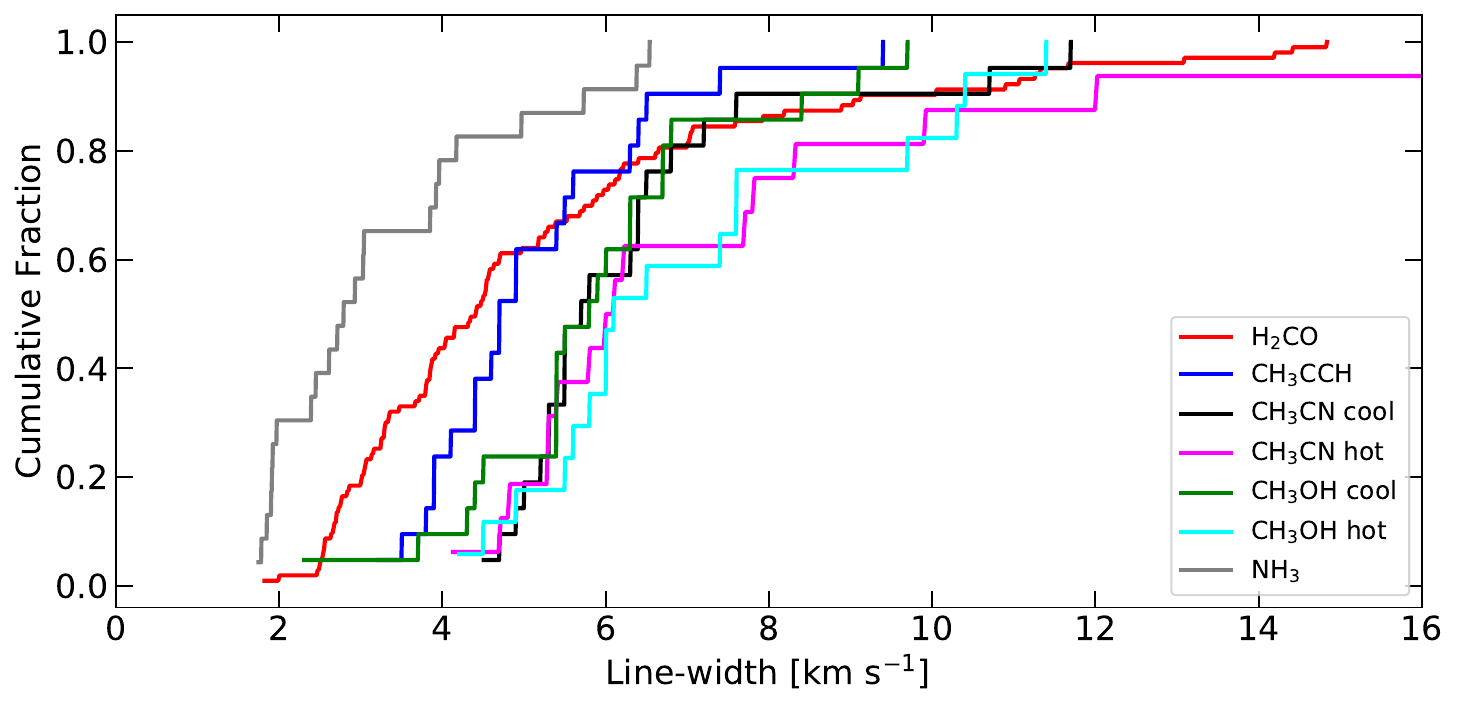}
                \caption{Cumulative distribution functions (CDFs) for temperatures (\textit{Top panel}), and linewidth (\textit{bottom panel}) for different \hii regions. Different colors represent various tracers and components, as indicated in the legend. $\rm NH_3$ properties are taken from \cite{2012A&A...544A.146W}. $\rm CH_3CCH$, $\rm CH_3OH$ (hot and cold), and $\rm CH_3CN$ (hot and cold) properties taken from \cite{2017A&A...603A..33G}, whereas dust properties taken from \cite{2018MNRAS.473.1059U}.}
                \label{fig:plot_cdf}
            \end{figure}

            In Fig.~\ref{fig:plot_cdf} we compare the gas kinetic temperature and linewidth obtained from the different gas tracers towards the \hii regions by \cite{2017A&A...603A..33G}. This shows that \phtco traces a wide range of the gas temperature and the traced temperature is hotter than molecules like $\rm NH_3$. This is based on the findings of \citet{2012A&A...544A.146W}, who reported NH$_3$ rotational temperatures in the range of 10–28\,K and kinetic temperatures between 12 and 35\,K. However, these authors note that the low-$E{\rm u}$ (NH$_3$ (1,1) and (2,2)) transitions used in their analysis exhibited different linewidths, indicating that the observed NH$_3$ transitions likely do not trace the same gas. Consequently, assuming equal beam-filling factors for these transitions represents only an approximation and may affect the radiative analysis and the subsequently derived NH$_3$-bearing gas temperatures. This conclusion is further supported by \citet{2017A&A...603A..33G}, who showed that these low-lying transitions of NH$_3$ are not an optimal tracers of warm gas. Similarly, \phtco traces hotter regions compared to that traced by $\rm CH_3CCH$. \cite{2017A&A...603A..33G} found two temperature components associated with the $\rm CH_3OH$ and $\rm CH_3CN$ namely, hot and cool components with an average of 21 and 184\,K, and 53 and 238\,K for \hii regions in their sample. The kinetic temperature derived using \phtco lies between the cool and hot component of $\rm CH_3OH$ and $\rm CH_3CN$ (see Fig.~\ref{fig:plot_cdf}). Figure~\ref{fig:plot_cdf} (bottom) shows Cumulative Distribution Function (CDF) comparisons of the molecular line width of different gas tracers. Although the Anderson-Darling test reveals that the distribution of linewidth of \phtco and $\rm CH_3CCH$, $\rm CH_3OH$ (cool component) shows no significant differences, that of the warm components show significant differences. The mean value of the linewidths of \phtco is greater than $\rm CH_3CCH$ and $\rm NH_3$ and is less than $\rm CH_3OH$ (hot component) and $\rm CH_3CN$ (hot component). In contrast, the linewidth of \phtco is similar to the linewidth of $\rm CH_3OH$ (hot component) and $\rm CH_3CN$ (hot component). If the observed linewidths of different molecular tracers were determined purely by thermal motion, they would depend directly on the gas temperature and inversely on the molecular mass. In reality, however, these linewidths include both thermal and non-thermal contributions, making direct comparisons between tracers more complex.
            
            \subsubsection{Chemical modeling}
            In this section, we aim to compare the position of \htco relative to the different Photodissociation Region (PDRs) tracers and dense gas tracers around \hii regions. We used an updated version (v7.0) of the Meudon PDR code \citep{2006ApJS..164..506L}\footnote{\url{https://ism.obspm.fr/}} to compare our observations with PDR models and investigate the chemical complexity surrounding an \hii region for a typical density-temperature structure. Using the Meudon code we ran models of constant thermal pressure (isobaric) across a 1D PDR in slab geometry, with a fixed pressure of 4.6$\times10^{7}$~K~cm$^{-3}$, determined by the average $n({\rm H_{2}})$ and $T_{\rm kin}$ as traced by \htco toward our sample (see Sect.~\ref{sec:dens} and  Sect. \ref{sec:tkin}). Combining the given density and temperature profiles with external sources of energy: the UV radiation field (G$_0$) and cosmic-ray ionization rate ($\zeta$), the code determines the abundance of each species by solving the  chemical reactions under steady-state conditions. 
            In addition to the external radiation fields described above, we included an internal stellar component—representing irradiation from host star(s)—and an isotropic ambient component. The isotropic field incorporates the interstellar radiation field with a value of 0.5 (in Habing units) on both the observer-facing and the backside of the cloud, as well as blackbody components that simulate dust infrared emission and the cosmic microwave background.
            A minimal external field is used such that illumination on the backside of the PDR is dominated by the stellar component. This isolates the cloud from external influences, focusing only on the effect of the embedded \hii region on the surrounding molecular material.
            
            We estimated the far-UV flux, G$_0$, assuming that a single massive star is responsible for generating the \hii region, from the massive stars that dominate the ionizing photon production in the sources studied. In Paper I, we derived the Lyman continuum photon flux for our sample using integrated radio flux measurements, finding a median value of $\rm 1.18\times10^{48}\,s^{-1}$. This flux corresponds to a stellar spectral type of O5.5V--O5V with an effective surface temperature of approximately 41,000\,K \citep{2005A&A...436.1049M}. We assumed that the distance between the star and the PDR boundary ($d_0$) is equal to the radius of \hii region, which in our sample has a median value of 0.26\,pc. This setup yields a stellar far-UV flux at the PDR surface of G$_0$ $\sim$ $\rm 5.3\times10^4$ in units of the Habing field \citep[$\rm 1.6\times10^{-3}\,erg\,cm^{-2}\,s^{-1}$; ][]{1968BAN....19..421H}. We note that this estimate of G$_0$ likely represents a lower limit, as the Lyman continuum flux derived from interferometric radio observations may underestimate the total radio flux, thus leading to an underestimation of the ionizing photon rate. In addition, absorption of UV photons by surrounding dust can further reduce the inferred value. Additionally, the relatively low spatial resolution (25$\arcsec$) of the RRL data used in Paper I could have led to an overestimated \hii region size. To explore this further, we also ran the PDR models using a smaller star–PDR distance of $d_0$ = 0.12\,pc, which is the 16$\rm ^{th}$ percentile of our sample \hii region size, this results in G$_0$ values of $\sim$ $\rm 2.5\times10^5$ in units of Habing field. Both sets of models were run for a fixed value of cosmic-ray ionization rate  of $\rm H_2$ molecules, $\zeta = 5\times 10^{-17}~$s$^{-1}$, that lies between estimations in diffuse and dense gas \citep{2015ApJ...800...40I, 2004A&A...417..993L}. We assumed a total visual extinction of $A_{\rm max}$ = 10\,mag for the cloud and adopted the extinction curve of HD 38087, \citep{1990ApJS...72..163F} with $R_{\rm V}$ = 5.6. For the modeling, we used an intermediate value for the column density to reddening ratio, $N_{\rm H}/E(B-V)$ = $\rm 1.05\times10^{22}\,cm^{-2}\,mag^{-1}$ \citep{2018A&A...615A.129J} and a dust-to-gas mass ratio of 0.01. \\

\begin{figure}
                \centering
                \includegraphics[width=1\linewidth]{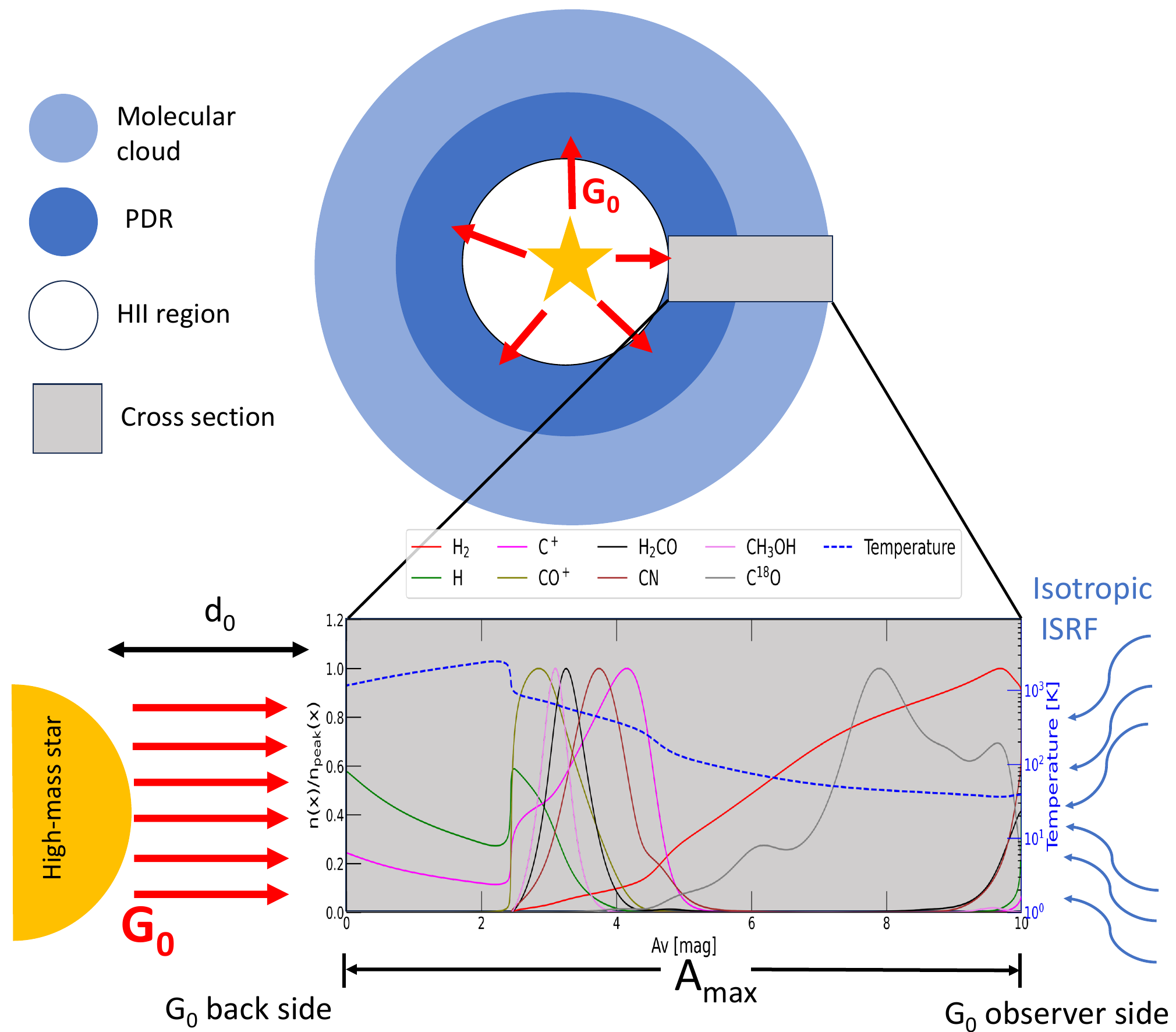}
                \caption{This schematic depicts a fully evolved \hii region in which accretion onto the central star has ceased, although the region remains embedded in its surrounding environment. Additionally, it outlines the various input parameters used in the Meudon PDR code and output molecular abundance as presented in Fig.~\ref{fig:sys_hii}.}
                \label{fig:PDR-cartoon}
            \end{figure}

            A schematic representation of the modeled profile across the \hii region associated PDR is presented in Fig.~\ref{fig:PDR-cartoon} while the results of the Meudon PDR code simulations for $d_0$ = 0.26\,pc (left panel) and $d_0$ = 0.12\,pc (right panel) are presented in Fig.~\ref{fig:sys_hii}. The central and bottom panels plot normalized molecular line abundances, n(x)/n$\rm_{peak}$(x), against $A_{\rm V}$ for PDR tracers \citep[$\rm c-C_3H_2$, $\rm C_2H$, CO$^+$, C$^+$;][]{2005ApJ...634.1133R, 2020A&A...644A.160K} and dense, molecular gas tracers \citep[$\rm CN$, $\rm CH_3OH$, $\rm C^{18}O$, $\rm H_2CO$;][]{2020A&A...644A.160K}, respectively, alongside the corresponding gas density ($n_{\rm H^+}$, $n_{\rm H}$ and $n_{\rm H_2}$) and $T_{\rm kin}$ profiles. Consistently between the two sets of models, the PDR tracers peak at low values of \nh $\sim$ $\rm 8\,\times\,10^3$ to $\rm 5.6\,\times\,10^5\,cm^{-3}$ ($A_{\rm V}$ = 2 to 5\,mag, yellow shaded region in Fig.~\ref{fig:sys_hii}) and \tkin $\sim$ 900 to 70\,K probing the dissociation front and translucent gas layers. However, as $A_v$ and \nh increase, the abundances of dense gas tracers like C$\rm ^{18}$O increase, peaking at $A_{\rm V} \sim 8$\,mag.

            Notably, we observed both \htco and CH$_3$OH exhibit enhanced abundances near the PDR interface, where the gas temperature reaches a few hundred kelvin and the density is around $10^{4-5}$\,cm$^{-3}$. This abundance, intermediate to that seen by typical PDR and dense gas tracers, is expected and provides insight into the chemistry of these two species. For instance, CH$_3$OH, the simplest complex organic molecule \citep[COM; ][]{2009ARA&A..47..427H} is formed efficiently in cold environments on grain surfaces by the continuous hydrogenation of CO \citep{2009A&A...505..629F} and by H-abstraction reactions of icy radicals \citep{2018MNRAS.479.2007A, 2020A&A...634A..52S, 2022ApJ...931L..33S}. CH$_3$OH is subsequently released into the gas phase via thermal sublimation or non-thermal desorption processes where the presence of elevated temperatures in the form of shocks, aids this release via sputtering. This explains the CH$_3$OH abundance peak seen at moderate $A_{\rm V}$'s and high temperatures, highlighting that the formation of CH$_3$OH only proceeds efficiently on grain surfaces and not in the gas phase. On the other hand, although \htco is an important precursor for the formation of CH$_3$OH on grains, unlike CH$_3$OH, H$_2$CO can also form in the cold gas-phase through neutral-neutral reactions between CH$_2$ and O$_2$ or  CH$_3$ and O \citep{2021A&A...656A.148R} or ion-neutral reactions between CH$_3^+$ and OH \citep{2007A&A...466.1197W}. This reinforces the role of \htco as an excellent tracer of gas kinetic temperature \citep{mangum}. In addition with increasing UV-radiation field  
            the PDR model produces higher abundances of \htco consistent with observation result towards massive star forming regions \citep{2014A&A...572A..63I, 2018A&A...611A...6T}. 
            
            Our Meudon PDR models (Fig.~\ref{fig:sys_hii}) reproduce \htco at gas densities consistent with our non-LTE PyRADEX results, however, at \tkin values of few $\times$ 100\,K. While a subset of \phtco velocity components shows high temperature (Fig.~\ref{fig:nh2_tkin_space}) the PDR models fail to reproduce significant amount of cooler \htco. This could be because we carried out single-pointing observations which represent averaged spectra toward each clump, the multi-component profiles likely arise from multiple molecular clumps within the beam, each with distinct velocities and physical conditions. Additional sources of uncertainty include the source specific \nh--\tkin profile assumed. Moreover, the broader, higher-temperature components probably trace gas affected by outflows, inflows, or shocks. In the following section, we examine the origin and nature of these `\htco Sub' components in detail.

        \subsection{Nature of secondary component (\htco Sub)}
            
            \begin{figure}
                \centering
                \includegraphics[width=0.95\linewidth]{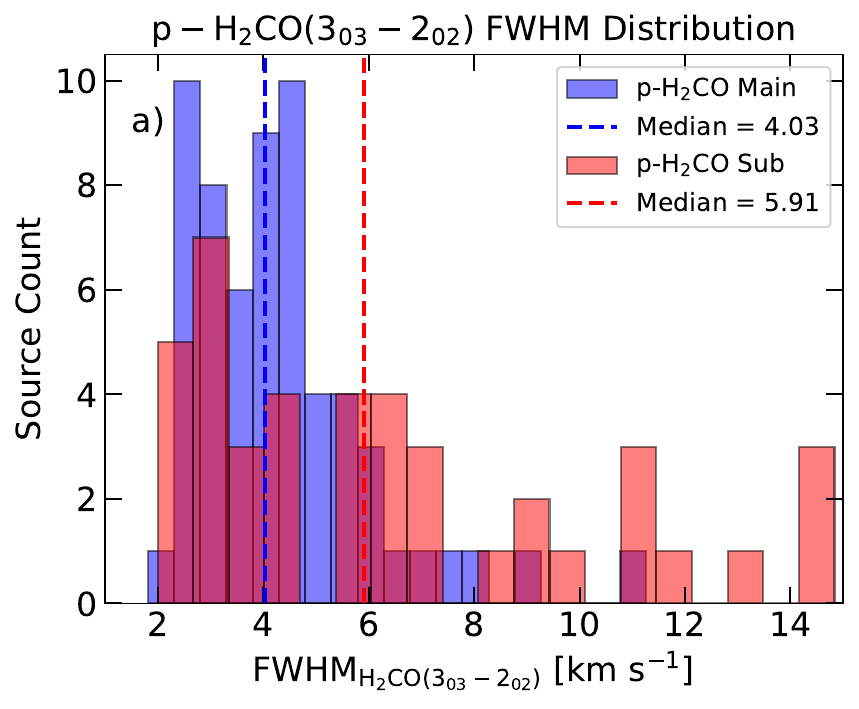}
                \includegraphics[width=0.95\linewidth]{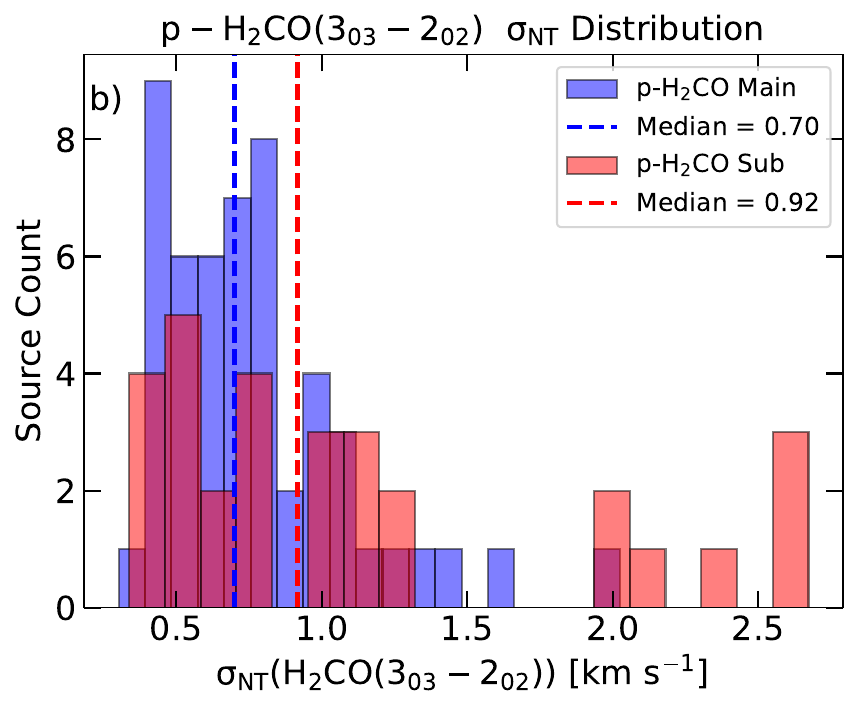}
                \includegraphics[width=0.95\linewidth]{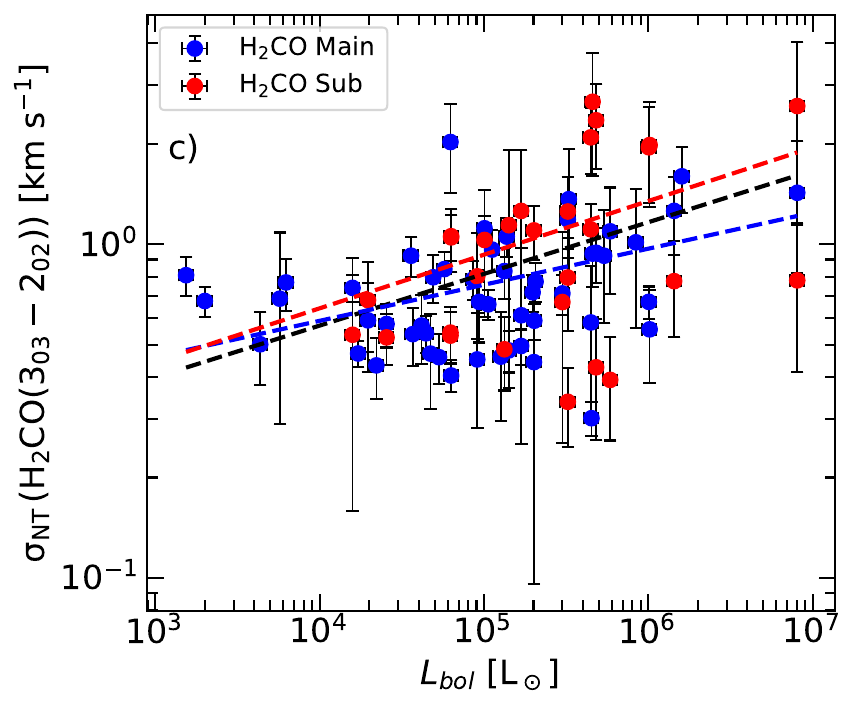}
                \caption{Distributions of (a) FWHM and (b) $\sigma_{\rm NT}$ of the \phtco \lal transition. The blue and red histograms represent the \htco Main and \htco Sub components, respectively, with the corresponding dashed lines marking their median values. Panel (c) shows the relation between $\sigma_{\rm NT}$ and $L_{\rm bol}$. The black, blue, and red dashed lines indicate the power-law fits for all components combined, the Main components, and the Sub components, respectively.}
                \label{fig:nt}
            \end{figure}

            \begin{table}
                \centering
                \caption{Clump bolometric Luminosity vs. $\sigma_{\rm NT}$ (\phtco \lal) non-thermal line width.}
                \label{tab:sigma_vs_Lbol}
                \begin{tabular}{l c c }
                \hline
                Sample & \multicolumn{2}{c}{$L_{\rm bol}-\sigma_{\rm NT}$(\phtco \lal)} \\
                \cline{2-3}
                 & Slope & $\log{\rm (Intercept)}$ \\
                \hline
                Full & $0.15\,(0.01)$ & $-0.86\,(0.03)$\\
                \htco Main & $0.1\,(0.01)$ & $-0.64\,(0.04)$\\
                \htco Sub & $0.16\,(0.004)$ & $-0.84\,(0.1)$\\ 
                \hline
                \end{tabular}
                \tablefoot{The format of the fits is $\log$($\sigma_{\rm NT}$) = Slope     $\times$ $\log$($L_{\rm bol}$) + $\log{\rm (Intercept)}$.}
                \end{table}
            The \phtco observations reveal multiple velocity components in a large fraction of the sources in our sample. We classify these components as `\htco Main' and `\htco Sub' based on their centroid velocities, as described in Sect.~\ref{sec:line_properties}. In this section, we investigate the nature and origin of the `\htco Sub' components. As discussed in Sect.~\ref{sec:pdr_code}, \htco forms in regions with layered density and temperature structures, which can produce multiple velocity components in the observed line profiles. The Meudon PDR models reproduce strong \htco emission from high--UV PDR (Fig.~\ref{fig:sys_hii}). Because these layers lie near the surface of \hii regions, they are subject to intense UV radiation, gas inflow/outflow, and turbulent motions, which broaden the observed lines.

            Figure~\ref{fig:nt}(a) shows the distribution of the \phtco \lal FWHM, indicating that the \htco Sub components exhibit systematically broader lines than the Main components, consistent with turbulence-driven broadening. Using the \tkin derived in Sect.~\ref{sec:tkin}, we calculated the thermal and non-thermal velocity dispersions for each component as
            \begin{equation}
                \sigma_{\rm T} = \sqrt{\frac{kT_{\rm kin}}{m_{\rm H_2CO}}},~~~\sigma_{\rm NT}=\sqrt{\left(\frac{\Delta v}{8\ln 2}\right)^2-\sigma_{\rm T}^2}
            \end{equation}
            where $k$ is the Boltzmann constant, $\Delta v$ is the \phtco \lal\ FWHM, and $m_{\rm H_2CO}$ is the mass of a formaldehyde molecule.  The thermal line width is significantly lower than the non-thermal line width for \htco with a median of 0.13 and 0.8\,\kms, respectively, indicating that non-thermal processes dominate the line broadening. This result is consistent with \citet{2018A&A...611A...6T}. Figure~\ref{fig:nt}(b) further shows that the \htco Sub components have higher non-thermal linewidths (median $\sim$0.92\,\kms) than the Main components (median $\sim$0.70\,\kms). This enhancement is likely due to the fact the \htco Sub components arise from turbulence and shock motions near the surfaces of \hii regions.
            
            The median sound speed ($a_{\rm s} = \sqrt{\frac{kT_{\rm kin}}{\mu m_{\rm H}}}$, where $\mu$ = 2.7 is the mean molecular weight for molecular clouds and $m_{\rm H}$ is the mass of the hydrogen atom) is $\sim$0.45\,\kms at a temperature of 66\,K. This gives the median Mach number (given as $M=\sigma_{\rm NT}/a_{\rm s}$) of 2.45 and 1.56 for the Sub and Main components, respectively, confirming that the \htco Sub components are dominated by supersonic non-thermal motions and trace strongly turbulent gas within our sample.

            We examined the relationship between the non-thermal line width of \phtco and the clump bolometric luminosity for the \htco transitions. Figure~\ref{fig:nt}(c) presents this correlation, where the non-thermal line width serves as a proxy for turbulent motion. We fitted a power-law relation for three cases: all components combined, \htco Main components, and \htco Sub components, with the fitting results summarized in Table~\ref{tab:sigma_vs_Lbol}. The non-thermal line width shows a positive correlation with $L_{\rm bol}$, indicating that turbulence in molecular clouds increases with luminosity. This trend suggests that more luminous sources are associated with more turbulent molecular environments \citep{2009A&A...507..369W}.
            
            The relationship between $\sigma_{\rm NT}$ of \phtco \lal and $L_{\rm bol}$ follows a power law of the form $\sigma_{\rm NT} \propto L_{\rm bol}^{0.1-0.16}$ (see Table~\ref{tab:sigma_vs_Lbol}). The slope is steeper for the \htco Sub component (0.16) than for the \htco Main component (0.10), supporting the interpretation that the \htco Sub emission originates from regions more influenced by massive stars, where turbulence is enhanced—likely near the ionization fronts of \hii regions.

            \subsection{Molecular pressure and confinement of \hii region}
\begin{figure*}
                \centering
                \includegraphics[width=0.45\linewidth]{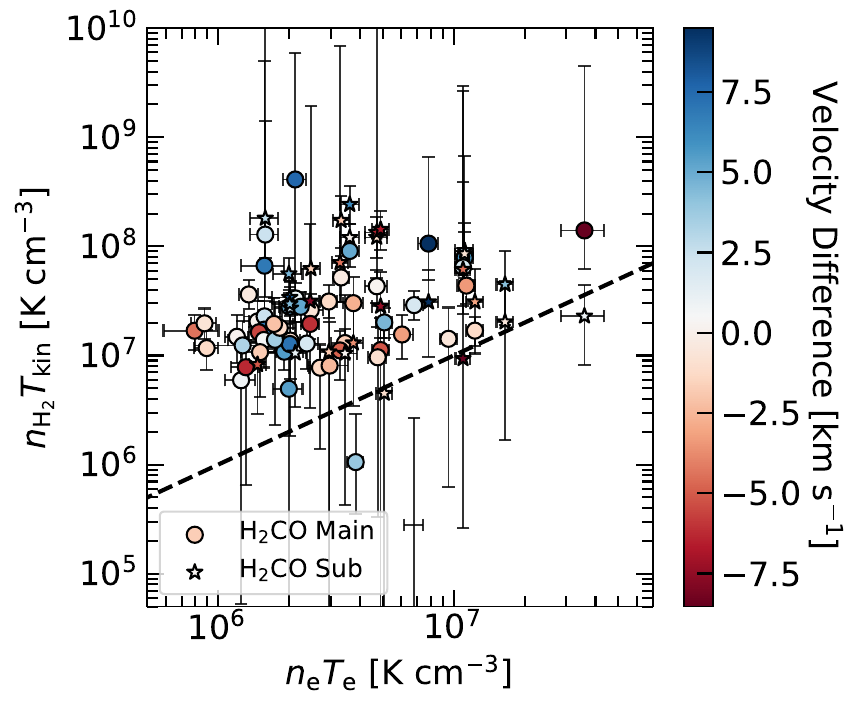}
                \includegraphics[width=0.37\linewidth]{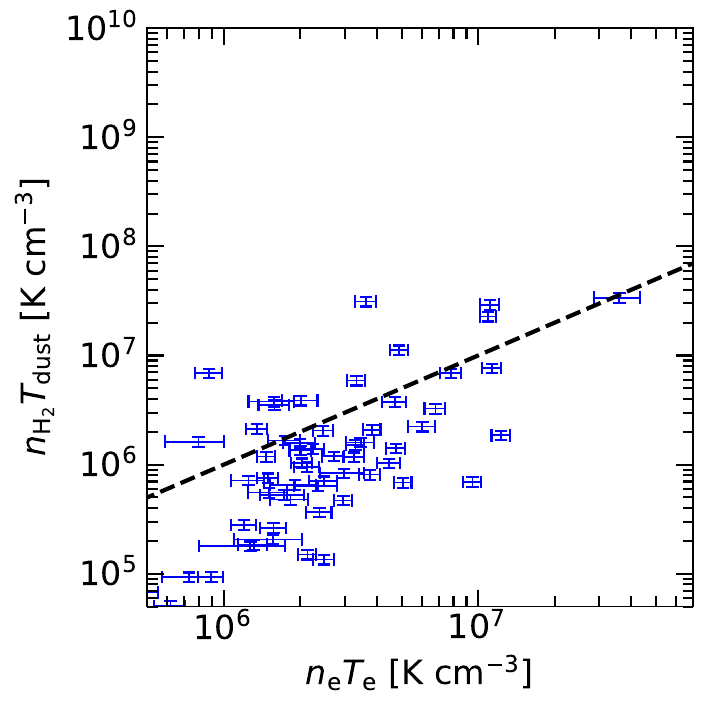}               
                \caption{Comparison between the thermal pressure of ionized and molecular gas, with the color bar indicating the velocity difference between the RRL and \phtco transitions. \textit{Left panel}: \nh and \tkin derived from \phtco in this work. \textit{Right panel}: \nh and $T_{\rm dust}$ estimated from the 870\,$\mu$m emission \citep{2018MNRAS.473.1059U,2022MNRAS.510.3389U}, representing the average clump pressure. The black dashed lines mark the slope of equal pressure. The plots are set to the same scale for visual comparison. }
                \label{fig:pressure}
            \end{figure*}

            \hii region expand under the ionized gas pressure but are impeded by the ambient molecular gas pressure. Figure~\ref{fig:pressure} (left) compares the thermal pressures of the ionized and molecular gas components. For nearly all sources in our sample, the molecular gas pressure exceeds that of the ionized gas. This implies that the molecular gas pressure can slow down the expansion of embedded \hii regions. Moreover showing that high molecular gas pressure persists regardless of the velocity offset between RRLs and \phtco transitions, further supporting the argument that molecular pressure can confine the expansion of embedded \hii regions.
            
            However, several caveats exist. Firstly, these \hii regions must still be embedded within their natal molecular clumps; our observations though limited by spectral resolution, try to fulfill this condition through our selection of \hii regions with sizes smaller than their parental ATLASGAL clump (Sect.~\ref{sec:source_selection}). Secondly, the molecular pressures derived from \htco should trace gas closer to the ionization front, this is supported by both observations and the Meudon PDR models (Sect.~\ref{sec:pdr_code}). A confirmation of the same awaits higher resolution \htco and ionized gas mapping. Thirdly, the ionized gas parameters ($n_e$, $T_e$) adopted from Paper~I represent spatial averages that may underestimate the true pressure at the ionization boundary. Figure~\ref{fig:pressure} (right) compares the ionized gas pressures with those derived from dust emission at 870\,$\mu$m \citep{2018MNRAS.473.1059U,2022MNRAS.510.3389U}, representing the average clump conditions. In this case, the ionized gas pressure often exceeds that of the molecular component. In contrast to the molecular gas, the pressures inferred from dust emission may not be representative of conditions in the vicinity of the expanding \hii regions. Resolving these apparent inconsistencies will require high-resolution observations of the ionized and molecular gas interfaces.
            
            Recently, \cite{2025arXiv250616700F} analyzed the expansion signatures of 35 \hii regions mapped in [CII] 158\,$\mu$m emission. Among these, 12 \hii regions were identified as candidates for expansion, and 8 of them were consistent with a spherical expansion model. They also found that thermal pressure and stellar winds are the primary drivers of \hii region expansion in their sample. Previous studies, including that of \cite{2011ApJS..194...32A} and Paper I, reported that roughly 50\% of cataloged Galactic \hii regions exhibit bubble-like morphologies at 8\,$\mu$m, based on GLIMPSE images \citep{2003PASP..115..953B}, which trace PDRs. In our sample, 23 sources exhibit bubble morphology including irregular bubbles and of the remaining 19 sources show compact morphology. To explore these processes in more detail, maps of PDRs and dense molecular gas tracers in the vicinity of expanding \hii regions are needed, along with radio continuum and RRL observations at optically thin frequencies. Such observations will allow a detailed analysis of both ionized and molecular gas, as well as high-resolution studies of the gas dynamics. This approach will provide stronger constraints on the formation, evolution, and feedback mechanisms of \hii regions.

        \section{Conclusion and summary} \label{sec:conclusion}
        We investigated the physical properties traced by molecular and ionized gas in a statistically significant sample of \hii regions associated with ATLASGAL dust clumps. We used observations carried out with the APEX 12\,m sub-mm telescope, of multiple transitions of \phtco associated with its $J$=3--2 and 4--3 transitions toward these sources. In our earlier study (Paper I), we examined the ionized gas properties of \hii regions using cm--RRLs from the GLOSTAR survey. In this work, we focused on the molecular gas properties of the embedded \hii regions and compared them with the ionized gas characteristics. Our main findings are summarized below:
        \begin{itemize}
            \item  We constrained the physical properties of massive clumps associated with \hii regions by deriving gas density and kinetic temperature for 52 clumps. We carried out non-LTE radiative transfer modeling of the six \phtco transitions studied, which serve as effective tracers of density and temperature in dense gas, using PyRADEX+MCMC. The resulting \tkin ranges from 33.7\,K to 265\,K, with a median value of $\sim$66\,K and \nh $\rm >10^5\,cm^{-3}$ across our sample, aligning with findings from previous studies.
            
            \item We analyzed the relationships among the physical properties of dust, molecular, and ionized gas by calculating the Spearman’s correlation coefficients and $p-$values for various parameters. Properties that directly depend on the central high-mass star, such as $L\rm _{bol}$ and $N\rm _{Lyc}$, exhibit stronger and more frequent correlations with other parameters. This suggests that the central high-mass star largely governs and influences the physical conditions of the surrounding dust, ionized, and molecular gas.
            
            \item By comparing \htco with other molecular gas tracers, we found that \htco traces denser gas located near the interface between the \hii region and the surrounding molecular cloud. The Meudon PDR models reproduced \htco abundances peaking near the ionized region's surface and at gas densities similar to those obtained from PyRADEX but at high values of \tkin. Understanding the origins of the cooler \htco components requires further investigation with more detailed modeling.   

            \item The nature of the `\htco Sub' components, reveals that these components are dominated by supersonic non-thermal motions and trace turbulent gas.  
            
            \item We investigated whether the surrounding molecular gas confines the expansion of \hii regions through excess pressure. Our analysis shows that, for most of our sources, the pressure of the neutral gas exceeds that of the ionized gas, regardless of the velocity difference between the ionized and molecular components. This finding suggests that the surrounding neutral molecular gas can slow down or hinder the expansion of \hii regions due to its higher pressure. However, there are several caveats, and a confirmation of this fining relies on future high resolution mapping of \hii regions and their surroundings in RRL and molecular lines. Such observations will soon be achievable with the unprecedented sensitivity and spatial resolution offered by next-generation radio facilities such as the ngVLA and SKA--Mid.    
        \end{itemize}

        The expansion and feedback of \hii regions significantly shape the interstellar medium, regulate star formation rates, and influence galactic evolution. Future studies should increase the sample size of embedded and expanding \hii regions to better explore the relationship between ionized and molecular gas. This will improve our understanding of how \hii regions evolve over time and impact their surrounding molecular environment. Placing tighter constraints on the connection between ionized and molecular gas across various evolutionary stages ranging from compact to classical diffuse \hii regions will be essential for understanding their development and potential role in triggering star formation.

        \section*{Data Availability}
        Tables~\ref{tab:nflash_line}, \ref{tab:sepia_line}, and \ref{tab:phy_prop} are only available in electronic form at the CDS via anonymous ftp to cdsarc.u-strasbg.fr (130.79.128.5) or via http://cdsweb.u-strasbg.fr/cgi-bin/qcat?J/A+A/.
	\begin{acknowledgements} 
        During the preparation of this article, we suffered the tragic loss of Prof. Dr Karl~M.~Menten, S.K.'s PhD supervisor and the principal investigator of the GLOSTAR survey. We dedicate this work to his enduring legacy. We are deeply indebted to the countless discussions with him that have inspired generations of radio astronomers. We will always miss his warmth, wisdom, boundless curiosity, generosity of spirit, kindness, and delightful sense of humor. We sincerely thank the referee for their positive, timely, and constructive feedback. We thank MPIfR and APEX staff for operating the observatory and collecting the data (\url{https://www.apex-telescope.org/ns/}).  This research made use of information from the ATLASGAL database at (\url{http://atlasgal.mpifr-bonn.mpg.de/cgi-bin/ATLASGAL_DATABASE.cgi}) supported by the MPIfR, Bonn. A.M.J. acknowledges support by the CRC1601 (SFB 1601 sub-project B2) funded by the DFG (German Research Foundation) - 500700252. S.K. and A.M.J. thanks the Max Planck Society for their support. Y.G. was supported by the Ministry of Science and Technology of China under the National Key R\&D Program under grant No. 2023YFA1608200, the National Natural Science Foundation of China (NSFC) under grant No. 12427901, and the Strategic Priority Research Program of the Chinese Academy of Sciences under grant No. XDB0800301. S.N. gratefully acknowledges the Collaborative Research Center 1601 (SFB 1601 sub-project B1) funded by the Deutsche Forschungsgemeinschaft (DFG, German Research Foundation) – 500700252. This research has made use of the SIMBAD database and the VizieR catalog, operated at CDS, Strasbourg, France. This research made use of Astropy (\url{http://www.astropy.org}), a community developed core Python package for Astronomy~\citep{2022ApJ...935..167A}. This document was prepared using the collaborative tool Overleaf available at:~\url{https://www.overleaf.com/}.
	\end{acknowledgements}
	
	\bibliographystyle{aa} 
	\bibliography{mybib.bib} 

@ARTICLE{2024A&A...689A..81K,
       author = {{Khan}, S. and {Rugel}, M.~R. and {Brunthaler}, A. and {Menten}, K.~M. and {Wyrowski}, F. and {Urquhart}, J.~S. and {Gong}, Y. and {Yang}, A.~Y. and {Nguyen}, H. and {Dokara}, R. and {Dzib}, S.~A. and {Medina}, S. -N.~X. and {Ortiz-Le{\'o}n}, G.~N. and {Pandian}, J.~D. and {Beuther}, H. and {Veena}, V.~S. and {Neupane}, S. and {Cheema}, A. and {Reich}, W. and {Roy}, N.},
        title = "{A global view on star formation: The GLOSTAR Galactic plane survey: X. Galactic H II region catalog using radio recombination lines}",
      journal = {\aap},
         year = 2024,
        month = sep,
       volume = {689},
          eid = {A81},
        pages = {A81},
          doi = {10.1051/0004-6361/202449390},
archivePrefix = {arXiv},
       eprint = {2407.05770},
 primaryClass = {astro-ph.GA},
       adsurl = {https://ui.adsabs.harvard.edu/abs/2024A&A...689A..81K}
}

@ARTICLE{2021A&A...651A..85B,
	author = {{Brunthaler}, A. and {Menten}, K.~M. and {Dzib}, S.~A. and {Cotton}, W.~D. and {Wyrowski}, F. and {Dokara}, R. and {Gong}, Y. and {Medina}, S. -N.~X. and {M{\"u}ller}, P. and {Nguyen}, H. and {Ortiz-Le{\'o}n}, G.~N. and {Reich}, W. and {Rugel}, M.~R. and {Urquhart}, J.~S. and {Winkel}, B. and {Yang}, A.~Y. and {Beuther}, H. and {Billington}, S. and {Carrasco-Gonzalez}, C. and {Csengeri}, T. and {Murugeshan}, C. and {Pandian}, J.~D. and {Roy}, N.},
	title = "{A global view on star formation: The GLOSTAR Galactic plane survey. I. Overview and first results for the Galactic longitude range 28{\textdegree} < l < 36{\textdegree}}",
	journal = {\aap},
	year = 2021,
	month = jul,
	volume = {651},
	eid = {A85},
	pages = {A85},
	doi = {10.1051/0004-6361/202039856},
	archivePrefix = {arXiv},
	eprint = {2106.00377},
	primaryClass = {astro-ph.GA},
	adsurl = {https://ui.adsabs.harvard.edu/abs/2021A&A...651A..85B}
}

@ARTICLE{2019A&A...627A.175M,
       author = {{Medina}, S. -N.~X. and {Urquhart}, J.~S. and {Dzib}, S.~A. and {Brunthaler}, A. and {Cotton}, B. and {Menten}, K.~M. and {Wyrowski}, F. and {Beuther}, H. and {Billington}, S.~J. and {Carrasco-Gonzalez}, C. and {Csengeri}, T. and {Gong}, Y. and {Hofner}, P. and {Nguyen}, H. and {Ortiz-Le{\'o}n}, G.~N. and {Ott}, J. and {Pandian}, J.~D. and {Roy}, N. and {Sarkar}, E. and {Wang}, Y. and {Winkel}, B.},
        title = "{GLOSTAR: Radio Source Catalog I. 28{\textdegree} < {\ensuremath{\ell}} < 36{\textdegree} and |b| < 1{\textdegree}}",
      journal = {\aap},
         year = 2019,
        month = jul,
       volume = {627},
          eid = {A175},
        pages = {A175},
          doi = {10.1051/0004-6361/201935249},
archivePrefix = {arXiv},
       eprint = {1905.09281},
 primaryClass = {astro-ph.GA},
       adsurl = {https://ui.adsabs.harvard.edu/abs/2019A&A...627A.175M}
}

@ARTICLE{2023A&A...680A..92Y,
       author = {{Yang}, A.~Y. and {Dzib}, S.~A. and {Urquhart}, J.~S. and {Brunthaler}, A. and {Medina}, S. -N.~X. and {Menten}, K.~M. and {Wyrowski}, F. and {Ortiz-Le{\'o}n}, G.~N. and {Cotton}, W.~D. and {Gong}, Y. and {Dokara}, R. and {Rugel}, M.~R. and {Beuther}, H. and {Pandian}, J.~D. and {Csengeri}, T. and {Veena}, V.~S. and {Roy}, N. and {Nguyen}, H. and {Winkel}, B. and {Ott}, J. and {Carrasco-Gonzalez}, C. and {Khan}, S. and {Cheema}, A.},
        title = "{A global view on star formation: The GLOSTAR Galactic plane survey. IX. Radio Source Catalog III: 2{\textdegree} < {\ensuremath{\ell}} < 28{\textdegree}, 36{\textdegree} < {\ensuremath{\ell}} < 40{\textdegree}, 56{\textdegree} < {\ensuremath{\ell}} < 60{\textdegree} and |b| < 1{\textdegree}, VLA B-configuration}",
      journal = {\aap},
         year = 2023,
        month = dec,
       volume = {680},
          eid = {A92},
        pages = {A92},
          doi = {10.1051/0004-6361/202347563},
archivePrefix = {arXiv},
       eprint = {2310.09777},
 primaryClass = {astro-ph.GA},
       adsurl = {https://ui.adsabs.harvard.edu/abs/2023A&A...680A..92Y}
}

@ARTICLE{2024A&A...689A.196M,
       author = {{Medina}, S. -N.~X. and {Dzib}, S.~A. and {Urquhart}, J.~S. and {Yang}, A.~Y. and {Brunthaler}, A. and {Menten}, K.~M. and {Wyrowski}, F. and {Cotton}, W.~D. and {Cheema}, A. and {Dokara}, R. and {Gong}, Y. and {Khan}, S. and {Nguyen}, H. and {Ortiz-Le{\'o}n}, G.~N. and {Rugel}, M.~R. and {Veena}, V.~S. and {Beuther}, H. and {Csengeri}, T. and {Pandian}, J.~D. and {Roy}, N.},
        title = "{A global view on star formation: The GLOSTAR Galactic plane survey: XI. Radio source catalog IV: 2{\textdegree} < {\ensuremath{\ell}} < 28{\textdegree}, 36{\textdegree} < {\ensuremath{\ell}} < 60{\textdegree} and |b| < 1{\textdegree}}",
      journal = {\aap},
         year = 2024,
        month = sep,
       volume = {689},
          eid = {A196},
        pages = {A196},
          doi = {10.1051/0004-6361/202449885},
archivePrefix = {arXiv},
       eprint = {2407.12585},
 primaryClass = {astro-ph.GA},
       adsurl = {https://ui.adsabs.harvard.edu/abs/2024A&A...689A.196M}
}

@ARTICLE{2023A&A...670A...9D,
       author = {{Dzib}, S.~A. and {Yang}, A.~Y. and {Urquhart}, J.~S. and {Medina}, S. -N.~X. and {Brunthaler}, A. and {Menten}, K.~M. and {Wyrowski}, F. and {Cotton}, W.~D. and {Dokara}, R. and {Ortiz-Le{\'o}n}, G.~N. and {Rugel}, M.~R. and {Nguyen}, H. and {Gong}, Y. and {Chakraborty}, A. and {Beuther}, H. and {Billington}, S.~J. and {Carrasco-Gonzalez}, C. and {Csengeri}, T. and {Hofner}, P. and {Ott}, J. and {Pandian}, J.~D. and {Roy}, N. and {Yanza}, V.},
        title = "{A global view on star formation: The GLOSTAR Galactic plane survey. VI. Radio Source Catalog II: 28{\textdegree} < {\ensuremath{\ell}} < 36{\textdegree} and |b| < 1{\textdegree}, VLA B-configuration}",
      journal = {\aap},
         year = 2023,
        month = feb,
       volume = {670},
          eid = {A9},
        pages = {A9},
          doi = {10.1051/0004-6361/202143019},
archivePrefix = {arXiv},
       eprint = {2210.00560},
 primaryClass = {astro-ph.GA},
       adsurl = {https://ui.adsabs.harvard.edu/abs/2023A&A...670A...9D}
}

@ARTICLE{2003PASP..115..953B,
       author = {{Benjamin}, Robert A. and {Churchwell}, E. and {Babler}, Brian L. and
         {Bania}, T.~M. and {Clemens}, Dan P. and {Cohen}, Martin and
         {Dickey}, John M. and {Indebetouw}, R{\'e}my and {Jackson}, James M. and
         {Kobulnicky}, Henry A. and {Lazarian}, Alex and {Marston}, A.~P. and
         {Mathis}, John S. and {Meade}, Marilyn R. and {Seager}, Sara and
         {Stolovy}, S.~R. and {Watson}, C. and {Whitney}, Barbara A. and
         {Wolff}, Michael J. and {Wolfire}, Mark G.},
        title = "{GLIMPSE. I. An SIRTF Legacy Project to Map the Inner Galaxy}",
      journal = {\pasp},
         year = 2003,
        month = aug,
       volume = {115},
       number = {810},
        pages = {953-964},
       adsurl = {https://ui.adsabs.harvard.edu/abs/2003PASP..115..953B}
}

@ARTICLE{2022MNRAS.510.3389U,
       author = {{Urquhart}, J.~S. and {Wells}, M.~R.~A. and {Pillai}, T. and {Leurini}, S. and {Giannetti}, A. and {Moore}, T.~J.~T. and {Thompson}, M.~A. and {Figura}, C. and {Colombo}, D. and {Yang}, A.~Y. and {K{\"o}nig}, C. and {Wyrowski}, F. and {Menten}, K.~M. and {Rigby}, A.~J. and {Eden}, D.~J. and {Ragan}, S.~E.},
        title = "{ATLASGAL - evolutionary trends in high-mass star formation}",
      journal = {\mnras},
         year = 2022,
        month = mar,
       volume = {510},
       number = {3},
        pages = {3389-3407},
          doi = {10.1093/mnras/stab3511},
archivePrefix = {arXiv},
       eprint = {2111.12816},
 primaryClass = {astro-ph.GA},
       adsurl = {https://ui.adsabs.harvard.edu/abs/2022MNRAS.510.3389U}
}

@ARTICLE{2018MNRAS.473.1059U,
       author = {{Urquhart}, J.~S. and {K{\"o}nig}, C. and {Giannetti}, A. and {Leurini}, S. and {Moore}, T.~J.~T. and {Eden}, D.~J. and {Pillai}, T. and {Thompson}, M.~A. and {Braiding}, C. and {Burton}, M.~G. and {Csengeri}, T. and {Dempsey}, J.~T. and {Figura}, C. and {Froebrich}, D. and {Menten}, K.~M. and {Schuller}, F. and {Smith}, M.~D. and {Wyrowski}, F.},
        title = "{ATLASGAL - properties of a complete sample of Galactic clumps}",
      journal = {\mnras},
         year = 2018,
        month = jan,
       volume = {473},
       number = {1},
        pages = {1059-1102},
          doi = {10.1093/mnras/stx2258},
archivePrefix = {arXiv},
       eprint = {1709.00392},
 primaryClass = {astro-ph.GA},
       adsurl = {https://ui.adsabs.harvard.edu/abs/2018MNRAS.473.1059U}
}

@ARTICLE{2009A&A...504..415S,
       author = {{Schuller}, F. and {Menten}, K.~M. and {Contreras}, Y. and {Wyrowski}, F. and {Schilke}, P. and {Bronfman}, L. and {Henning}, T. and {Walmsley}, C.~M. and {Beuther}, H. and {Bontemps}, S. and {Cesaroni}, R. and {Deharveng}, L. and {Garay}, G. and {Herpin}, F. and {Lefloch}, B. and {Linz}, H. and {Mardones}, D. and {Minier}, V. and {Molinari}, S. and {Motte}, F. and {Nyman}, L. -{\r{A}}. and {Reveret}, V. and {Risacher}, C. and {Russeil}, D. and {Schneider}, N. and {Testi}, L. and {Troost}, T. and {Vasyunina}, T. and {Wienen}, M. and {Zavagno}, A. and {Kovacs}, A. and {Kreysa}, E. and {Siringo}, G. and {Wei{\ss}}, A.},
        title = "{ATLASGAL - The APEX telescope large area survey of the galaxy at 870 {\ensuremath{\mu}}m}",
      journal = {\aap},
         year = 2009,
        month = sep,
       volume = {504},
       number = {2},
        pages = {415-427},
          doi = {10.1051/0004-6361/200811568},
archivePrefix = {arXiv},
       eprint = {0903.1369},
 primaryClass = {astro-ph.GA},
       adsurl = {https://ui.adsabs.harvard.edu/abs/2009A&A...504..415S}
}

@ARTICLE{2018A&A...611A...6T,
       author = {{Tang}, X.~D. and {Henkel}, C. and {Wyrowski}, F. and {Giannetti}, A. and {Menten}, K.~M. and {Csengeri}, T. and {Leurini}, S. and {Urquhart}, J.~S. and {K{\"o}nig}, C. and {G{\"u}sten}, R. and {Lin}, Y.~X. and {Zheng}, X.~W. and {Esimbek}, J. and {Zhou}, J.~J.},
        title = "{ATLASGAL-selected massive clumps in the inner Galaxy. VI. Kinetic temperature and spatial density measured with formaldehyde}",
      journal = {\aap},
         year = 2018,
        month = mar,
       volume = {611},
          eid = {A6},
        pages = {A6},
          doi = {10.1051/0004-6361/201732168},
archivePrefix = {arXiv},
       eprint = {1711.10012},
 primaryClass = {astro-ph.GA},
       adsurl = {https://ui.adsabs.harvard.edu/abs/2018A&A...611A...6T}
}

@ARTICLE{2014A&A...565A..75C,
       author = {{Csengeri}, T. and {Urquhart}, J.~S. and {Schuller}, F. and {Motte}, F. and {Bontemps}, S. and {Wyrowski}, F. and {Menten}, K.~M. and {Bronfman}, L. and {Beuther}, H. and {Henning}, Th. and {Testi}, L. and {Zavagno}, A. and {Walmsley}, M.},
        title = "{The ATLASGAL survey: a catalog of dust condensations in the Galactic plane}",
      journal = {\aap},
         year = 2014,
        month = may,
       volume = {565},
          eid = {A75},
        pages = {A75},
          doi = {10.1051/0004-6361/201322434},
archivePrefix = {arXiv},
       eprint = {1312.0937},
 primaryClass = {astro-ph.GA},
       adsurl = {https://ui.adsabs.harvard.edu/abs/2014A&A...565A..75C}
}

@ARTICLE{2014A&A...572A..63I,
       author = {{Immer}, K. and {Galv{\'a}n-Madrid}, R. and {K{\"o}nig}, C. and {Liu}, H.~B. and {Menten}, K.~M.},
        title = "{Diversity of chemistry and excitation conditions in the high-mass star forming complex W33}",
      journal = {\aap},
         year = 2014,
        month = dec,
       volume = {572},
          eid = {A63},
        pages = {A63},
          doi = {10.1051/0004-6361/201423780},
archivePrefix = {arXiv},
       eprint = {1409.7125},
 primaryClass = {astro-ph.GA},
       adsurl = {https://ui.adsabs.harvard.edu/abs/2014A&A...572A..63I}
}

@ARTICLE{2018A&A...609A..16T,
       author = {{Tang}, X.~D. and {Henkel}, C. and {Menten}, K.~M. and {Wyrowski}, F. and {Brinkmann}, N. and {Zheng}, X.~W. and {Gong}, Y. and {Lin}, Y.~X. and {Esimbek}, J. and {Zhou}, J.~J. and {Yuan}, Y. and {Li}, D.~L. and {He}, Y.~X.},
        title = "{Kinetic temperature of massive star-forming molecular clumps measured with formaldehyde. III. The Orion molecular cloud 1}",
      journal = {\aap},
         year = 2018,
        month = jan,
       volume = {609},
          eid = {A16},
        pages = {A16},
          doi = {10.1051/0004-6361/201731849},
archivePrefix = {arXiv},
       eprint = {1709.07694},
 primaryClass = {astro-ph.GA},
       adsurl = {https://ui.adsabs.harvard.edu/abs/2018A&A...609A..16T}
}

@ARTICLE{2007A&A...468..627V,
       author = {{van der Tak}, F.~F.~S. and {Black}, J.~H. and {Sch{\"o}ier}, F.~L. and {Jansen}, D.~J. and {van Dishoeck}, E.~F.},
        title = "{A computer program for fast non-LTE analysis of interstellar line spectra. With diagnostic plots to interpret observed line intensity ratios}",
      journal = {\aap},
         year = 2007,
        month = jun,
       volume = {468},
       number = {2},
        pages = {627-635},
          doi = {10.1051/0004-6361:20066820},
archivePrefix = {arXiv},
       eprint = {0704.0155},
 primaryClass = {astro-ph},
       adsurl = {https://ui.adsabs.harvard.edu/abs/2007A&A...468..627V}
}

@ARTICLE{2013MNRAS.432.2573W,
       author = {{Wiesenfeld}, L. and {Faure}, A.},
        title = "{Rotational quenching of H$_{2}$CO by molecular hydrogen: cross-sections, rates and pressure broadening}",
      journal = {\mnras},
         year = 2013,
        month = jul,
       volume = {432},
       number = {3},
        pages = {2573-2578},
          doi = {10.1093/mnras/stt616},
archivePrefix = {arXiv},
       eprint = {1304.4804},
 primaryClass = {astro-ph.SR},
       adsurl = {https://ui.adsabs.harvard.edu/abs/2013MNRAS.432.2573W}
}

@ARTICLE{2017A&A...608A.144Y,
       author = {{Yang}, C. and {Omont}, A. and {Beelen}, A. and {Gao}, Y. and {van der Werf}, P. and {Gavazzi}, R. and {Zhang}, Z. -Y. and {Ivison}, R. and {Lehnert}, M. and {Liu}, D. and {Oteo}, I. and {Gonz{\'a}lez-Alfonso}, E. and {Dannerbauer}, H. and {Cox}, P. and {Krips}, M. and {Neri}, R. and {Riechers}, D. and {Baker}, A.~J. and {Micha{\l}owski}, M.~J. and {Cooray}, A. and {Smail}, I.},
        title = "{Molecular gas in the Herschel-selected strongly lensed submillimeter galaxies at z   2-4 as probed by multi-J CO lines}",
      journal = {\aap},
         year = 2017,
        month = dec,
       volume = {608},
          eid = {A144},
        pages = {A144},
          doi = {10.1051/0004-6361/201731391},
archivePrefix = {arXiv},
       eprint = {1709.04740},
 primaryClass = {astro-ph.GA},
       adsurl = {https://ui.adsabs.harvard.edu/abs/2017A&A...608A.144Y}
}

@ARTICLE{mangum,
       author = {{Mangum}, Jeffrey G. and {Wootten}, Alwyn},
        title = "{Formaldehyde as a Probe of Physical Conditions in Dense Molecular Clouds}",
      journal = {\apjs},
         year = 1993,
        month = nov,
       volume = {89},
        pages = {123},
          doi = {10.1086/191841},
       adsurl = {https://ui.adsabs.harvard.edu/abs/1993ApJS...89..123M}
}

@ARTICLE{2017A&A...603A..33G,
       author = {{Giannetti}, A. and {Leurini}, S. and {Wyrowski}, F. and {Urquhart}, J. and {Csengeri}, T. and {Menten}, K.~M. and {K{\"o}nig}, C. and {G{\"u}sten}, R.},
        title = "{ATLASGAL-selected massive clumps in the inner Galaxy. V. Temperature structure and evolution}",
      journal = {\aap},
         year = 2017,
        month = jul,
       volume = {603},
          eid = {A33},
        pages = {A33},
          doi = {10.1051/0004-6361/201630048},
archivePrefix = {arXiv},
       eprint = {1703.08485},
 primaryClass = {astro-ph.GA},
       adsurl = {https://ui.adsabs.harvard.edu/abs/2017A&A...603A..33G}
}

@ARTICLE{2012A&A...544A.146W,
       author = {{Wienen}, M. and {Wyrowski}, F. and {Schuller}, F. and {Menten}, K.~M. and {Walmsley}, C.~M. and {Bronfman}, L. and {Motte}, F.},
        title = "{Ammonia from cold high-mass clumps discovered in the inner Galactic disk by the ATLASGAL survey}",
      journal = {\aap},
         year = 2012,
        month = aug,
       volume = {544},
          eid = {A146},
        pages = {A146},
          doi = {10.1051/0004-6361/201118107},
archivePrefix = {arXiv},
       eprint = {1208.4848},
 primaryClass = {astro-ph.GA},
       adsurl = {https://ui.adsabs.harvard.edu/abs/2012A&A...544A.146W}
}

@ARTICLE{2012A&A...542L...3K,
       author = {{Klein}, B. and {Hochg{\"u}rtel}, S. and {Kr{\"a}mer}, I. and {Bell}, A. and {Meyer}, K. and {G{\"u}sten}, R.},
        title = "{High-resolution wide-band fast Fourier transform spectrometers}",
      journal = {\aap},
         year = 2012,
        month = jun,
       volume = {542},
          eid = {L3},
        pages = {L3},
          doi = {10.1051/0004-6361/201218864},
archivePrefix = {arXiv},
       eprint = {1203.3972},
 primaryClass = {astro-ph.IM},
       adsurl = {https://ui.adsabs.harvard.edu/abs/2012A&A...542L...3K}
}

@ARTICLE{1994A&A...291..943O,
       author = {{Ossenkopf}, V. and {Henning}, Th.},
        title = "{Dust opacities for protostellar cores.}",
      journal = {\aap},
         year = 1994,
        month = nov,
       volume = {291},
        pages = {943-959},
       adsurl = {https://ui.adsabs.harvard.edu/abs/1994A&A...291..943O}
}

@ARTICLE{2011ApJ...742...58M,
       author = {{McCauley}, Patrick I. and {Mangum}, Jeffrey G. and {Wootten}, Alwyn},
        title = "{Formaldehyde Densitometry of Galactic Star-forming Regions Using the H$_{2}$CO {}3$_{12}$-{}3$_{13}$ and {}4$_{13}$-{}4$_{14}$ Transitions}",
      journal = {\apj},
         year = 2011,
        month = nov,
       volume = {742},
       number = {1},
          eid = {58},
        pages = {58},
          doi = {10.1088/0004-637X/742/1/58},
archivePrefix = {arXiv},
       eprint = {1108.3719},
 primaryClass = {astro-ph.GA},
       adsurl = {https://ui.adsabs.harvard.edu/abs/2011ApJ...742...58M}
}

@ARTICLE{2013ApJ...766..108M,
       author = {{Mangum}, Jeffrey G. and {Darling}, Jeremy and {Henkel}, Christian and {Menten}, Karl M.},
        title = "{Formaldehyde Densitometry of Starburst Galaxies: Density-independent Global Star Formation}",
      journal = {\apj},
         year = 2013,
        month = apr,
       volume = {766},
       number = {2},
          eid = {108},
        pages = {108},
          doi = {10.1088/0004-637X/766/2/108},
archivePrefix = {arXiv},
       eprint = {1302.3526},
 primaryClass = {astro-ph.CO},
       adsurl = {https://ui.adsabs.harvard.edu/abs/2013ApJ...766..108M}
}

@ARTICLE{2017A&A...598A..30T,
       author = {{Tang}, X.~D. and {Henkel}, C. and {Menten}, K.~M. and {Zheng}, X.~W. and {Esimbek}, J. and {Zhou}, J.~J. and {Yeh}, C.~C. and {K{\"o}nig}, C. and {Yuan}, Y. and {He}, Y.~X. and {Li}, D.~L.},
        title = "{Kinetic temperature of massive star forming molecular clumps measured with formaldehyde}",
      journal = {\aap},
         year = 2017,
        month = feb,
       volume = {598},
          eid = {A30},
        pages = {A30},
          doi = {10.1051/0004-6361/201629694},
archivePrefix = {arXiv},
       eprint = {1610.05548},
 primaryClass = {astro-ph.GA},
       adsurl = {https://ui.adsabs.harvard.edu/abs/2017A&A...598A..30T}
}

@ARTICLE{1977ApJ...214..725E,
       author = {{Elmegreen}, B.~G. and {Lada}, C.~J.},
        title = "{Sequential formation of subgroups in OB associations.}",
      journal = {\apj},
         year = 1977,
        month = jun,
       volume = {214},
        pages = {725-741},
          doi = {10.1086/155302},
       adsurl = {https://ui.adsabs.harvard.edu/abs/1977ApJ...214..725E}
}

@ARTICLE{2003A&A...399.1135D,
       author = {{Deharveng}, L. and {Zavagno}, A. and {Salas}, L. and {Porras}, A. and {Caplan}, J. and {Cruz-Gonz{\'a}lez}, I.},
        title = "{Sequential star formation at the periphery of the H II regions Sh 217 and Sh 219}",
      journal = {\aap},
         year = 2003,
        month = mar,
       volume = {399},
        pages = {1135-1145},
          doi = {10.1051/0004-6361:20021841},
archivePrefix = {arXiv},
       eprint = {astro-ph/0212383},
 primaryClass = {astro-ph},
       adsurl = {https://ui.adsabs.harvard.edu/abs/2003A&A...399.1135D}
}

@ARTICLE{2006ApJS..164..506L,
       author = {{Le Petit}, Franck and {Nehm{\'e}}, Cyrine and {Le Bourlot}, Jacques and {Roueff}, Evelyne},
        title = "{A Model for Atomic and Molecular Interstellar Gas: The Meudon PDR Code}",
      journal = {\apjs},
         year = 2006,
        month = jun,
       volume = {164},
       number = {2},
        pages = {506-529},
          doi = {10.1086/503252},
archivePrefix = {arXiv},
       eprint = {astro-ph/0602150},
 primaryClass = {astro-ph},
       adsurl = {https://ui.adsabs.harvard.edu/abs/2006ApJS..164..506L}
}

@ARTICLE{2005A&A...436.1049M,
       author = {{Martins}, F. and {Schaerer}, D. and {Hillier}, D.~J.},
        title = "{A new calibration of stellar parameters of Galactic O stars}",
      journal = {\aap},
         year = 2005,
        month = jun,
       volume = {436},
       number = {3},
        pages = {1049-1065},
          doi = {10.1051/0004-6361:20042386},
archivePrefix = {arXiv},
       eprint = {astro-ph/0503346},
 primaryClass = {astro-ph},
       adsurl = {https://ui.adsabs.harvard.edu/abs/2005A&A...436.1049M}
}

@ARTICLE{1968BAN....19..421H,
       author = {{Habing}, H.~J.},
        title = "{The interstellar radiation density between 912 A and 2400 A}",
      journal = {\bain},
         year = 1968,
        month = jan,
       volume = {19},
        pages = {421},
       adsurl = {https://ui.adsabs.harvard.edu/abs/1968BAN....19..421H}
}

@ARTICLE{1990ApJS...72..163F,
       author = {{Fitzpatrick}, Edward L. and {Massa}, Derck},
        title = "{An Analysis of the Shapes of Ultraviolet Extinction Curves. III. an Atlas of Ultraviolet Extinction Curves}",
      journal = {\apjs},
         year = 1990,
        month = jan,
       volume = {72},
        pages = {163},
          doi = {10.1086/191413},
       adsurl = {https://ui.adsabs.harvard.edu/abs/1990ApJS...72..163F}
}

@ARTICLE{2018A&A...615A.129J,
       author = {{Joblin}, C. and {Bron}, E. and {Pinto}, C. and {Pilleri}, P. and {Le Petit}, F. and {Gerin}, M. and {Le Bourlot}, J. and {Fuente}, A. and {Berne}, O. and {Goicoechea}, J.~R. and {Habart}, E. and {K{\"o}hler}, M. and {Teyssier}, D. and {Nagy}, Z. and {Montillaud}, J. and {Vastel}, C. and {Cernicharo}, J. and {R{\"o}llig}, M. and {Ossenkopf-Okada}, V. and {Bergin}, E.~A.},
        title = "{Structure of photodissociation fronts in star-forming regions revealed by Herschel observations of high-J CO emission lines}",
      journal = {\aap},
         year = 2018,
        month = jul,
       volume = {615},
          eid = {A129},
        pages = {A129},
          doi = {10.1051/0004-6361/201832611},
archivePrefix = {arXiv},
       eprint = {1801.03893},
 primaryClass = {astro-ph.GA},
       adsurl = {https://ui.adsabs.harvard.edu/abs/2018A&A...615A.129J}
}

@ARTICLE{2011ApJ...736..149G,
       author = {{Ginsburg}, Adam and {Darling}, Jeremy and {Battersby}, Cara and {Zeiger}, Ben and {Bally}, John},
        title = "{Galactic H$_{2}$CO Densitometry. I. Pilot Survey of Ultracompact H II Regions and Methodology}",
      journal = {\apj},
         year = 2011,
        month = aug,
       volume = {736},
       number = {2},
          eid = {149},
        pages = {149},
          doi = {10.1088/0004-637X/736/2/149},
archivePrefix = {arXiv},
       eprint = {1106.1430},
 primaryClass = {astro-ph.GA},
       adsurl = {https://ui.adsabs.harvard.edu/abs/2011ApJ...736..149G}
}

@ARTICLE{2011A&A...532A.127D,
       author = {{Du}, Z.~M. and {Zhou}, J.~J. and {Esimbek}, J. and {Han}, X.~H. and {Zhang}, C.~P.},
        title = "{A H$_{2}$CO and H110{\ensuremath{\alpha}} survey of H ii regions with the 25-m radio telescope of Nanshan Station}",
      journal = {\aap},
         year = 2011,
        month = aug,
       volume = {532},
          eid = {A127},
        pages = {A127},
          doi = {10.1051/0004-6361/201116506},
       adsurl = {https://ui.adsabs.harvard.edu/abs/2011A&A...532A.127D}
}

@ARTICLE{1980A&AS...40..379D,
       author = {{Downes}, D. and {Wilson}, T.~L. and {Bieging}, J. and {Wink}, J.},
        title = "{H110alpha and H2CO survey of galactic radio sources.}",
      journal = {\aaps},
         year = 1980,
        month = jun,
       volume = {40},
        pages = {379-394},
       adsurl = {https://ui.adsabs.harvard.edu/abs/1980A&AS...40..379D}
}

@ARTICLE{1983A&A...127..388H,
       author = {{Henkel}, C. and {Wilson}, T.~L. and {Walmsley}, C.~M. and {Pauls}, T.},
        title = "{Formaldehyde towards compact HII regions : densities and isotope ratios.}",
      journal = {\aap},
         year = 1983,
        month = nov,
       volume = {127},
        pages = {388-394},
       adsurl = {https://ui.adsabs.harvard.edu/abs/1983A&A...127..388H}
}

@ARTICLE{1990ApJ...348..542M,
       author = {{Mangum}, Jeffrey G. and {Wootten}, Alwyn and {Loren}, Robert B. and {Wadiak}, E. James},
        title = "{Observations of the Formaldehyde Emission in Orion-KL: Abundances, Distribution, and Kinematics of the Dense Gas in the Orion Molecular Ridge}",
      journal = {\apj},
         year = 1990,
        month = jan,
       volume = {348},
        pages = {542},
          doi = {10.1086/168262},
       adsurl = {https://ui.adsabs.harvard.edu/abs/1990ApJ...348..542M}
}

@ARTICLE{2020A&A...634A..52S,
       author = {{Simons}, M.~A.~J. and {Lamberts}, T. and {Cuppen}, H.~M.},
        title = "{Formation of COMs through CO hydrogenation on interstellar grains}",
      journal = {\aap},
         year = 2020,
        month = feb,
       volume = {634},
          eid = {A52},
        pages = {A52},
          doi = {10.1051/0004-6361/201936522},
archivePrefix = {arXiv},
       eprint = {2001.04895},
 primaryClass = {astro-ph.SR},
       adsurl = {https://ui.adsabs.harvard.edu/abs/2020A&A...634A..52S}
}

@ARTICLE{2008A&A...481..345M,
       author = {{Molinari}, S. and {Pezzuto}, S. and {Cesaroni}, R. and {Brand}, J. and {Faustini}, F. and {Testi}, L.},
        title = "{The evolution of the spectral energy distribution in massive young stellar objects}",
      journal = {\aap},
         year = 2008,
        month = apr,
       volume = {481},
       number = {2},
        pages = {345-365},
          doi = {10.1051/0004-6361:20078661},
       adsurl = {https://ui.adsabs.harvard.edu/abs/2008A&A...481..345M}
}

@ARTICLE{2016ApJ...826L...8M,
       author = {{Molinari}, S. and {Merello}, M. and {Elia}, D. and {Cesaroni}, R. and {Testi}, L. and {Robitaille}, T.},
        title = "{Calibration of Evolutionary Diagnostics in High-mass Star Formation}",
      journal = {\apjl},
         year = 2016,
        month = jul,
       volume = {826},
       number = {1},
          eid = {L8},
        pages = {L8},
          doi = {10.3847/2041-8205/826/1/L8},
archivePrefix = {arXiv},
       eprint = {1604.06192},
 primaryClass = {astro-ph.GA},
       adsurl = {https://ui.adsabs.harvard.edu/abs/2016ApJ...826L...8M}
}

@INPROCEEDINGS{2005IAUS..227..111K,
       author = {{Kurtz}, Stan},
        title = "{Hypercompact HII regions}",
    booktitle = {Massive Star Birth: A Crossroads of Astrophysics},
         year = 2005,
       editor = {{Cesaroni}, R. and {Felli}, M. and {Churchwell}, E. and {Walmsley}, M.},
       series = {IAU Symposium},
       volume = {227},
        month = jan,
        pages = {111-119},
          doi = {10.1017/S1743921305004424},
       adsurl = {https://ui.adsabs.harvard.edu/abs/2005IAUS..227..111K}
}

@ARTICLE{2004ApJ...605..285S,
       author = {{Sewilo}, M. and {Churchwell}, E. and {Kurtz}, S. and {Goss}, W.~M. and {Hofner}, P.},
        title = "{Broad Radio Recombination Lines from Hypercompact H II Regions}",
      journal = {\apj},
         year = 2004,
        month = apr,
       volume = {605},
       number = {1},
        pages = {285-299},
          doi = {10.1086/382268},
       adsurl = {https://ui.adsabs.harvard.edu/abs/2004ApJ...605..285S}
}

@ARTICLE{2013A&A...550A.135A,
       author = {{Ao}, Y. and {Henkel}, C. and {Menten}, K.~M. and {Requena-Torres}, M.~A. and {Stanke}, T. and {Mauersberger}, R. and {Aalto}, S. and {M{\"u}hle}, S. and {Mangum}, J.},
        title = "{The thermal state of molecular clouds in the Galactic center: evidence for non-photon-driven heating}",
      journal = {\aap},
         year = 2013,
        month = feb,
       volume = {550},
          eid = {A135},
        pages = {A135},
          doi = {10.1051/0004-6361/201220096},
archivePrefix = {arXiv},
       eprint = {1211.7142},
 primaryClass = {astro-ph.GA},
       adsurl = {https://ui.adsabs.harvard.edu/abs/2013A&A...550A.135A}
}

@ARTICLE{2014A&A...563A..97G,
       author = {{Gerner}, T. and {Beuther}, H. and {Semenov}, D. and {Linz}, H. and {Vasyunina}, T. and {Bihr}, S. and {Shirley}, Y.~L. and {Henning}, Th.},
        title = "{Chemical evolution in the early phases of massive star formation. I}",
      journal = {\aap},
         year = 2014,
        month = mar,
       volume = {563},
          eid = {A97},
        pages = {A97},
          doi = {10.1051/0004-6361/201322541},
archivePrefix = {arXiv},
       eprint = {1401.6382},
 primaryClass = {astro-ph.SR},
       adsurl = {https://ui.adsabs.harvard.edu/abs/2014A&A...563A..97G}
}

@ARTICLE{2011ApJS..194...32A,
       author = {{Anderson}, L.~D. and {Bania}, T.~M. and {Balser}, Dana S. and {Rood}, Robert T.},
        title = "{The Green Bank Telescope H II Region Discovery Survey. II. The Source Catalog}",
      journal = {\apjs},
         year = 2011,
        month = jun,
       volume = {194},
       number = {2},
          eid = {32},
        pages = {32},
          doi = {10.1088/0067-0049/194/2/32},
archivePrefix = {arXiv},
       eprint = {1103.5085},
 primaryClass = {astro-ph.GA},
       adsurl = {https://ui.adsabs.harvard.edu/abs/2011ApJS..194...32A}
}

@ARTICLE{2022ApJ...935..167A,
       author = {{Astropy Collaboration} and {Price-Whelan}, Adrian M. and {Lim}, Pey Lian and {Earl}, Nicholas and {Starkman}, Nathaniel and {Bradley}, Larry and {Shupe}, David L. and {Patil}, Aarya A. and {Corrales}, Lia and {Brasseur}, C.~E. and {N{\"o}the}, Maximilian and {Donath}, Axel and {Tollerud}, Erik and {Morris}, Brett M. and {Ginsburg}, Adam and {Vaher}, Eero and {Weaver}, Benjamin A. and {Tocknell}, James and {Jamieson}, William and {van Kerkwijk}, Marten H. and {Robitaille}, Thomas P. and {Merry}, Bruce and {Bachetti}, Matteo and {G{\"u}nther}, H. Moritz and {Aldcroft}, Thomas L. and {Alvarado-Montes}, Jaime A. and {Archibald}, Anne M. and {B{\'o}di}, Attila and {Bapat}, Shreyas and {Barentsen}, Geert and {Baz{\'a}n}, Juanjo and {Biswas}, Manish and {Boquien}, M{\'e}d{\'e}ric and {Burke}, D.~J. and {Cara}, Daria and {Cara}, Mihai and {Conroy}, Kyle E. and {Conseil}, Simon and {Craig}, Matthew W. and {Cross}, Robert M. and {Cruz}, Kelle L. and {D'Eugenio}, Francesco and {Dencheva}, Nadia and {Devillepoix}, Hadrien A.~R. and {Dietrich}, J{\"o}rg P. and {Eigenbrot}, Arthur Davis and {Erben}, Thomas and {Ferreira}, Leonardo and {Foreman-Mackey}, Daniel and {Fox}, Ryan and {Freij}, Nabil and {Garg}, Suyog and {Geda}, Robel and {Glattly}, Lauren and {Gondhalekar}, Yash and {Gordon}, Karl D. and {Grant}, David and {Greenfield}, Perry and {Groener}, Austen M. and {Guest}, Steve and {Gurovich}, Sebastian and {Handberg}, Rasmus and {Hart}, Akeem and {Hatfield-Dodds}, Zac and {Homeier}, Derek and {Hosseinzadeh}, Griffin and {Jenness}, Tim and {Jones}, Craig K. and {Joseph}, Prajwel and {Kalmbach}, J. Bryce and {Karamehmetoglu}, Emir and {Ka{\l}uszy{\'n}ski}, Miko{\l}aj and {Kelley}, Michael S.~P. and {Kern}, Nicholas and {Kerzendorf}, Wolfgang E. and {Koch}, Eric W. and {Kulumani}, Shankar and {Lee}, Antony and {Ly}, Chun and {Ma}, Zhiyuan and {MacBride}, Conor and {Maljaars}, Jakob M. and {Muna}, Demitri and {Murphy}, N.~A. and {Norman}, Henrik and {O'Steen}, Richard and {Oman}, Kyle A. and {Pacifici}, Camilla and {Pascual}, Sergio and {Pascual-Granado}, J. and {Patil}, Rohit R. and {Perren}, Gabriel I. and {Pickering}, Timothy E. and {Rastogi}, Tanuj and {Roulston}, Benjamin R. and {Ryan}, Daniel F. and {Rykoff}, Eli S. and {Sabater}, Jose and {Sakurikar}, Parikshit and {Salgado}, Jes{\'u}s and {Sanghi}, Aniket and {Saunders}, Nicholas and {Savchenko}, Volodymyr and {Schwardt}, Ludwig and {Seifert-Eckert}, Michael and {Shih}, Albert Y. and {Jain}, Anany Shrey and {Shukla}, Gyanendra and {Sick}, Jonathan and {Simpson}, Chris and {Singanamalla}, Sudheesh and {Singer}, Leo P. and {Singhal}, Jaladh and {Sinha}, Manodeep and {Sip{\H{o}}cz}, Brigitta M. and {Spitler}, Lee R. and {Stansby}, David and {Streicher}, Ole and {{\v{S}}umak}, Jani and {Swinbank}, John D. and {Taranu}, Dan S. and {Tewary}, Nikita and {Tremblay}, Grant R. and {de Val-Borro}, Miguel and {Van Kooten}, Samuel J. and {Vasovi{\'c}}, Zlatan and {Verma}, Shresth and {de Miranda Cardoso}, Jos{\'e} Vin{\'\i}cius and {Williams}, Peter K.~G. and {Wilson}, Tom J. and {Winkel}, Benjamin and {Wood-Vasey}, W.~M. and {Xue}, Rui and {Yoachim}, Peter and {Zhang}, Chen and {Zonca}, Andrea and {Astropy Project Contributors}},
        title = "{The Astropy Project: Sustaining and Growing a Community-oriented Open-source Project and the Latest Major Release (v5.0) of the Core Package}",
      journal = {\apj},
         year = 2022,
        month = aug,
       volume = {935},
       number = {2},
          eid = {167},
        pages = {167},
          doi = {10.3847/1538-4357/ac7c74},
archivePrefix = {arXiv},
       eprint = {2206.14220},
 primaryClass = {astro-ph.IM},
       adsurl = {https://ui.adsabs.harvard.edu/abs/2022ApJ...935..167A}
}

@ARTICLE{2025arXiv250616700F,
       author = {{Faerber}, Timothy and {Anderson}, Loren D. and {Luisi}, Matteo and {Bonne}, Lars and {Schneider}, Nicola and {Ossenkopf-Okada}, Volker and {Tielens}, Alexander and {Simon}, Robert and {R{\"o}llig}, Markus},
        title = "{Expansion Signatures in 35 HII Regions traced by SOFIA [CII] Emission}",
      journal = {arXiv e-prints},
         year = 2025,
        month = jun,
          eid = {arXiv:2506.16700},
        pages = {arXiv:2506.16700},
          doi = {10.48550/arXiv.2506.16700},
archivePrefix = {arXiv},
       eprint = {2506.16700},
 primaryClass = {astro-ph.GA},
       adsurl = {https://ui.adsabs.harvard.edu/abs/2025arXiv250616700F}
}

@ARTICLE{2018ARA&A..56...41M,
       author = {{Motte}, Fr{\'e}d{\'e}rique and {Bontemps}, Sylvain and {Louvet}, Fabien},
        title = "{High-Mass Star and Massive Cluster Formation in the Milky Way}",
      journal = {\araa},
         year = 2018,
        month = sep,
       volume = {56},
        pages = {41-82},
          doi = {10.1146/annurev-astro-091916-055235},
archivePrefix = {arXiv},
       eprint = {1706.00118},
 primaryClass = {astro-ph.GA},
       adsurl = {https://ui.adsabs.harvard.edu/abs/2018ARA&A..56...41M}
}

@ARTICLE{1977JChPh..67.1576F,
       author = {{Fabricant}, Barbara and {Krieger}, David and {Muenter}, J.~S.},
        title = "{Molecular beam electric resonance study of formaldehyde, thioformaldehyde, and ketene}",
      journal = {\jcp},
         year = 1977,
        month = aug,
       volume = {67},
       number = {4},
        pages = {1576-1586},
          doi = {10.1063/1.434988},
       adsurl = {https://ui.adsabs.harvard.edu/abs/1977JChPh..67.1576F}
}

@ARTICLE{2023A&A...678A.130G,
       author = {{Gong}, Y. and {Ortiz-Le{\'o}n}, G.~N. and {Rugel}, M.~R. and {Menten}, K.~M. and {Brunthaler}, A. and {Wyrowski}, F. and {Henkel}, C. and {Beuther}, H. and {Dzib}, S.~A. and {Urquhart}, J.~S. and {Yang}, A.~Y. and {Pandian}, J.~D. and {Dokara}, R. and {Veena}, V.~S. and {Nguyen}, H. and {Medina}, S. -N.~X. and {Cotton}, W.~D. and {Reich}, W. and {Winkel}, B. and {M{\"u}ller}, P. and {Skretas}, I. and {Csengeri}, T. and {Khan}, S. and {Cheema}, A.},
        title = "{A global view on star formation: The GLOSTAR Galactic plane survey. VIII. Formaldehyde absorption in Cygnus X}",
      journal = {\aap},
         year = 2023,
        month = oct,
       volume = {678},
          eid = {A130},
        pages = {A130},
          doi = {10.1051/0004-6361/202346102},
archivePrefix = {arXiv},
       eprint = {2308.00993},
 primaryClass = {astro-ph.GA},
       adsurl = {https://ui.adsabs.harvard.edu/abs/2023A&A...678A.130G}
}

@ARTICLE{2006A&A...454L..13G,
       author = {{G{\"u}sten}, R. and {Nyman}, L. {\r{A}}. and {Schilke}, P. and {Menten}, K. and {Cesarsky}, C. and {Booth}, R.},
        title = "{The Atacama Pathfinder EXperiment (APEX) - a new submillimeter facility for southern skies -}",
      journal = {\aap},
         year = 2006,
        month = aug,
       volume = {454},
       number = {2},
        pages = {L13-L16},
          doi = {10.1051/0004-6361:20065420},
       adsurl = {https://ui.adsabs.harvard.edu/abs/2006A&A...454L..13G}
}

@ARTICLE{2022A&A...668A...2M,
       author = {{Meledin}, D. and {Lapkin}, I. and {Fredrixon}, M. and {Sundin}, E. and {Ferm}, S. -E. and {Pavolotsky}, A. and {Strandberg}, M. and {Desmaris}, V. and {L{\'o}pez}, C. and {Bergman}, P. and {Olberg}, M. and {Conway}, J. and {Torstensson}, K. and {Dur{\'a}n}, C. and {Montenegro-Montes}, F.~M. and {De Breuck}, C. and {Belitsky}, V.},
        title = "{SEPIA345: A 345 GHz dual polarization heterodyne receiver channel for SEPIA at the APEX telescope}",
      journal = {\aap},
         year = 2022,
        month = dec,
       volume = {668},
          eid = {A2},
        pages = {A2},
          doi = {10.1051/0004-6361/202244211},
       adsurl = {https://ui.adsabs.harvard.edu/abs/2022A&A...668A...2M}
}

@ARTICLE{2001A&A...370L..49M,
       author = {{M{\"u}ller}, H.~S.~P. and {Thorwirth}, S. and {Roth}, D.~A. and {Winnewisser}, G.},
        title = "{The Cologne Database for Molecular Spectroscopy, CDMS}",
      journal = {\aap},
         year = 2001,
        month = apr,
       volume = {370},
        pages = {L49-L52},
          doi = {10.1051/0004-6361:20010367},
       adsurl = {https://ui.adsabs.harvard.edu/abs/2001A&A...370L..49M}
}

@ARTICLE{2019JOSS....4.1864F,
       author = {{Foreman-Mackey}, Daniel and {Farr}, Will and {Sinha}, Manodeep and {Archibald}, Anne and {Hogg}, David and {Sanders}, Jeremy and {Zuntz}, Joe and {Williams}, Peter and {Nelson}, Andrew and {de Val-Borro}, Miguel and {Erhardt}, Tobias and {Pashchenko}, Ilya and {Pla}, Oriol},
        title = "{emcee v3: A Python ensemble sampling toolkit for affine-invariant MCMC}",
      journal = {The Journal of Open Source Software},
         year = 2019,
        month = nov,
       volume = {4},
       number = {43},
          eid = {1864},
        pages = {1864},
          doi = {10.21105/joss.01864},
archivePrefix = {arXiv},
       eprint = {1911.07688},
 primaryClass = {astro-ph.IM},
       adsurl = {https://ui.adsabs.harvard.edu/abs/2019JOSS....4.1864F}
}

@ARTICLE{2019ApJ...885..131R,
       author = {{Reid}, M.~J. and {Menten}, K.~M. and {Brunthaler}, A. and {Zheng}, X.~W. and {Dame}, T.~M. and {Xu}, Y. and {Li}, J. and {Sakai}, N. and {Wu}, Y. and {Immer}, K. and {Zhang}, B. and {Sanna}, A. and {Moscadelli}, L. and {Rygl}, K.~L.~J. and {Bartkiewicz}, A. and {Hu}, B. and {Quiroga-Nu{\~n}ez}, L.~H. and {van Langevelde}, H.~J.},
        title = "{Trigonometric Parallaxes of High-mass Star-forming Regions: Our View of the Milky Way}",
      journal = {\apj},
         year = 2019,
        month = nov,
       volume = {885},
       number = {2},
          eid = {131},
        pages = {131},
          doi = {10.3847/1538-4357/ab4a11},
archivePrefix = {arXiv},
       eprint = {1910.03357},
 primaryClass = {astro-ph.GA},
       adsurl = {https://ui.adsabs.harvard.edu/abs/2019ApJ...885..131R}
}

@ARTICLE{2005ApJ...634.1133R,
       author = {{Rizzo}, J.~R. and {Fuente}, A. and {Garc{\'\i}a-Burillo}, S.},
        title = "{Insights into the Carbon Chemistry of Monoceros R2}",
      journal = {\apj},
         year = 2005,
        month = dec,
       volume = {634},
       number = {2},
        pages = {1133-1145},
          doi = {10.1086/497128},
archivePrefix = {arXiv},
       eprint = {astro-ph/0508311},
 primaryClass = {astro-ph},
       adsurl = {https://ui.adsabs.harvard.edu/abs/2005ApJ...634.1133R}
}

@ARTICLE{2020A&A...644A.160K,
       author = {{Kim}, W. -J. and {Wyrowski}, F. and {Urquhart}, J.~S. and {P{\'e}rez-Beaupuits}, J.~P. and {Pillai}, T. and {Tiwari}, M. and {Menten}, K.~M.},
        title = "{ATLASGAL-selected massive clumps in the inner Galaxy. VIII. Chemistry of photodissociation regions}",
      journal = {\aap},
         year = 2020,
        month = dec,
       volume = {644},
          eid = {A160},
        pages = {A160},
          doi = {10.1051/0004-6361/202039059},
archivePrefix = {arXiv},
       eprint = {2009.14238},
 primaryClass = {astro-ph.GA},
       adsurl = {https://ui.adsabs.harvard.edu/abs/2020A&A...644A.160K}
}

@ARTICLE{2009ARA&A..47..427H,
       author = {{Herbst}, Eric and {van Dishoeck}, Ewine F.},
        title = "{Complex Organic Interstellar Molecules}",
      journal = {\araa},
         year = 2009,
        month = sep,
       volume = {47},
       number = {1},
        pages = {427-480},
          doi = {10.1146/annurev-astro-082708-101654},
       adsurl = {https://ui.adsabs.harvard.edu/abs/2009ARA&A..47..427H}
}

@ARTICLE{2009A&A...505..629F,
       author = {{Fuchs}, G.~W. and {Cuppen}, H.~M. and {Ioppolo}, S. and {Romanzin}, C. and {Bisschop}, S.~E. and {Andersson}, S. and {van Dishoeck}, E.~F. and {Linnartz}, H.},
        title = "{Hydrogenation reactions in interstellar CO ice analogues. A combined experimental/theoretical approach}",
      journal = {\aap},
         year = 2009,
        month = oct,
       volume = {505},
       number = {2},
        pages = {629-639},
          doi = {10.1051/0004-6361/200810784},
       adsurl = {https://ui.adsabs.harvard.edu/abs/2009A&A...505..629F}
}

@ARTICLE{2018MNRAS.479.2007A,
       author = {{{\'A}lvarez-Barcia}, S. and {Russ}, P. and {K{\"a}stner}, J. and {Lamberts}, T.},
        title = "{Hydrogen transfer reactions of interstellar complex organic molecules}",
      journal = {\mnras},
         year = 2018,
        month = sep,
       volume = {479},
       number = {2},
        pages = {2007-2015},
          doi = {10.1093/mnras/sty1478},
archivePrefix = {arXiv},
       eprint = {1806.02062},
 primaryClass = {astro-ph.GA},
       adsurl = {https://ui.adsabs.harvard.edu/abs/2018MNRAS.479.2007A}
}

@ARTICLE{2022ApJ...931L..33S,
       author = {{Santos}, Julia C. and {Chuang}, Ko-Ju and {Lamberts}, Thanja and {Fedoseev}, Gleb and {Ioppolo}, Sergio and {Linnartz}, Harold},
        title = "{First Experimental Confirmation of the CH$_{3}$O + H$_{2}$CO {\textrightarrow} CH$_{3}$OH + HCO Reaction: Expanding the CH$_{3}$OH Formation Mechanism in Interstellar Ices}",
      journal = {\apjl},
         year = 2022,
        month = jun,
       volume = {931},
       number = {2},
          eid = {L33},
        pages = {L33},
          doi = {10.3847/2041-8213/ac7158},
archivePrefix = {arXiv},
       eprint = {2205.12284},
 primaryClass = {astro-ph.GA},
       adsurl = {https://ui.adsabs.harvard.edu/abs/2022ApJ...931L..33S}
}

@ARTICLE{2021A&A...656A.148R,
       author = {{Ramal-Olmedo}, Juan C. and {Menor-Salv{\'a}n}, C{\'e}sar A. and {Fortenberry}, Ryan C.},
        title = "{Mechanisms for gas-phase molecular formation of neutral formaldehyde (H$_{2}$CO) in cold astrophysical regions}",
      journal = {\aap},
         year = 2021,
        month = dec,
       volume = {656},
          eid = {A148},
        pages = {A148},
          doi = {10.1051/0004-6361/202141616},
       adsurl = {https://ui.adsabs.harvard.edu/abs/2021A&A...656A.148R}
}

@ARTICLE{2007A&A...466.1197W,
       author = {{Woodall}, J. and {Ag{\'u}ndez}, M. and {Markwick-Kemper}, A.~J. and {Millar}, T.~J.},
        title = "{The UMIST database for astrochemistry 2006}",
      journal = {\aap},
         year = 2007,
        month = may,
       volume = {466},
       number = {3},
        pages = {1197-1204},
          doi = {10.1051/0004-6361:20064981},
       adsurl = {https://ui.adsabs.harvard.edu/abs/2007A&A...466.1197W}
}

@ARTICLE{2015ApJ...800...40I,
       author = {{Indriolo}, Nick and {Neufeld}, D.~A. and {Gerin}, M. and {Schilke}, P. and {Benz}, A.~O. and {Winkel}, B. and {Menten}, K.~M. and {Chambers}, E.~T. and {Black}, John H. and {Bruderer}, S. and {Falgarone}, E. and {Godard}, B. and {Goicoechea}, J.~R. and {Gupta}, H. and {Lis}, D.~C. and {Ossenkopf}, V. and {Persson}, C.~M. and {Sonnentrucker}, P. and {van der Tak}, F.~F.~S. and {van Dishoeck}, E.~F. and {Wolfire}, Mark G. and {Wyrowski}, F.},
        title = "{Herschel Survey of Galactic OH$^{+}$, H$_{2}$O$^{+}$, and H$_{3}$O$^{+}$: Probing the Molecular Hydrogen Fraction and Cosmic-Ray Ionization Rate}",
      journal = {\apj},
         year = 2015,
        month = feb,
       volume = {800},
       number = {1},
          eid = {40},
        pages = {40},
          doi = {10.1088/0004-637X/800/1/40},
archivePrefix = {arXiv},
       eprint = {1412.1106},
 primaryClass = {astro-ph.GA},
       adsurl = {https://ui.adsabs.harvard.edu/abs/2015ApJ...800...40I}
}

@ARTICLE{2004A&A...417..993L,
       author = {{Le Petit}, F. and {Roueff}, E. and {Herbst}, E.},
        title = "{H$_{3}$$^{+}$ and other species in the diffuse cloud towards {\ensuremath{\zeta}} Persei: A new detailed model}",
      journal = {\aap},
         year = 2004,
        month = apr,
       volume = {417},
        pages = {993-1002},
          doi = {10.1051/0004-6361:20035629},
       adsurl = {https://ui.adsabs.harvard.edu/abs/2004A&A...417..993L}
}

@ARTICLE{2005JMoSt.742..215M,
       author = {{M{\"u}ller}, Holger S.~P. and {Schl{\"o}der}, Frank and {Stutzki}, J{\"u}rgen and {Winnewisser}, Gisbert},
        title = "{The Cologne Database for Molecular Spectroscopy, CDMS: a useful tool for astronomers and spectroscopists}",
      journal = {Journal of Molecular Structure},
         year = 2005,
        month = may,
       volume = {742},
       number = {1-3},
        pages = {215-227},
          doi = {10.1016/j.molstruc.2005.01.027},
       adsurl = {https://ui.adsabs.harvard.edu/abs/2005JMoSt.742..215M}
}

@ARTICLE{2016JMoSp.327...95E,
       author = {{Endres}, Christian P. and {Schlemmer}, Stephan and {Schilke}, Peter and {Stutzki}, J{\"u}rgen and {M{\"u}ller}, Holger S.~P.},
        title = "{The Cologne Database for Molecular Spectroscopy, CDMS, in the Virtual Atomic and Molecular Data Centre, VAMDC}",
      journal = {Journal of Molecular Spectroscopy},
         year = 2016,
        month = sep,
       volume = {327},
        pages = {95-104},
          doi = {10.1016/j.jms.2016.03.005},
       adsurl = {https://ui.adsabs.harvard.edu/abs/2016JMoSp.327...95E}
}

@ARTICLE{2009A&A...507..369W,
       author = {{Wang}, K. and {Wu}, Y.~F. and {Ran}, L. and {Yu}, W.~T. and {Miller}, M.},
        title = "{The relation between $^{\{13}$\}CO J = 2-1 line width in molecular clouds and bolometric luminosity of associated IRAS sources}",
      journal = {\aap},
         year = 2009,
        month = nov,
       volume = {507},
       number = {1},
        pages = {369-376},
          doi = {10.1051/0004-6361/200811104},
archivePrefix = {arXiv},
       eprint = {0909.3312},
 primaryClass = {astro-ph.SR},
       adsurl = {https://ui.adsabs.harvard.edu/abs/2009A&A...507..369W}
}

@ARTICLE{2021A&A...655A..12T,
       author = {{Tang}, X.~D. and {Henkel}, C. and {Menten}, K.~M. and {Gong}, Y. and {Chen}, C. -H.~R. and {Li}, D.~L. and {Lee}, M. -Y. and {Mangum}, J.~G. and {Ao}, Y.~P. and {M{\"u}hle}, S. and {Aalto}, S. and {Garc{\'\i}a-Burillo}, S. and {Mart{\'\i}n}, S. and {Viti}, S. and {Muller}, S. and {Costagliola}, F. and {Asiri}, H. and {Levshakov}, S.~A. and {Spaans}, M. and {Ott}, J. and {Impellizzeri}, C.~M.~V. and {Fukui}, Y. and {He}, Y.~X. and {Esimbek}, J. and {Zhou}, J.~J. and {Zheng}, X.~W. and {Zhao}, X. and {Li}, J.~S.},
        title = "{Kinetic temperature of massive star-forming molecular clumps measured with formaldehyde. IV. The ALMA view of N113 and N159W in the LMC}",
      journal = {\aap},
         year = 2021,
        month = nov,
       volume = {655},
          eid = {A12},
        pages = {A12},
          doi = {10.1051/0004-6361/202141804},
archivePrefix = {arXiv},
       eprint = {2108.10519},
 primaryClass = {astro-ph.GA},
       adsurl = {https://ui.adsabs.harvard.edu/abs/2021A&A...655A..12T}
}

@INPROCEEDINGS{2011EAS....51...45E,
       author = {{Elmegreen}, B.~G.},
        title = "{Triggered Star Formation}",
    booktitle = {EAS Publications Series},
         year = 2011,
       editor = {{Charbonnel}, Corinne and {Montmerle}, Thierry},
       series = {EAS Publications Series},
       volume = {51},
        month = nov,
    publisher = {EDP},
        pages = {45-58},
          doi = {10.1051/eas/1151004},
archivePrefix = {arXiv},
       eprint = {1101.3112},
 primaryClass = {astro-ph.GA},
       adsurl = {https://ui.adsabs.harvard.edu/abs/2011EAS....51...45E}
}

@ARTICLE{2012MNRAS.421..408T,
       author = {{Thompson}, M.~A. and {Urquhart}, J.~S. and {Moore}, T.~J.~T. and {Morgan}, L.~K.},
        title = "{The statistics of triggered star formation: an overdensity of massive young stellar objects around Spitzer bubbles}",
      journal = {\mnras},
         year = 2012,
        month = mar,
       volume = {421},
       number = {1},
        pages = {408-418},
          doi = {10.1111/j.1365-2966.2011.20315.x},
archivePrefix = {arXiv},
       eprint = {1111.0972},
 primaryClass = {astro-ph.GA},
       adsurl = {https://ui.adsabs.harvard.edu/abs/2012MNRAS.421..408T}
}

@ARTICLE{2010A&A...523A...6D,
       author = {{Deharveng}, L. and {Schuller}, F. and {Anderson}, L.~D. and {Zavagno}, A. and {Wyrowski}, F. and {Menten}, K.~M. and {Bronfman}, L. and {Testi}, L. and {Walmsley}, C.~M. and {Wienen}, M.},
        title = "{A gallery of bubbles. The nature of the bubbles observed by Spitzer and what ATLASGAL tells us about the surrounding neutral material}",
      journal = {\aap},
         year = 2010,
        month = nov,
       volume = {523},
          eid = {A6},
        pages = {A6},
          doi = {10.1051/0004-6361/201014422},
archivePrefix = {arXiv},
       eprint = {1008.0926},
 primaryClass = {astro-ph.GA},
       adsurl = {https://ui.adsabs.harvard.edu/abs/2010A&A...523A...6D}
}

@ARTICLE{1979A&AS...35...23A,
       author = {{Altenhoff}, W.~J. and {Downes}, D. and {Pauls}, T. and {Schraml}, J.},
        title = "{Survey of the galactic plane at 4.875 GHz.}",
      journal = {\aaps},
         year = 1979,
        month = jan,
       volume = {35},
        pages = {23},
       adsurl = {https://ui.adsabs.harvard.edu/abs/1979A&AS...35...23A}
}

@ARTICLE{1970A&A.....4..357R,
       author = {{Reifenstein}, E.~C. and {Wilson}, T.~L. and {Burke}, B.~F. and {Mezger}, P.~G. and {Altenhoff}, W.~J.},
        title = "{A Survey of H1O9{\ensuremath{\alpha}} Recombination Line Emission in Galactic HII Regions of the Northern Sky}",
      journal = {\aap},
         year = 1970,
        month = mar,
       volume = {4},
        pages = {357-377},
       adsurl = {https://ui.adsabs.harvard.edu/abs/1970A&A.....4..357R}
}

@ARTICLE{1989ApJS...71..469L,
       author = {{Lockman}, Felix J.},
        title = "{A Survey of Radio H II Regions in the Northern Sky}",
      journal = {\apjs},
         year = 1989,
        month = nov,
       volume = {71},
        pages = {469},
          doi = {10.1086/191383},
       adsurl = {https://ui.adsabs.harvard.edu/abs/1989ApJS...71..469L}
}

@ARTICLE{1997ApJ...488..224K,
       author = {{Kuchar}, Thomas A. and {Clark}, Frank O.},
        title = "{All-Sky 4.85 GHz Flux Measurements of H II Regions}",
      journal = {\apj},
         year = 1997,
        month = oct,
       volume = {488},
       number = {1},
        pages = {224-233},
          doi = {10.1086/304697},
       adsurl = {https://ui.adsabs.harvard.edu/abs/1997ApJ...488..224K}
}

@ARTICLE{2018A&A...615A.103K,
       author = {{Kalcheva}, I.~E. and {Hoare}, M.~G. and {Urquhart}, J.~S. and {Kurtz}, S. and {Lumsden}, S.~L. and {Purcell}, C.~R. and {Zijlstra}, A.~A.},
        title = "{The coordinated radio and infrared survey for high-mass star formation. III. A catalogue of northern ultra-compact H II regions}",
      journal = {\aap},
         year = 2018,
        month = jul,
       volume = {615},
          eid = {A103},
        pages = {A103},
          doi = {10.1051/0004-6361/201832734},
archivePrefix = {arXiv},
       eprint = {1803.09334},
 primaryClass = {astro-ph.GA},
       adsurl = {https://ui.adsabs.harvard.edu/abs/2018A&A...615A.103K}
}

@ARTICLE{2019A&A...623A.105G,
       author = {{Gao}, X.~Y. and {Reich}, P. and {Hou}, L.~G. and {Reich}, W. and {Han}, J.~L.},
        title = "{A Sino-German {\ensuremath{\lambda}}6 cm polarisation survey of the Galactic plane. IX. H II regions}",
      journal = {\aap},
         year = 2019,
        month = mar,
       volume = {623},
          eid = {A105},
        pages = {A105},
          doi = {10.1051/0004-6361/201834092},
archivePrefix = {arXiv},
       eprint = {1901.00631},
 primaryClass = {astro-ph.GA},
       adsurl = {https://ui.adsabs.harvard.edu/abs/2019A&A...623A.105G}
}

@ARTICLE{2014ApJS..212....1A,
       author = {{Anderson}, L.~D. and {Bania}, T.~M. and {Balser}, Dana S. and {Cunningham}, V. and {Wenger}, T.~V. and {Johnstone}, B.~M. and {Armentrout}, W.~P.},
        title = "{The WISE Catalog of Galactic H II Regions}",
      journal = {\apjs},
         year = 2014,
        month = may,
       volume = {212},
       number = {1},
          eid = {1},
        pages = {1},
          doi = {10.1088/0067-0049/212/1/1},
archivePrefix = {arXiv},
       eprint = {1312.6202},
 primaryClass = {astro-ph.GA},
       adsurl = {https://ui.adsabs.harvard.edu/abs/2014ApJS..212....1A}
}

@ARTICLE{2007A&A...461...11U,
       author = {{Urquhart}, J.~S. and {Busfield}, A.~L. and {Hoare}, M.~G. and {Lumsden}, S.~L. and {Clarke}, A.~J. and {Moore}, T.~J.~T. and {Mottram}, J.~C. and {Oudmaijer}, R.~D.},
        title = "{The RMS survey. Radio observations of candidate massive YSOs in the southern hemisphere}",
      journal = {\aap},
         year = 2007,
        month = jan,
       volume = {461},
       number = {1},
        pages = {11-23},
          doi = {10.1051/0004-6361:20065837},
archivePrefix = {arXiv},
       eprint = {astro-ph/0605738},
 primaryClass = {astro-ph},
       adsurl = {https://ui.adsabs.harvard.edu/abs/2007A&A...461...11U}
}

@ARTICLE{2009A&A...501..539U,
       author = {{Urquhart}, J.~S. and {Hoare}, M.~G. and {Purcell}, C.~R. and {Lumsden}, S.~L. and {Oudmaijer}, R.~D. and {Moore}, T.~J.~T. and {Busfield}, A.~L. and {Mottram}, J.~C. and {Davies}, B.},
        title = "{The RMS survey. 6 cm continuum VLA observations towards candidate massive YSOs in the northern hemisphere}",
      journal = {\aap},
         year = 2009,
        month = jul,
       volume = {501},
       number = {2},
        pages = {539-551},
          doi = {10.1051/0004-6361/200912108},
archivePrefix = {arXiv},
       eprint = {0905.1174},
 primaryClass = {astro-ph.GA},
       adsurl = {https://ui.adsabs.harvard.edu/abs/2009A&A...501..539U}
}

@ARTICLE{2013MNRAS.435..400U,
       author = {{Urquhart}, J.~S. and {Thompson}, M.~A. and {Moore}, T.~J.~T. and {Purcell}, C.~R. and {Hoare}, M.~G. and {Schuller}, F. and {Wyrowski}, F. and {Csengeri}, T. and {Menten}, K.~M. and {Lumsden}, S.~L. and {Kurtz}, S. and {Walmsley}, C.~M. and {Bronfman}, L. and {Morgan}, L.~K. and {Eden}, D.~J. and {Russeil}, D.},
        title = "{ATLASGAL - properties of compact H II regions and their natal clumps}",
      journal = {\mnras},
         year = 2013,
        month = oct,
       volume = {435},
       number = {1},
        pages = {400-428},
          doi = {10.1093/mnras/stt1310},
archivePrefix = {arXiv},
       eprint = {1307.4105},
 primaryClass = {astro-ph.GA},
       adsurl = {https://ui.adsabs.harvard.edu/abs/2013MNRAS.435..400U}
}
	
	\begin{appendix}
    \section{Additional figures}
            \begin{figure*}[ht!]
                \centering
                \includegraphics[width=01.\linewidth]{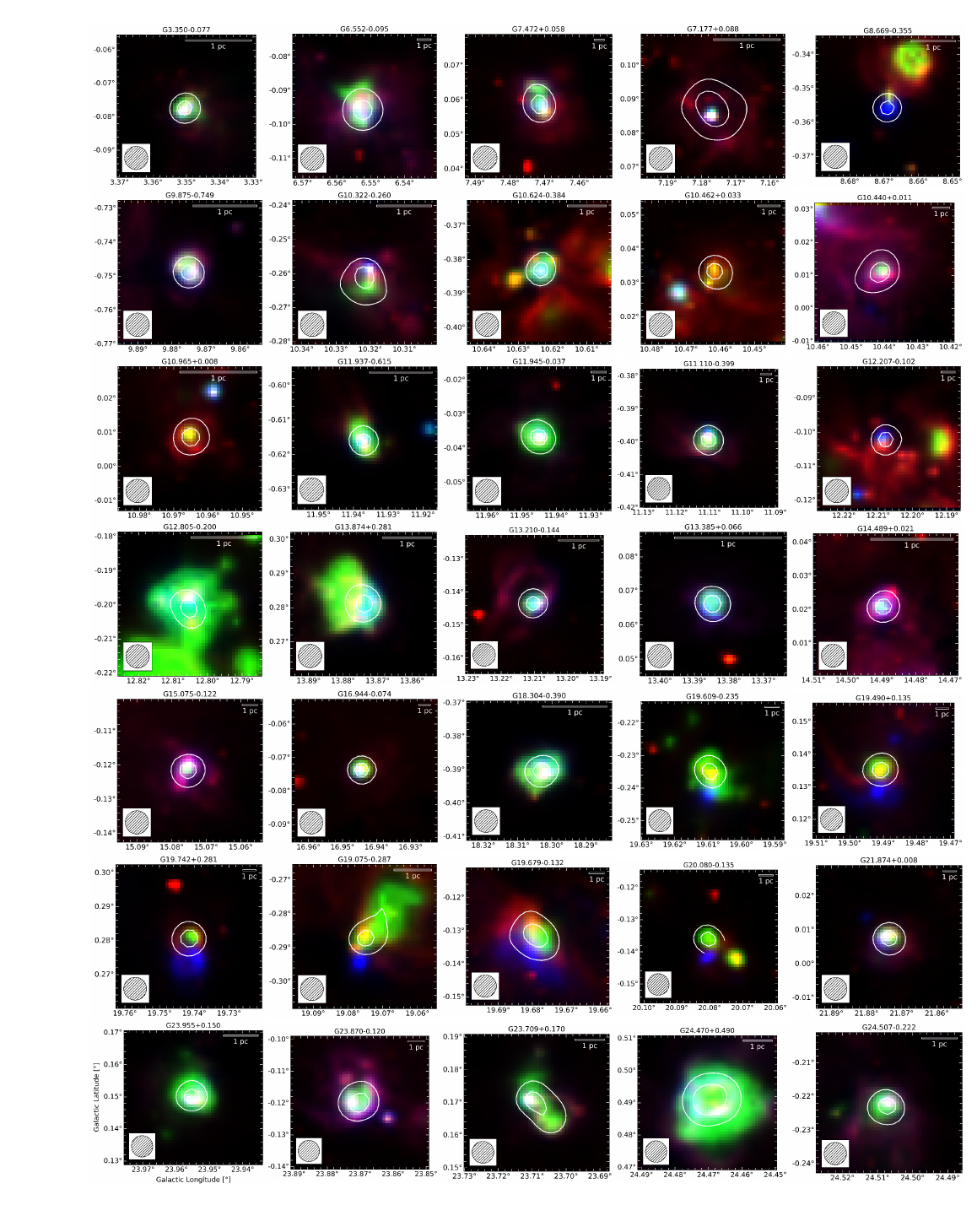}
                \caption{Three-color composite images of the targets of this study. The images combine mid--infrared data from Herschel 70\,$\mu$m (red), MIPSGAL 24\,$\mu$m (green), and GLIMPSE 8.0\,$\mu$m (blue), overlaid with the GLOSTAR 5\,GHz radio continuum contours in white. The filled white circle in corner represents the 25\arcsec~beam of the radio continuum data from Paper~I.}
                \label{fig:rgb_image}
            \end{figure*}

            \begin{figure*}[ht!]\ContinuedFloat
                \centering
                \includegraphics[width=01.\linewidth]{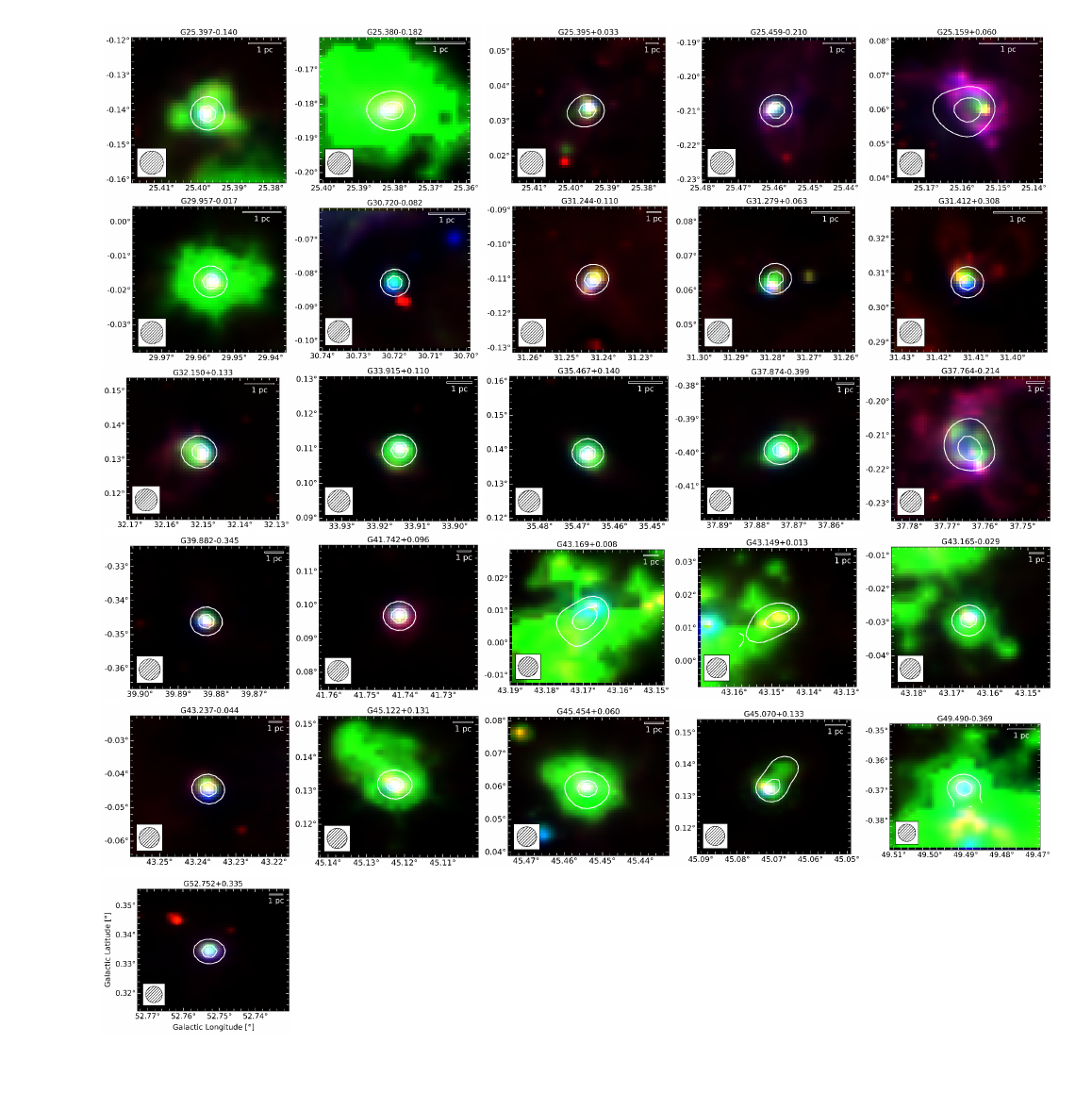}
                \caption{continued.}
            \end{figure*}
            \begin{figure*}
                \includegraphics[width=8cm]{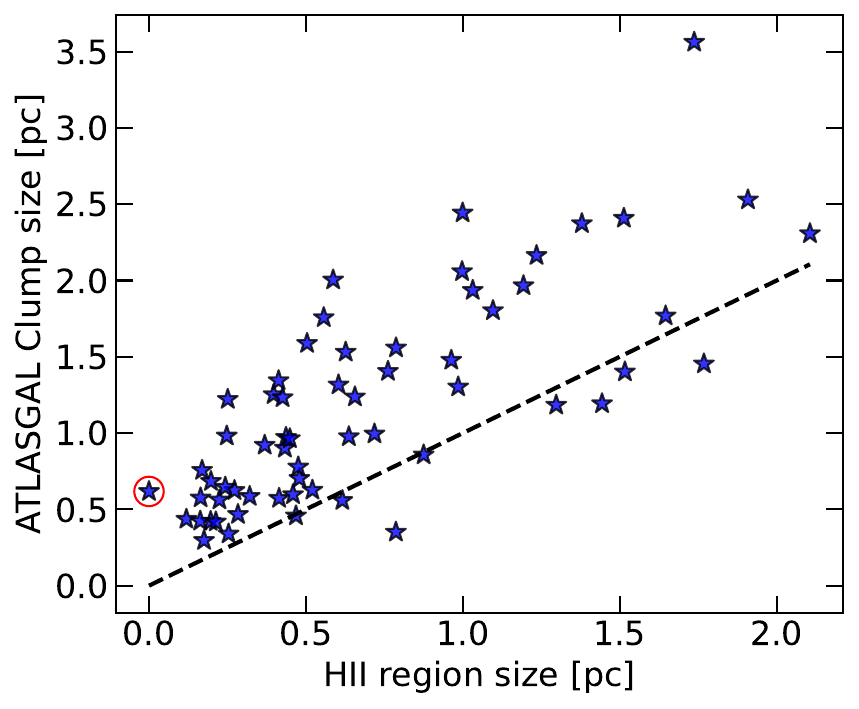}
                \caption{Comparison of physical size of \hii regions and associated ATLASGAL clump size taken from \cite{2014A&A...565A..75C}. The black dashed line shows the line of equality. The red circle represents only unresolved \hii region, G030.720$-$0.082, in our sample.}
                \label{fig:atlas_hii_size}
            \end{figure*}
            \begin{figure*}
               \centering
                \includegraphics[width=0.45\linewidth]{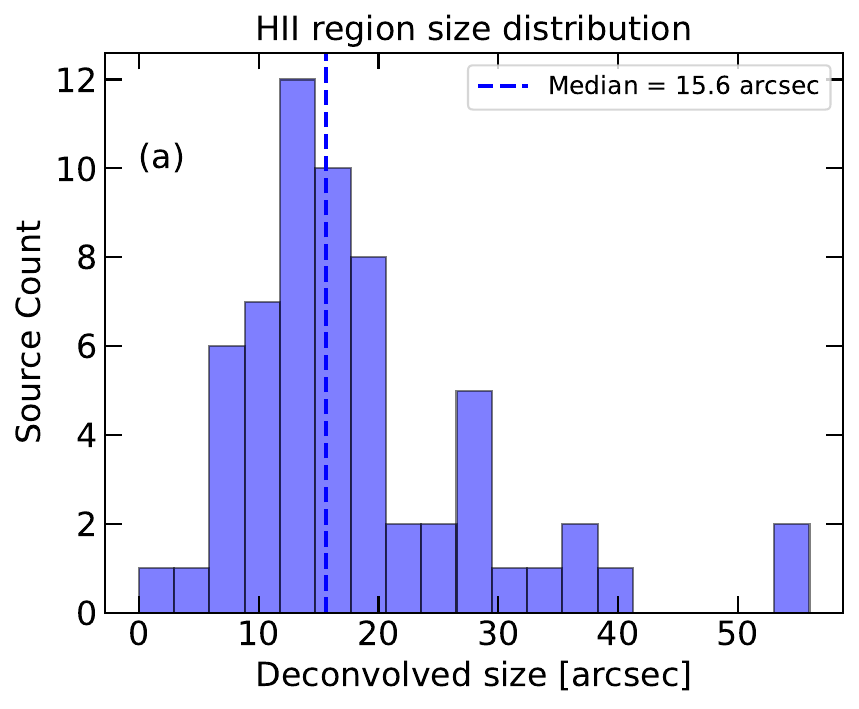}
                \includegraphics[width=0.45\linewidth]{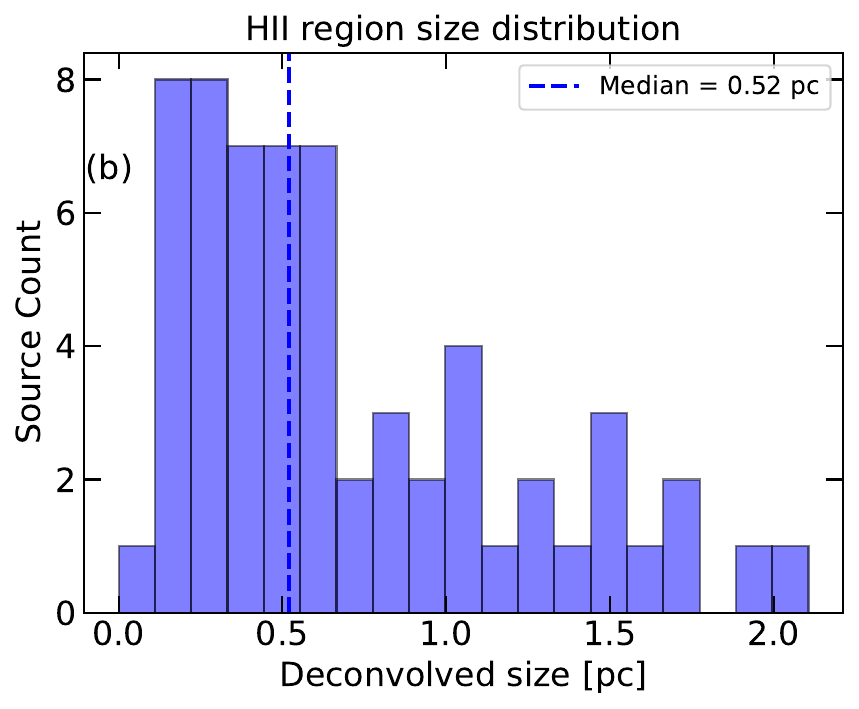}
                \caption{Distribution of (a) angular and (b) physical deconvolved FHWM size of \hii region. The vertical dashed line shows the median value of the distribution.}
                \label{fig:size_component}                
            \end{figure*}
            
            \begin{figure*}
                \centering
                \includegraphics[width=0.45\linewidth]{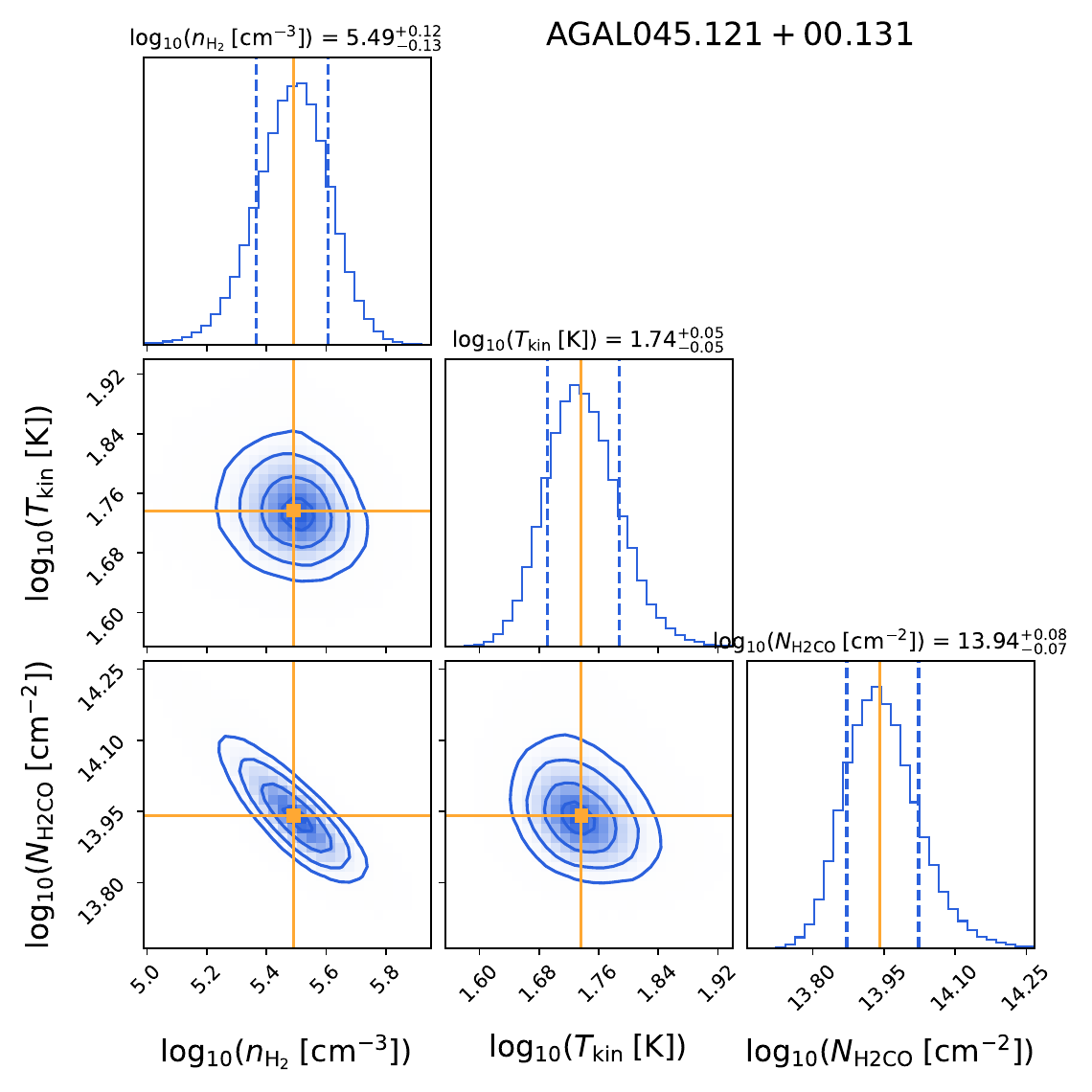}
                \includegraphics[width=0.45\linewidth]{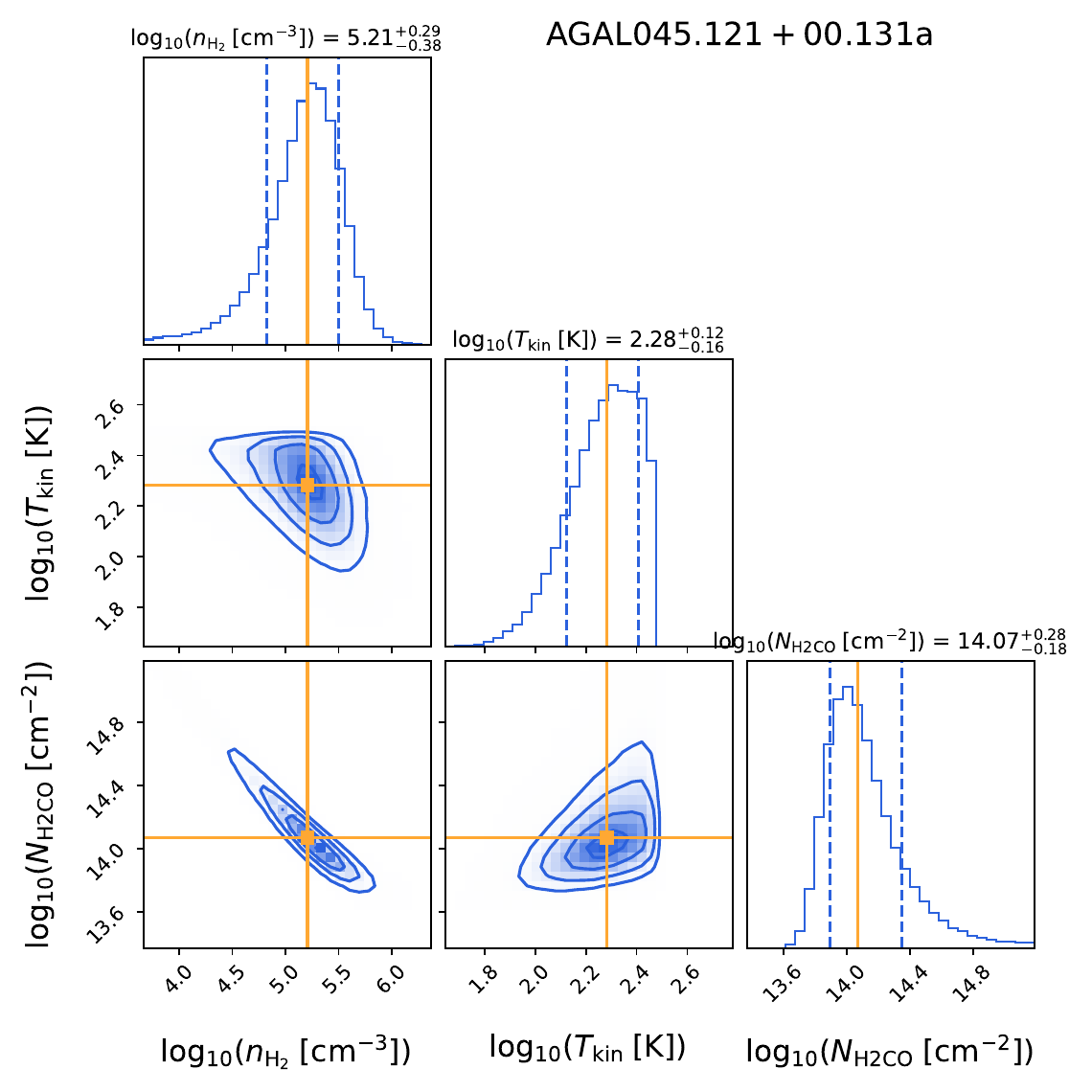}
                \caption{Example corner plot for AGAL045.121+00.131. To constrain physical properties of the molecular clouds we used the non-LTE RADEX modelling with MCMC algorithm, here we show the corresponding corner plot for each Gaussian component displaying the one-dimensional (1D) histogram of the posterior distribution for each parameter (kinetic temperature, $T_{\rm kin}$, H$\rm _2$ density, \nh, and \phtco column density, \cd). The histograms represent the probability density functions (PDFs) of the Monte Carlo fit parameters, while the blue curves represent the scatter plot of the joint distribution between the two parameters. The orange solid lines indicate the median of the PDFs, representing the most likely fit parameters, and the dotted blue lines correspond to the 1$\sigma$ confidence intervals. The sub-title of each corner plot shows most likely value of the parameters with upper and lower 1$\sigma$ error.}
                \label{fig:corner_plot_eg}
            \end{figure*}

            \begin{figure*}[ht!]
                \centering
                \includegraphics[width=0.75\linewidth]{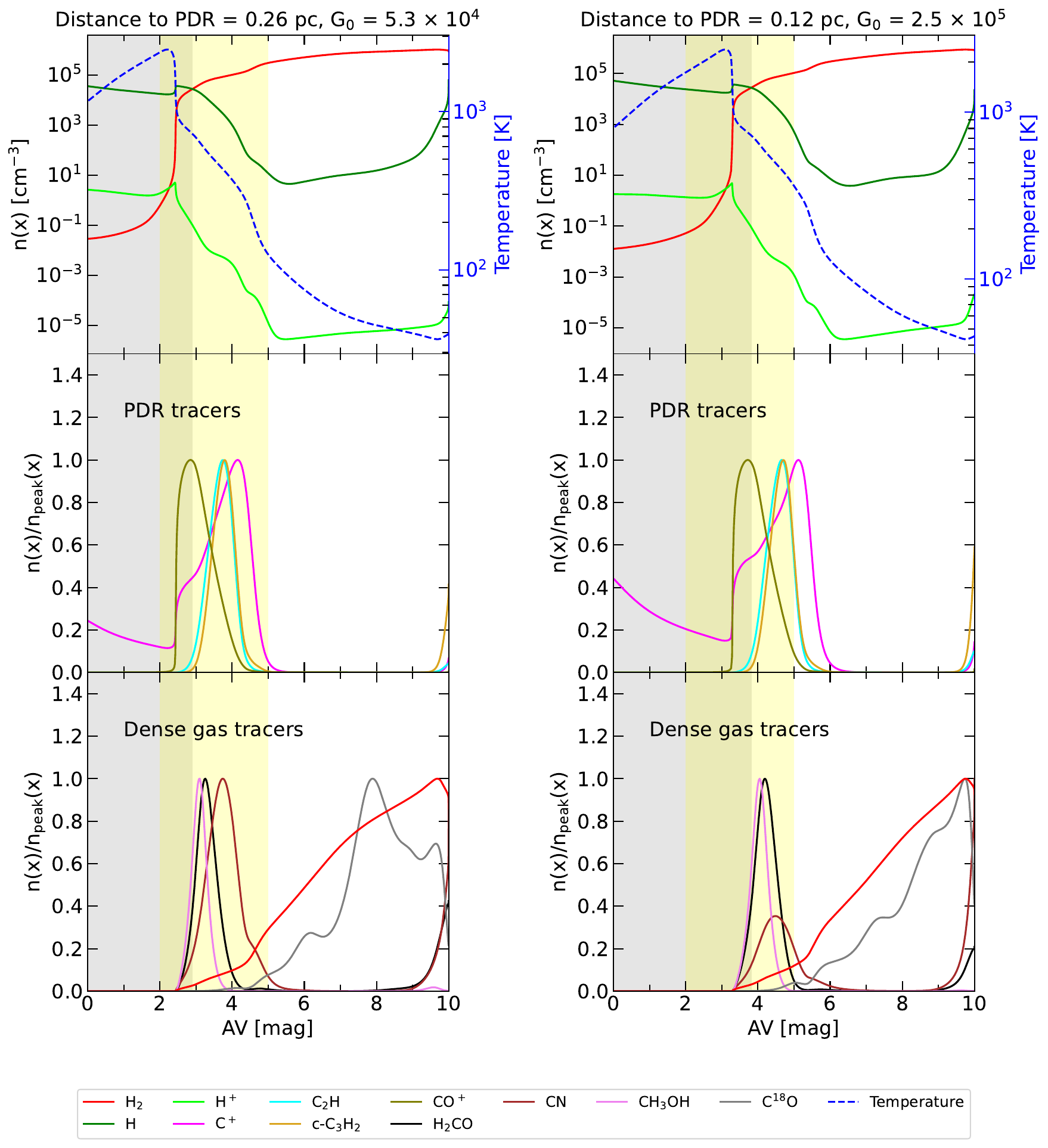}
                \caption{The figure shows the Meudon PDR model results for a constant thermal pressure of $P_{\rm th}$ $\sim$ $\rm 4.6\times10^{7}\,K\,cm^{-3}$. (\textit{Left panel:}) This model assumes a central star of spectral type O5V, with the distance between the stellar surface and the PDR surface set to $d_0$ = 0.26\,pc, resulting in a stellar far-UV flux of G$_0$ $\sim$ $\rm 5.3\times10^4$ in units of Habing field (see text). The \textit{top plot} shows the evolution of hydrogen species -- H$_2$, H, and H$^+$, as well as the gas temperature as a function of visual extinction $A_{\rm V}$. The \textit{middle and bottom panels} present the spatial density profiles of key PDR and dense gas tracer  molecules, where each density has been normalized to its maximum value within the cloud. (\textit{Right panel:}) This case uses the same stellar type (O5V) but assumes a smaller distance of $d_0$ = 0.12\,pc between the star and the PDR surface, leading to a higher far-UV flux of G$_0$ $\sim$ $\rm 2.5\times10^5$ in Habing units. The \textit{left and right plots} display the same quantities as in the top panel for this more compact configuration.}
                \label{fig:sys_hii}
            \end{figure*} 

    \section{Effect of distance on source selection}\label{sec:dist_eff}
        \begin{figure*}
            \centering
            \includegraphics[width=0.45\linewidth]{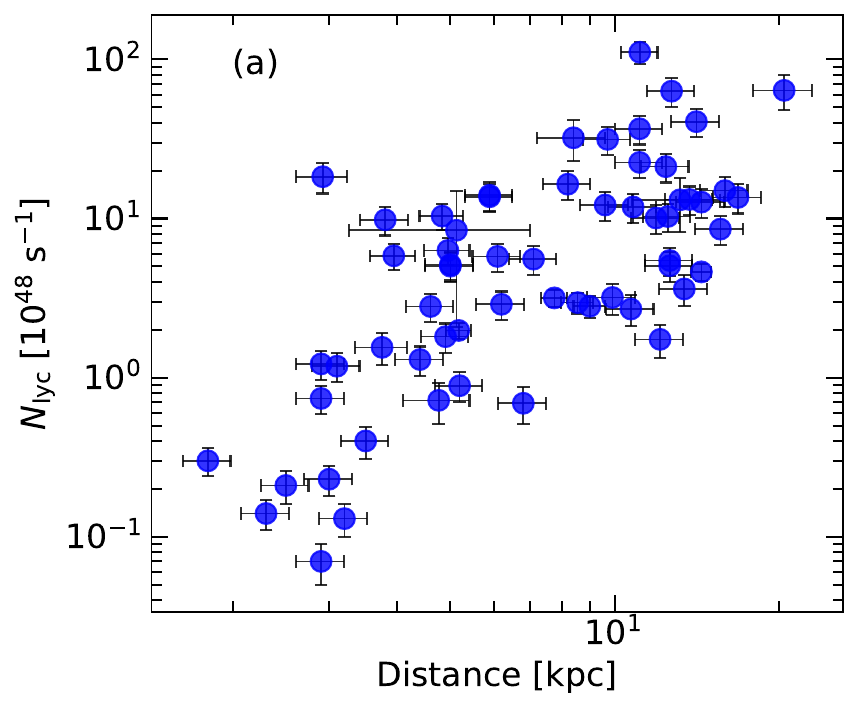}
            \includegraphics[width=0.45\linewidth]{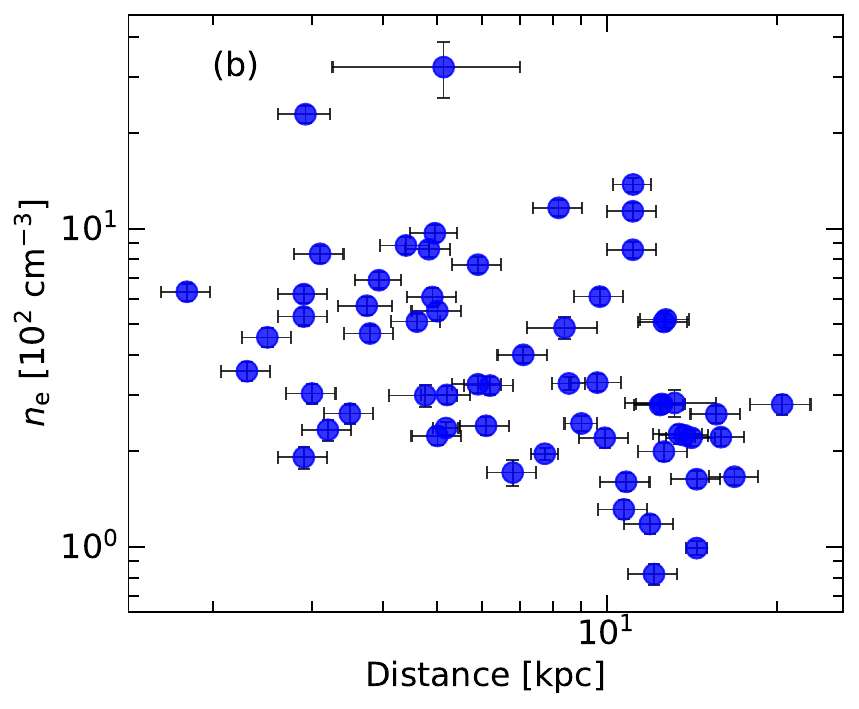}
            \includegraphics[width=0.45\linewidth]{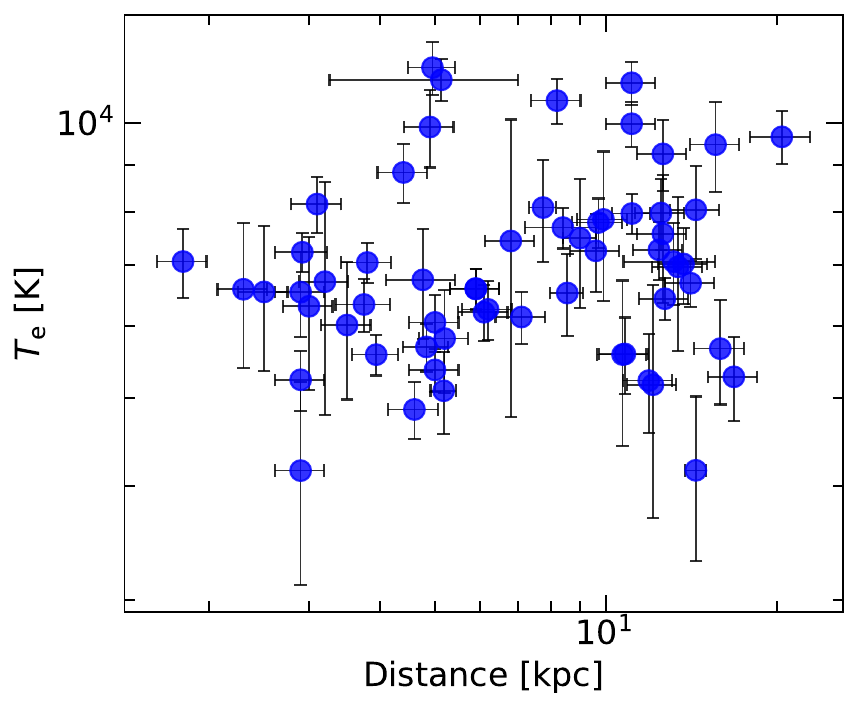}
            \caption{Relation between the distances and (a) $N_{\rm Lyc}$, (b) $n_{\rm e}$, and (c) $T_{\rm e}$ for our sample.}
            \label{fig:phy_prop_vs_dist}
        \end{figure*}

        In Sect.~\ref{sec:source_selection}, we outlined the criteria used to select the sources, and Fig.~\ref{fig:dist} illustrates their distance distribution. To compare the physical conditions of the ionized and dense molecular gas, we adopted ionized gas parameters from Paper~I. Because these parameters vary with distance, we examined the distance dependence within our sample. Figure~\ref{fig:phy_prop_vs_dist} shows the trends of $N_{\rm Lyc}$, $n_{\rm e}$, and $T_{\rm e}$ as a function of distance. While $n_{\rm e}$ exhibits only a weak dependence and $T_{\rm e}$ remains largely independent of distance, $N_{\rm Lyc}$ increases systematically, reflecting its strong distance dependence. This spread in $N_{\rm Lyc}$ indicates that our sample encompasses \hii regions powered by a wide range of spectral type of central OB star. Overall, the distance variation introduces no significant bias in the derived ionized gas properties but rather ensures a diverse and representative sample.

    \section{Tables}

    \begin{table*}[ht!]
        \centering
        \caption{Line fitting parameters of the \phtco lines observed with nFLASH230.}
        \label{tab:nflash_line}
        \resizebox{\linewidth}{!}{%
        \begin{tabular}{ccccc|ccc|ccc|ccc}
            \hline \hline
             GLOSTAR & ATLASGAL & Comp\tablefootmark{1} & $\eta_{\rm nFLASH230}$ & $\upsilon_{\rm lsr}$ &
            \multicolumn{3}{c|}{\phtco ($\rm3_{0,3}-2_{0,2}$)} &
            \multicolumn{3}{c|}{\phtco ($\rm3_{2,2}-2_{2,1}$)} &
            \multicolumn{3}{c}{\phtco ($\rm3_{2,1}-2_{2,0}$)} \\
            \cline{6-8} \cline{9-11} \cline{12-14}
            & & & & &
            $T_{\rm peak}$ & $I$ & FWHM &
            $T_{\rm peak}$ & $I$ & FWHM &
            $T_{\rm peak}$ & $I$ & FWHM \\
            & & & & [km\,s$^{-1}$] &
            [K]& [K\,km\,s$^{-1}$] & [km\,s$^{-1}$] &
            [K]& [K\,km\,s$^{-1}$] & [km\,s$^{-1}$] &
            [K]& [K\,km\,s$^{-1}$] & [km\,s$^{-1}$] \\
            \hline \hline
            G3.350$-$0.077 & AGAL003.351-00.077 & a & 0.55 & 7.47 & 0.38 $\pm$ 0.03 & 1.27 $\pm$ 0.16 & 3.12 $\pm$ 0.28 &   &   &   &   &   &   \\ 
            G3.350$-$0.077 & AGAL003.351-00.077 &   & 0.55 & 10.77 & 0.90 $\pm$ 0.02 & 3.99 $\pm$ 0.08 & 4.15 $\pm$ 0.6 & 0.14 $\pm$ 0.02 & 0.61 $\pm$ 0.07 & 4.15 $\pm$ 0.6 & 0.20 $\pm$ 0.02 & 0.90 $\pm$ 0.07 & 4.15 $\pm$ 0.6\\
            G6.552$-$0.095 & AGAL006.551-00.097 &   & 0.65 & 12.94 & 0.49 $\pm$ 0.01 & 1.69 $\pm$ 0.08 & 3.25 $\pm$ 0.12 & 0.07 $\pm$ 0.01 & 0.20 $\pm$ 0.06 & 2.69 $\pm$ 0.66 & 0.09 $\pm$ 0.02 & 0.26 $\pm$ 0.07 & 2.61 $\pm$ 0.54\\
            G6.552$-$0.095 & AGAL006.551-00.097 & a & 0.65 & 17.06 & 0.78 $\pm$ 0.02 & 2.48 $\pm$ 0.08 & 3.00 $\pm$ 0.07 & 0.13 $\pm$ 0.01 & 0.52 $\pm$ 0.08 & 3.83 $\pm$ 0.46 & 0.13 $\pm$ 0.01 & 0.57 $\pm$ 0.10 & 4.29 $\pm$ 0.58\\
            G7.472+0.058 & AGAL007.471+00.059 &   & 0.63 & -14.41 & 0.72 $\pm$ 0.02 & 3.33 $\pm$ 0.13 & 4.31 $\pm$ 0.13 & 0.13 $\pm$ 0.01 & 0.62 $\pm$ 0.10 & 4.51 $\pm$ 0.56 & 0.14 $\pm$ 0.01 & 0.61 $\pm$ 0.10 & 4.03 $\pm$ 0.48\\
            G7.177+0.088 & AGAL007.178+00.086 &   & 0.46 & 6.26 & 0.55 $\pm$ 0.02 & 2.65 $\pm$ 0.08 & 4.53 $\pm$ 0.6 & 0.09 $\pm$ 0.02 & 0.45 $\pm$ 0.07 & 4.53 $\pm$ 0.6 & 0.13 $\pm$ 0.02 & 0.64 $\pm$ 0.07 & 4.53 $\pm$ 0.6\\
            G8.669$-$0.355 & AGAL008.671-00.356 &   & 0.49 & 33.92 & 1.61 $\pm$ 0.12 & 7.84 $\pm$ 0.65 & 4.58 $\pm$ 0.16 & 0.27 $\pm$ 0.61 & 2.45 $\pm$ 6.27 & 8.48 $\pm$ 10.25 & 0.22 $\pm$ 0.04 & 3.32 $\pm$ 0.72 & 14.14 $\pm$ 1.60\\
            G8.669$-$0.355 & AGAL008.671-00.356 & a & 0.49 & 34.43 & 3.03 $\pm$ 0.13 & 8.71 $\pm$ 0.43 & 2.70 $\pm$ 0.07 & 1.12 $\pm$ 0.67 & 3.68 $\pm$ 2.96 & 3.10 $\pm$ 1.66 & 1.37 $\pm$ 0.05 & 4.92 $\pm$ 0.25 & 3.37 $\pm$ 0.13\\
            G8.669$-$0.355 & AGAL008.671-00.356 & b & 0.49 & 36.13 & 1.67 $\pm$ 0.04 & 17.89 $\pm$ 0.53 & 10.05 $\pm$ 0.16 & 0.96 $\pm$ 0.34 & 4.47 $\pm$ 2.19 & 4.38 $\pm$ 1.46 & 1.02 $\pm$ 0.04 & 4.83 $\pm$ 0.27 & 4.47 $\pm$ 0.18\\
            G9.875$-$0.749 & AGAL009.879-00.751 & a & 0.75 & 28.95 & 0.24 $\pm$ 0.03 & 2.35 $\pm$ 0.31 & 9.13 $\pm$ 0.64 &   &   &   & 0.06 $\pm$ 0.04 & 0.25 $\pm$ 0.17 & 3.65 $\pm$ 1.46\\
            \hline \hline
        \end{tabular}}
        \tablefoot{\tablefootmark{1}{a,b,c: indicates the secondary velocity components.} \\ The table presents the first ten sources from our sample. The entire table is accessible at the CDS.}
    \end{table*}

    \begin{table*}[ht!]
        \centering
        \caption{Line fitting parameters of the \phtco lines observed with SEPIA345.}
        \label{tab:sepia_line}
        \resizebox{\linewidth}{!}{%
        \begin{tabular}{ccccc|ccc|ccc|ccc}
            \hline \hline
             GLOSTAR & ATLASGAL & Comp\tablefootmark{1} & $\eta_{\rm SEPIA345}$ & $\upsilon_{\rm lsr}$ &
            \multicolumn{3}{c|}{\phtco ($\rm4_{0,4}-3_{0,3}$)} &
            \multicolumn{3}{c|}{\phtco ($\rm4_{2,3}-3_{2,2}$)} &
            \multicolumn{3}{c}{\phtco ($\rm4_{2,2}-3_{2,1}$)} \\
            \cline{6-8} \cline{9-11} \cline{12-14}
            & & & & &
            $T_{\rm peak}$ & $I$ & FWHM &
            $T_{\rm peak}$ & $I$ & FWHM &
            $T_{\rm peak}$ & $I$ & FWHM \\
            & & & & [km\,s$^{-1}$] &
            [K]& [K\,km\,s$^{-1}$] & [km\,s$^{-1}$] &
            [K]& [K\,km\,s$^{-1}$] & [km\,s$^{-1}$] &
            [K]& [K\,km\,s$^{-1}$] & [km\,s$^{-1}$] \\
            \hline \hline

        G3.350$-$0.077 & AGAL003.351-00.077 & a & 0.68 & 7.47 & 0.17 $\pm$ 0.02 & 0.64 $\pm$ 0.09 & 3.49 $\pm$ 0.34 & 0.06 $\pm$ 0.01 & 0.25 $\pm$ 0.09 & 3.91 $\pm$ 1.09 &  &  & \\
        G3.350$-$0.077 & AGAL003.351-00.077 &  & 0.68 & 10.77 & 0.48 $\pm$ 0.01 & 2.13 $\pm$ 0.05 & 4.15 $\pm$ 0.6 & 0.11 $\pm$ 0.01 & 0.49 $\pm$ 0.04 & 4.15 $\pm$ 0.6 & 0.11 $\pm$ 0.01 & 0.48 $\pm$ 0.04 & 4.15 $\pm$ 0.6 \\
        G6.552$-$0.095 & AGAL006.551-00.097 &  & 0.76 & 12.94 & 0.33 $\pm$ 0.01 & 1.0 $\pm$ 0.06 & 2.87 $\pm$ 0.13 & 0.06 $\pm$ 0.01 & 0.21 $\pm$ 0.06 & 3.5 $\pm$ 0.79 & 0.06 $\pm$ 0.01 & 0.2 $\pm$ 0.06 & 2.94 $\pm$ 0.71 \\
        G6.552$-$0.095 & AGAL006.551-00.097 & a & 0.76 & 17.06 & 0.36 $\pm$ 0.01 & 1.14 $\pm$ 0.06 & 2.96 $\pm$ 0.12& 0.09 $\pm$ 0.01 & 0.31 $\pm$ 0.06 & 3.11 $\pm$ 0.45 & 0.09 $\pm$ 0.01 & 0.35 $\pm$ 0.07 & 3.64 $\pm$ 0.59 \\
        G7.472+0.058 & AGAL007.471+00.059 &  & 0.75 & -14.41 & 0.37 $\pm$ 0.01 & 1.48 $\pm$ 0.07 & 3.74 $\pm$ 0.13 & 0.11 $\pm$ 0.01 & 0.34 $\pm$ 0.06 & 3.04 $\pm$ 0.39 & 0.09 $\pm$ 0.01 & 0.43 $\pm$ 0.07 & 4.67 $\pm$ 0.56 \\
        G7.177+0.088 & AGAL007.178+00.086 &  & 0.6 & 6.26 & 0.31 $\pm$ 0.01 & 1.47 $\pm$ 0.06 & 4.53 $\pm$ 0.6 &    &    &    & 0.09 $\pm$ 0.01 & 0.41 $\pm$ 0.06 & 4.53 $\pm$ 0.6 \\
        G8.669$-$0.355 & AGAL008.671-00.356 &  & 0.63 & 33.92 & 0.61 $\pm$ 0.07 & 5.64 $\pm$ 0.72 & 8.7 $\pm$ 0.53 & 0.34 $\pm$ 0.04 & 3.45 $\pm$ 0.51 & 9.43 $\pm$ 0.68 & 0.22 $\pm$ 0.03 & 2.57 $\pm$ 0.45 & 11.04 $\pm$ 0.99 \\
        G8.669$-$0.355 & AGAL008.671-00.356 & a & 0.63 & 34.43 & 3.73 $\pm$ 0.07 & 13.19 $\pm$ 0.34 & 3.32 $\pm$ 0.06 & 1.47 $\pm$ 0.05 & 4.88 $\pm$ 0.23 & 3.11 $\pm$ 0.1 & 1.53 $\pm$ 0.04 & 5.07 $\pm$ 0.18 & 3.1 $\pm$ 0.08 \\
        G8.669$-$0.355 & AGAL008.671-00.356 & b & 0.63 & 36.13 & 1.1 $\pm$ 0.04 & 8.1 $\pm$ 0.36 & 6.94 $\pm$ 0.19 & 0.55 $\pm$ 0.03 & 2.26 $\pm$ 0.17 & 3.88 $\pm$ 0.21 & 0.63 $\pm$ 0.03 & 2.96 $\pm$ 0.18 & 4.42 $\pm$ 0.19 \\
        G9.875$-$0.749 & AGAL009.879-00.751 & a & 0.84  &  &  &  &  &  &  &  &  &  &  \\ 
        
        \hline \hline 
        \end{tabular}}
        \tablefoot{\tablefootmark{1}{a,b,c: indicates the secondary velocity components.} \\ The table presents the first ten sources from our sample. The entire table is accessible at the CDS.}
   \end{table*}

    \begin{table*}[ht!]
        \centering
        \caption{Derived physical properties of molecular clouds.}
        \label{tab:phy_prop}
        \resizebox{\linewidth}{!}{\begin{tabular}{cccccccc}
        \hline \hline \\
            GLOSTAR & ATLASGAL & Comp\tablefootmark{1} & $T_{\rm kin}$ & $log10(n_{\rm H_2})$ & $log10(N_{\rm para-H_2CO})$ & $log10(X_{\rm para-H_2CO})$ & $log10(N_{\rm H_2})$ \\
            &  &  & [K] & [cm$^{-3}$] & [cm$^{-2}$] &  & [cm$^{-2}$] \\ \\
        \hline \hline \\

        G3.350$-$0.077 & AGAL003.351$-$00.077 & a & 111.25$\rm ^{+1.89}_{-0.49}$ & 4.97$\rm ^{+0.51}_{-0.73}$ & 13.43$\rm ^{+0.6}_{-0.36}$ & -8.59 & 22.02 \\ \\
        G3.350$-$0.077 & AGAL003.351$-$00.077 &  & 47.2$\rm ^{+1.1}_{-0.92}$ & 5.44$\rm ^{+0.12}_{-0.13}$ & 13.69$\rm ^{+0.09}_{-0.08}$ & -8.32 & 22.02 \\ \\
        G6.552$-$0.095 & AGAL006.551$-$00.097 &  & 40.24$\rm ^{+1.15}_{-0.86}$ & 5.71$\rm ^{+0.18}_{-0.18}$ & 13.05$\rm ^{+0.09}_{-0.08}$ & -9.12 & 22.17 \\ \\
        G6.552$-$0.095 & AGAL006.551$-$00.097 & a & 59.25$\rm ^{+1.21}_{-0.87}$ & 5.14$\rm ^{+0.19}_{-0.43}$ & 13.57$\rm ^{+0.35}_{-0.14}$ & -8.61 & 22.17 \\ \\
        G7.472+0.058 & AGAL007.471+00.059 &  & 52.11$\rm ^{+1.23}_{-0.88}$ & 5.17$\rm ^{+0.2}_{-0.72}$ & 13.74$\rm ^{+0.61}_{-0.15}$ & -8.55 & 22.29 \\ \\
        G7.177+0.088 & AGAL007.178+00.086 &  & 54.77$\rm ^{+1.15}_{-0.88}$ & 5.49$\rm ^{+0.14}_{-0.16}$ & 13.52$\rm ^{+0.1}_{-0.08}$ & -7.25 & 20.77 \\ \\
        G8.669$-$0.355 & AGAL008.671$-$00.356 &  & 145.45$\rm ^{+1.42}_{-0.72}$ & 5.33$\rm ^{+0.27}_{-0.43}$ & 14.53$\rm ^{+0.28}_{-0.14}$ & -7.68 & 22.21 \\ \\
        G8.669$-$0.355 & AGAL008.671$-$00.356 & a & 121.81$\rm ^{+1.42}_{-0.67}$ & 5.94$\rm ^{+0.8}_{-0.24}$ & 14.09$\rm ^{+0.07}_{-0.03}$ & -8.13 & 22.21 \\ \\
        G8.669$-$0.355 & AGAL008.671$-$00.356 & b &  &  &  &  & 22.21 \\ \\
        G9.875$-$0.749 & AGAL009.879$-$00.751 & a &  &  &  &  & 21.63 \\ \\

        \hline \hline 
        \end{tabular}}
        \tablefoot{\tablefootmark{1}{a,b,c: indicates the secondary velocity components.} \\ The table presents the first ten sources from our sample. The entire table is accessible at the CDS.}
   \end{table*}

        \section{Optical depth}\label{sec:optical_depth}  
\begin{figure*}
                \centering
                \includegraphics[width=0.45\linewidth]{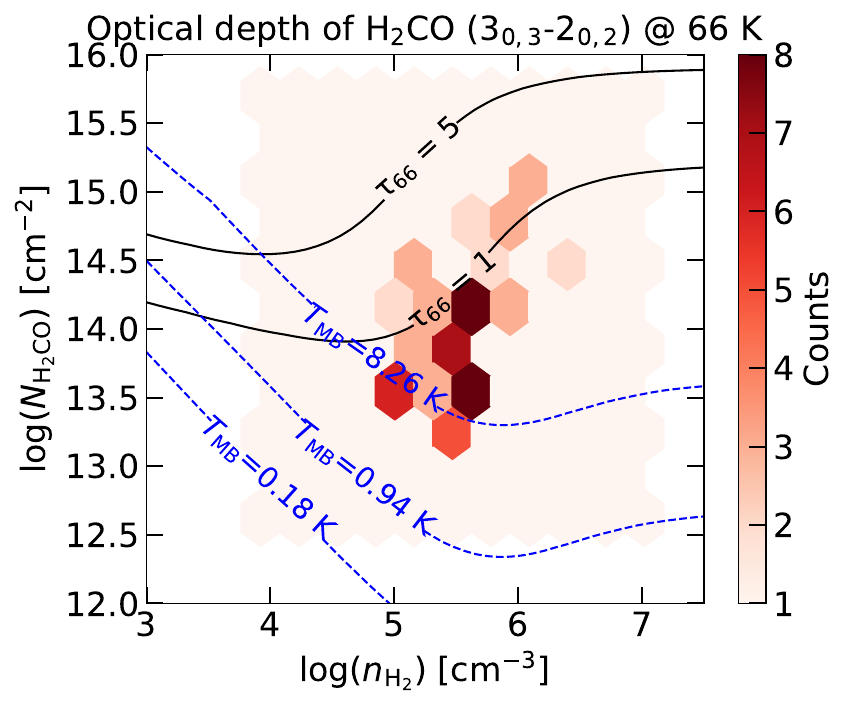}
                \includegraphics[width=0.45\linewidth]{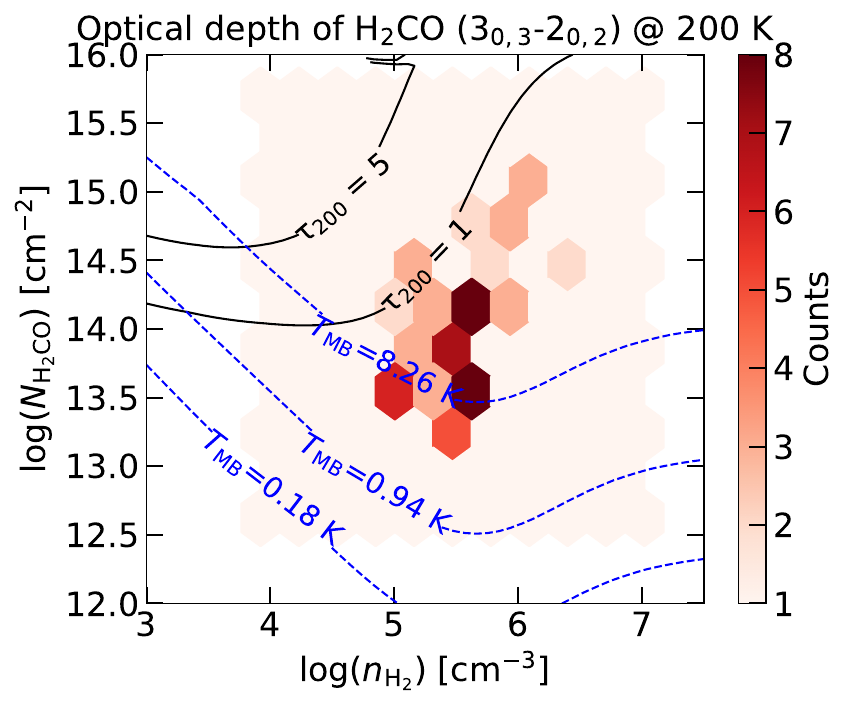}

                \caption{The color hexagonally binned distributions display the PyRADEX+MCMC constrained \nh and \cd of our sources. The black solid curves show the optical depth of $\tau = 1$ and 5 for \phtco \lal transition on top of \nh and \cd space, assuming fixed kinetic temperature of 66\,K (the median temperature of our sample; top panel) and 200\,K (higher temperature; bottom temperature). The blue dashed curves indicate the contour corresponding to the brightest, average, and faintest observed intensities for \phtco \lal transition (see the text).}
                \label{fig:tau}
            \end{figure*}

        Since much of the analysis presented above assumes optically thin conditions, we here examine the validity of this assumption. To assess the influence of optical depth on the derived quantities, we display in Fig.~\ref{fig:tau} the PyRADEX+MCMC reproduced optical depths, $\tau$, for the observed \phtco \lal transitions (which has the lowest value of $E_{\rm u}$ and the brightest observed transition) across the range of gas densities and \phtco column densities probed but for fixed values of \tkin, at median \tkin of 66\,K (Fig.~\ref{fig:tau} left) and at a higher \tkin of 200\,K (Fig.~\ref{fig:tau} right).
            
        The background color scale represents the counts visualized over hexagonally binned distributions across the derived \nh -- \cd space for our sample. The solid black curves mark contours of the optical depth equal to 1 and 5 at \tkin of 66 and 200\,K. The blue dashed curves represent the beam filling factor corrected line intensities of the \phtco \lal, corresponding to the brightest ($T_{\rm MB}$ =8.26\,K), median ($T_{\rm MB}$ =0.94\,K), and the lowest ($T_{\rm MB}$ =0.18\,K) detection in our sample. We find that all the observed \phtco transitions are optically thin ($\rm \tau < 1$) at column densities \cd $\rm <~10^{14}~cm^{-2}$ and spatial densities \nh = $\rm 10^{3-8}~cm^{-3}$. We also notice that the transition becomes very optically thick ($\tau>5$), at column densities higher than $\rm ~5\times10^{15}~cm^{-2}$. The spatial density and column density derived using the observed \phtco line intensities, shown as colored hexagonal bins in Fig.~\ref{fig:tau}, lie predominantly in or near the optically thin region ($\tau<1$). Among the observed \phtco transitions, the low-lying \phtco \lal line is the brightest, while the other transitions, which have higher, upper energy levels ($E_{\rm u} > 34$\,K; see Table~\ref{tab:transitions}), are expected to have much lower optical depths ($\tau \ll 1$). This indicates that optical depth effects have only a minimal impact on the derived spatial densities and kinetic temperatures from the \phtco transitions. Furthermore, as was found by \citet{2018A&A...611A...6T}, varying the kinetic temperature only weakly affects the optical depth of the \phtco lines as evident from Fig.~\ref{fig:tau}. 
            
	\end{appendix}
	
\end{document}